\documentclass[aps,pra,floatfix,onecolumn,tightenlines,superscriptaddress,notitlepage]{revtex4-1}
\usepackage{graphicx}
\usepackage{amsmath}
\usepackage{latexsym}
\usepackage{graphicx}
\usepackage{multirow}
\usepackage{url}
\usepackage[bookmarks, colorlinks=true, pdftitle={Macroscopic quantum resonators},
            pdftex
            ]{hyperref}

\newcommand{\ci}{\mathrm{i}}

\newcommand{\glei}[1]{(\ref{#1})}

\newcommand{\rkl}[1]{\left( #1\right)}
\newcommand{\ekl}[1]{\left[ #1\right]}

\begin{document}

\title{Macroscopic quantum resonators (MAQRO): 2015 Update}

\author{Rainer Kaltenbaek}
\email[Corresponding author: ]{rainer.kaltenbaek@univie.ac.at}
\affiliation{Vienna Center for Quantum Science and Technology, University of Vienna, Austria}

\author{Markus Arndt}
\affiliation{Vienna Center for Quantum Science and Technology, University of Vienna, Austria}

\author{Markus Aspelmeyer}
\affiliation{Vienna Center for Quantum Science and Technology, University of Vienna, Austria}

\author{Peter F. Barker}
\affiliation{Department of Physics and Astronomy, University College London, United Kingdom}

\author{Angelo Bassi}
\affiliation{Department of Physics, University of Trieste, and INFN, Italy}

\author{James Bateman}
\affiliation{Physics and Astronomy, University of Southampton, United Kingdom}

\author{Kai Bongs}
\affiliation{School of Physics and Astronomy, University of Birmingham, United Kingdom}

\author{Sougato Bose}
\affiliation{Department of Physics and Astronomy, University College London, United Kingdom}

\author{Claus Braxmaier}
\affiliation{Center of Applied Space Technology and Micro Gravity (ZARM), University of Bremen, Germany}
\affiliation{German Aerospace Center (DLR), Institute for Space Systems, Bremen, Germany}

\author{\v{C}aslav Brukner}
\affiliation{Vienna Center for Quantum Science and Technology, University of Vienna, Austria}
\affiliation{Institute of Quantum Optics and Quantum Information (IQOQI), Austrian Academy of Sciences, Austria}

\author{Bruno Christophe}
\affiliation{ONERA, The French Aerospace Lab, Ch{\^a}tillon, France}

\author{Michael Chwalla}
\affiliation{Airbus Defence \& Space Friedrichshafen, Germany}

\author{Pierre-Fran\c{c}ois Cohadon}
\affiliation{Laboratoire Kastler Brossel, UPMC-Sorbonne Universit{\'e}s, CNRS, ENS-PSL Research University, Coll{\`e}ge de France, Paris, France}

\author{Adrian M. Cruise}
\affiliation{School of Physics and Astronomy, University of Birmingham, United Kingdom}

\author{Catalina Curceanu}
\affiliation{Laboratori Nazionali di Frascati dell'INFN, Italy}

\author{Kishan Dholakia}
\affiliation{School of Physics and Astronomy, University of St. Andrews, United Kingdom}

\author{Klaus D\"{o}ringshoff}
\affiliation{Institut f\"ur Physik, Humboldt-Universit\"at zu Berlin, Germany}

\author{Wolfgang Ertmer}
\affiliation{Institut f\"ur Quantenoptik, Leibniz Universit\"at Hannover, Germany}

\author{Jan Gieseler}
\affiliation{Photonics Laboratory, ETH Zurich, Zurich, Switzerland}

\author{Norman G\"{u}rlebeck}
\affiliation{Center of Applied Space Technology and Micro Gravity (ZARM), University of Bremen, Germany}

\author{Gerald Hechenblaikner}
\affiliation{European Southern Observatory, Munich, Germany}

\author{Antoine Heidmann}
\affiliation{Laboratoire Kastler Brossel, UPMC-Sorbonne Universit{\'e}s, CNRS, ENS-PSL Research University, Coll{\`e}ge de France, Paris, France}

\author{Sven Herrmann}
\affiliation{Center of Applied Space Technology and Micro Gravity (ZARM), University of Bremen, Germany}

\author{Sabine Hossenfelder}
\affiliation{Nordita, KTH Royal Institute of Technology and Stockholm University, Sweden}

\author{Ulrich Johann}
\affiliation{Airbus Defence \& Space Friedrichshafen, Immenstaad, Germany}

\author{Nikolai Kiesel}
\affiliation{Vienna Center for Quantum Science and Technology, University of Vienna, Austria}

\author{Myungshik Kim}
\affiliation{QOLS, Blackett Laboratory, Imperial College London, United Kingdom}

\author{Claus L\"{a}mmerzahl}
\affiliation{Center of Applied Space Technology and Micro Gravity (ZARM), University of Bremen, Germany}

\author{Astrid Lambrecht}
\affiliation{Laboratoire Kastler Brossel, UPMC-Sorbonne Universit{\'e}s, CNRS, ENS-PSL Research University, Coll{\`e}ge de France, Paris, France}

\author{Michael Mazilu}
\affiliation{School of Physics and Astronomy, University of St. Andrews, United Kingdom}

\author{Gerard J. Milburn}
\affiliation{ARC Centre for Engineered Quantum Systems, University of Queensland, Australia}

\author{Holger M\"{u}ller}
\affiliation{Department of Physics, University of California, Berkeley, USA}

\author{Lukas Novotny}
\affiliation{Photonics Laboratory, ETH Zurich, Zurich, Switzerland}

\author{Mauro Paternostro}
\affiliation{Centre for Theoretical Atomic, Molecular and Optical Physics, School of Mathematics and Physics, Queen's University, Belfast, United Kingdom}

\author{Achim Peters}
\affiliation{Institut f\"ur Physik, Humboldt-Universit\"at zu Berlin, Germany}

\author{Igor Pikovski}
\affiliation{ITAMP, Harvard-Smithsonian Center for Astrophysics,  USA}

\author{Andr{\'e} Pilan-Zanoni}
\affiliation{Airbus Defence \& Space Friedrichshafen, Germany}

\author{Ernst M. Rasel}
\affiliation{Institut f\"ur Quantenoptik, Leibniz Universit\"at Hannover, Germany}

\author{Serge Reynaud}
\affiliation{Laboratoire Kastler Brossel, UPMC-Sorbonne Universit{\'e}s, CNRS, ENS-PSL Research University, Coll{\`e}ge de France, Paris, France}

\author{C. Jess Riedel}
\affiliation{Perimeter Institute for Theoretical Physics, Canada}

\author{Manuel Rodrigues}
\affiliation{ONERA, The French Aerospace Lab, Ch{\^a}tillon, France}

\author{Lo\"\i{}c Rondin}
\affiliation{Photonics Laboratory, ETH Zurich, Zurich, Switzerland}

\author{Albert Roura}
\affiliation{Institute f{\"u}r Quantenphysik, Universit{\"a}t Ulm, Germany}

\author{Wolfgang P. Schleich}
\affiliation{Institute f{\"u}r Quantenphysik, Universit{\"a}t Ulm, Germany}
\affiliation{Texas A \& M University Institute for Advanced Study, Institute for Quantum Science and Engineering and Department of Physics and Astronomy, Texas A \& M University, USA}

\author{J\"{o}rg Schmiedmayer}
\affiliation{Vienna Center for Quantum Science and Technology, Technical University of Vienna, Austria}

\author{Thilo Schuldt}
\affiliation{German Aerospace Center (DLR), Institute for Space Systems, Bremen, Germany}

\author{Keith C. Schwab}
\affiliation{Applied Physics, California Institute of Technology, USA}

\author{Martin Tajmar}
\affiliation{Institute of Aerospace Engineering, Technische Universit{\"a}t Dresden, Germany}

\author{Guglielmo M. Tino}
\affiliation{Dipartimento di Fisica e Astronomia and LENS, Universit{\'a} di Firenze, INFN, Italy}

\author{Hendrik Ulbricht}
\affiliation{Physics and Astronomy, University of Southampton, United Kingdom}

\author{Rupert Ursin}
\affiliation{Institute of Quantum Optics and Quantum Information (IQOQI), Austrian Academy of Sciences, Austria}

\author{Vlatko Vedral}
\affiliation{Atomic and Laser Physics, Clarendon Laboratory, University of Oxford, United Kingdom}
\affiliation{Center for Quantum Technologies, National University of Singapore, Republic of Singapore}

\collaboration{MAQRO Consortium, names after first author sorted alphabetically}
\noaffiliation

\begin{abstract}
Do the laws of quantum physics still hold for macroscopic objects -- this is at the heart of Schr\"odinger's cat paradox -- or do gravitation or yet unknown effects set a limit for massive particles? What is the fundamental relation between quantum physics and gravity? Ground-based experiments addressing these questions may soon face limitations due to limited free-fall times and the quality of vacuum and microgravity. The proposed mission MAQRO may overcome these limitations and allow addressing those fundamental questions. MAQRO harnesses recent developments in quantum optomechanics, high-mass matter-wave interferometry as well as state-of-the-art space technology to push macroscopic quantum experiments towards their ultimate performance limits and to open new horizons for applying quantum technology in space. The main scientific goal of MAQRO is to probe the vastly unexplored ``quantum-classical'' transition for increasingly massive objects, testing the predictions of quantum theory for truly macroscopic objects in a size and mass regime unachievable in ground-based experiments. The hardware for the mission will largely be based on available space technology. Here, we present the MAQRO proposal submitted in response to the (M4) Cosmic Vision call of the European Space Agency for a medium-size mission opportunity with a possible launch in 2025.
\end{abstract}

\maketitle

\section{Executive Summary}
\label{sec::summary}
\subsection{What are the fundamental physical laws of the universe?}
The laws of quantum physics challenge our understanding of the nature of physical reality and of space-time, suggesting the necessity of radical revisions of their underlying concepts. Experimental tests of quantum phenomena, such as quantum superpositions involving massive macroscopic objects, provide novel insights into those fundamental questions.  MAQRO allows entering a new parameter regime of macroscopic quantum physics addressing some of the most important questions in our current understanding of the basic laws of gravity and of quantum physics of macroscopic bodies.

\subsection{Fundamental science and technology pathfinder}
The main scientific objective of MAQRO is to test the predictions of quantum theory in a hitherto inaccessible regime of quantum superpositions of macroscopic objects that contain up to $10^{10}$ atoms. This is achieved by combining techniques from quantum optomechanics, matter-wave interferometry and from optical trapping of dielectric particles. MAQRO will test quantum physics in a parameter regime orders of magnitude beyond existing ground-based experimental tests -- a realm where alternative theoretical models predict noticeable deviations from the laws of quantum physics\cite{Bassi2003a,Bassi2005a,Bassi2013a}. These models have been suggested to harmonize the paradoxical quantum phenomena both with the classical macroscopic world\cite{Ghirardi1986a,Gisin1989a,Pearle1989a,Ghirardi1990a} and with notions of Minkowski space-time\cite{Diosi1984a,Penrose1996a,Diosi2007a}. MAQRO will, therefore, enable a direct investigation of the underlying nature of quantum reality and space-time, and it may pave the way towards testing the ultimate limit of matter-wave interference posed by space-time fluctuations\cite{Jaekel1994a,Lamine2006a}. Recent works showed that MAQRO might even allow testing certain models of dark matter\cite{Riedel2013a,Bateman2015a}. In contrast to collapse models, even standard quantum theory, in the presence of gravitation, predicts decoherence for spatially extended, massive superpositions\cite{Zych2011a,Pikovski2013a}. While this is not applicable in a microgravity setting, ground-based tests in this direction may benefit from the technology development necessary for MAQRO.

By pushing the limits of state-of-the-art experiments and by harnessing the space environment for achieving the requirements of high-precision quantum experiments, MAQRO may prove a pathfinder for quantum technology in space. For example, quantum optomechanics is already proving a useful tool in high-precision experiments on Earth\cite{Abbot2009a}. MAQRO may open the door for using such technology in future space missions.

\subsection{A unique environment for macroscopic quantum experiments}
In ground-based experiments, the ultimate limitations for observing macroscopic quantum superpositions are vibrations, gravitational field-gradients, and decoherence through interaction with the environment. Such interactions comprise, e.g., collisions with background gas as well as scattering, emission and absorption of blackbody radiation. The spacecraft design of MAQRO allows operating the experimental platform in an environment offering a unique combination of microgravity ($\lesssim 10^{-9}\,$g), low pressure ($\lesssim 10^{-13}\,$Pa) and low temperature ($\lesssim 20\,$K). This allows sufficiently suppressing quantum decoherence for the effects of alternative theoretical models to become experimentally accessible, and to observe the evolution of macroscopic superpositions over free-fall times of about $100\,$s.

\subsection{The case for space}
The main reasons for performing MAQRO in space are the required quality of the microgravity environment ($\lesssim 10^{-9}\,$g), the long free-fall times ($100\,$s), the high number of data points required (up to $\sim 10^4$ per measurement run), and the combination of low pressure ($\lesssim 10^{-13}\,$Pa) and low temperature ($\lesssim 20\,$K) while having full optical access. These conditions cannot be fulfilled with ground-based experiments.

\subsection{Technological heritage for MAQRO}
MAQRO benefits from recent developments in space technology. In particular, MAQRO relies on technological heritage from LISA Pathfinder (LPF)\cite{Armano2010a}, the LISA technology package (LTP)\cite{Anza2005a}, GAIA\cite{Lindegren2007a}, GOCE\cite{Drinkwater2007a,Marque2010a}, Microscope\cite{Touboul2001a,Liorzou2014a} and the James Webb Space Telescope (JWST)\cite{Lightsey2012a}. The spacecraft, launcher, ground segment and orbit (L1/L2) are identical to LPF. 

The most apparent modifications to the LPF design are an external, passively cooled optical instrument thermally shielded from the spacecraft, and the use of two capacitive inertial sensors from ONERA technology. In addition, the propulsion system will be mounted differently to achieve the required low vacuum level at the external subsystem, and to achieve low thruster noise in one spatial direction. The additional optical instruments and the external platform will reach TRL~5 at the start of the BCD phases. For all other elements, the TRL is 6-9 because of the technological heritage from LPF and other missions. 

\subsection{Alternative mission scenarios}
Implicit strengths of MAQRO are its relatively low weight and power consumption such that MAQRO's scientific instrument can, in principle, be combined on the same spacecraft with other missions that have similar requirements in precision and orbit. An example could be sun-observation instruments benefiting from an L1 orbit. Another example could be a combination with the ASTROD I mission or similar mission concepts fulfilling the orbit requirements of MAQRO. 

\subsection{Technological Readiness \& the MAQRO consortium}
Since its original proposal as an M3 mission in 2010\cite{Kaltenbaek2012b}, MAQRO has made significant progress in technology development\cite{Kaltenbaek2013b} and in its support within the scientific community. In 2013, we formed the MAQRO consortium, now consisting of more than 30 groups from the UK, Germany, Italy, France, Austria, Switzerland, the US, Australia and Sweden.

MAQRO benefits from significant technological progress made since 2010. The TRLs of several core technologies increased from initial concepts to TRL~3-5. In particular, research groups within the MAQRO consortium have successfully demonstrated cavity cooling of trapped nanospheres\cite{Kiesel2013a,Millen2014b}, feedback cooling \cite{Li2011a,Gieseler2012a}, optical trapping of nanospheres in high vacuum\cite{Gieseler2013a}, and hybrid optical \& Paul trapping of nanospheres\cite{Millen2014a,Millen2014b}. Moreover, optomechanical cooling close to the quantum ground state was successfully demonstrated\cite{OConnell2010a,Teufel2011a,Chan2011a}. Detailed thermal studies of the MAQRO shield design showed the feasibility of achieving the temperature and vacuum technical requirements of MAQRO\cite{Hechenblaikner2014a}. A more detailed thermal study showed even better results\cite{Pilan-Zanoni2015a}. A collaboration of the University of Vienna, the University of Bremen and Airbus Defence \&{}Space, successfully implemented a high-finesse, adhesively bonded optical cavity using space-proof glue and ultra-low-expansion (ULE) material\cite{Kaltenbaek2015a}. The same technology is currently in use to implement a high-finesse test cavity with the same specifications as needed for MAQRO. Based on recent theoretical studies\cite{Bateman2014a}, the design of MAQRO was adapted for preparing macroscopic superpositions with state-of-the-art non-linear-optics and laser technology\cite{Lin2013a} also benefiting from recent advances in the single-mode transmission of deep-UV light\cite{Gebert2014a}. In this way, a central drawback of the initial MAQRO proposal (the need for low power, extremely short-wavelength light) could be resolved.

\section{Science Case}
\label{sec::science}
Do the laws of quantum physics remain applicable without modification even up to the macroscopic level? This question lies at the heart of Schr\"odinger's famous gedankenexperiment (thought experiment) of a dead-and-alive cat\cite{Schroedinger1935a}. Matter-wave experiments have confirmed the predictions of quantum physics from the microscopic level of electrons\cite{Davisson1927a,Thomson1927a}, atoms and small molecules\cite{Estermann1930a} up to massive molecules with up to $10^4$ atomic mass units (amu)\cite{Eibenberger2013a}. Still, experiments are orders of magnitude from where alternative theories predict deviations from quantum physics\cite{Bassi2013a,Adler2009a}.

Using ever more massive test particles on Earth may soon face principal limitations because of the limited free-fall times as well as the limited quality of microgravity environments achievable on Earth. Currently, it is assumed that this limit will be reached for interferometric experiments with particles in the mass range between $10^6\,$amu and $10^8\,$amu\cite{Bateman2014a}. These limitations may be overcome by harnessing space as an experimental environment for high-mass matter-wave interferometry\cite{Kaltenbaek2012b}. At the same time, quantum optomechanics provides novel tools for quantum-state preparation and high-sensitivity measurements\cite{Aspelmeyer2014a}. The mission proposal MAQRO combines these aspects in order to test the foundations of quantum physics in a parameter regime many orders of magnitude beyond current ground-based experiments, in particular, for particle masses in the range between $10^8\,$amu and $10^{11}\,$amu. This way, MAQRO will not only significantly extend the parameter range over which quantum physics can be tested. It will also allow for decisive tests of a number of alternative theories, denoted as ``collapse models'' predicting notable deviations from the predictions of quantum theory within the parameter regime tested.

An important feature of MAQRO is that the parameter range covered has some overlap with experiments that should be achievable on ground even before a possible launch of MAQRO. This allows cross-checking the performance of MAQRO and to provide a fail-safe in case the predictions of quantum physics should fail already for masses between $10^6\,$amu and $10^8\,$amu. In this case, MAQRO would not allow for observing matter-wave interference due to the presence of strong, non-quantum decoherence. For this reason, the MAQRO instrument is designed for allowing three modes of operation for testing quantum physics over a wide parameter range – even in the presence of strong decoherence:
\begin{itemize}
\setlength{\itemsep}{0pt}
\setlength{\parskip}{0pt}
\item \textbf{Non-interferometric tests of collapse models}\\
The stochastic momentum transfer in collapse models can lead to heating of the centre-of-mass motion of trapped nanospheres\cite{Collett2003a,Bahrami2014a}. This can, in principle, be observed by comparing the measured noise spectra with theoretical predictions\cite{Kaltenbaek2013a}.
\item \textbf{Deviations from quantum physics in wave-packet expansion}\\
As in the frequency-based non-interferometric approach above, this method is based on the stochastic momentum transfer due to collapse mechanisms. In particular, the momentum transfer leads to a random walk resulting in an increased rate for the expansion of wavepackets\cite{Collett2003a,Kaltenbaek2013a,Bera2015a}.
\item \textbf{High-mass matter-wave interferometry}\\
This central experiment of MAQRO is based on the original M3 proposal\cite{Kaltenbaek2012b}. It has been adapted for harnessing the successful technique of Talbot-Lau interferometry, which currently holds the mass record for matter-wave interferometry\cite{Eibenberger2013a}. The goal is to observe matter-wave interferometry with particles of varying size and mass, comparing the interference visibility the predictions of quantum theory and the predictions of alternative theoretical models.
\end{itemize}
In particular, the non-interferometric tests and observing wave-packet expansion will allow for performing tests in the presence of comparatively strong decoherence mechanisms. If these two tests show agreement with the predictions of quantum physics, MAQRO's scientific instrument can then be used for performing matter-wave interferometry to test for smaller deviations from quantum physics.

\subsection{Non-interferometric tests of quantum physics}
\label{subsec::sci:noninterf}
The vast majority of the proposals for the test of collapse models put forward so far is based on interferometric approaches in which massive systems are prepared in large spatial quantum superposition states. In order for such tests to be effective, the superposition has to be sufficiently stable in time to allow for the performance of the necessary measurements. Needless to say, these are extremely demanding requirements from a practical viewpoint. Matter-wave interferometry and cavity quantum optomechanics are generally considered as potentially winning technological platforms in this context, and considerable efforts have been made towards the development of suited experimental configurations using levitated spheres or gas-phase molecular or metallic-cluster beams. Alternatively, one might adopt a radically different approach and think of non-interferometric strategies to achieve the goal of a successful test.

MAQRO offers the opportunity for exploring one such possibility by addressing the influences that collapse models (or in general, any non-linear effect on quantum systems) have on the spectrum of light interacting with a radiation-pressure-driven mechanical oscillator in a cavity-optomechanics setting. The overarching goal of this part of MAQRO is to affirm and consolidate novel approaches to the revelation of deviations from standard quantum mechanics in ways that are experimentally viable and open up unforeseen perspectives in the quest at the center of the MAQRO endeavours. 

A benchmark in this sense will be provided by the assessment of the CSL model through a non-interferometric approach. In particular, we will take advantage of the fact that the inclusion of the CSL mechanism in the dynamics of a harmonic oscillator results in an extra line-broadening effect that can be made visible from its density noise spectrum. By bypassing the necessity of preparing, manipulating, and sustaining the quantum superposition state of a massive object, the proposed scheme would be helpful in bringing the goal of observing collapse-induced effects closer to the current experimental capabilities.

\begin{figure}[htb]
 \begin{center}
  \includegraphics[width=0.55\linewidth]{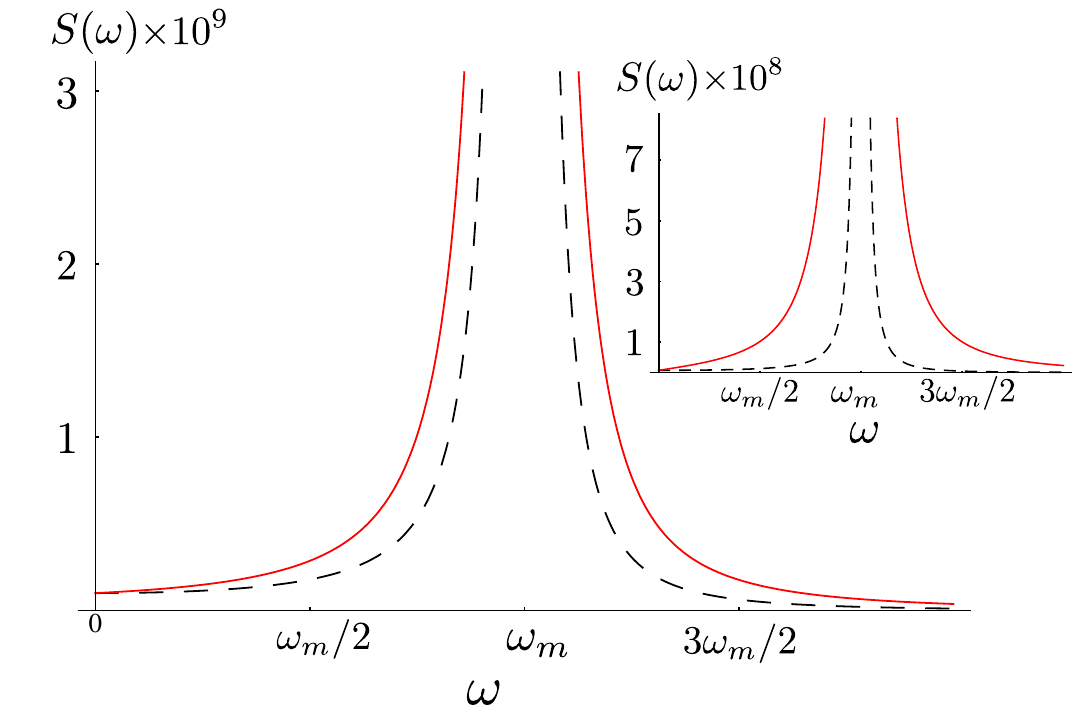}
  \caption{\textbf{Broadening of noise power spectra.} Comparison between the density noise spectrum of the mechanical position fluctuation operators with (solid red line) and without (dashed black line) the influence of the CSL mechanism obtained using Adler's estimate of the CSL coupling strength and a mechanical oscillator of 15 ng. The inset shows an analogous study for $m=150\,$ng (figure from Ref.~\cite{Bahrami2014a})\label{fig::NSbroadening}.}
  \end{center}
\end{figure}

The equation of motion of the optomechanical system (regardless of its embodiment) in the presence of the CSL mechanism can be cast in the form given in equation \glei{equ::noninterf1}
\begin{equation}
\frac{\partial}{\partial t} \hat{\mathcal{O}} = \frac{\mathrm{i}}{\hbar} \left[ \hat{H}, \hat{\mathcal{O}} \right] + \frac{\mathrm{i}}{\hbar} \left[ \hat{V}_t, \hat{\mathcal{O}} \right] + \hat{\mathcal{N}},\label{equ::noninterf1}
\end{equation}
where $\hat{\mathcal{O}}$ is an operator of the system, $\hat{H}$ is the Hamiltonian of the mechanical oscillator coupled to the cavity light field, $\hat{\mathcal{N}}$ embodies all the relevant sources of quantum noise affecting the system, and $\hat{V}_t$ is a stochastic linear potential (linked directly to the position of the harmonic oscillator) that accounts for the effective action of the CSL mechanism\cite{Bahrami2014a}. It can be shown that such a potential is zero-mean and delta-correlated, and thus embodies a source of white noise that adds up to the relevant noise mechanisms affecting the optomechanical system, namely the damping of the optical cavity and the Brownian motion (occurring at temperature $T$) of the mechanical oscillator. A lengthy calculation based on the study, in the frequency domain, of the fluctuation operators of both the optical and mechanical system, leads to the following expression for the density noise spectrum of the mechanical system's position fluctuation:
\begin{equation}
S(\omega) = \frac{2 \alpha^2_s \hbar^2 \kappa \chi^2 \left(\Delta^2 + \kappa^2 + \omega^2\right) + \hbar m \omega \left[\left(\Delta^2 + \kappa^2 - \omega^2\right)^2 + 4 \kappa^2 \omega^2 \right] \left[\gamma_m \mathrm{coth}(\beta \omega) + \mathcal{Y}\right]}{\left\vert 2 \alpha^2_s \Delta\,\hbar \chi^2 + m \left(\omega^2 - \omega^2_m - \mathrm{i} \gamma_m \omega \right) \left[\Delta^2 + \left(\kappa+\mathrm{i} \omega\right)^2\right] \right\vert^2},\label{equ::noninterf2}
\end{equation}
where $\alpha_s$ is the steady-state amplitude of the cavity field, $\kappa$ is the cavity damping rate, $\chi$ is the optomechanical coupling rate. $\Delta$ is the detuning between the cavity field and an external pump, $m$ is the mass of the mechanical oscillator, $\gamma_m$ is the mechanical damping rate, $\omega_m$ is the mechanical frequency, and $\beta$ is the inverse temperature of the system. Finally, we have introduced:
\begin{equation}
\mathcal{Y} = \lambda \sqrt{\frac{\hbar}{m \omega_m}},
\end{equation}
where $\lambda$ is the CSL coefficient. In our numerical simulations of the observability of the effects, we have used the value of such parameter achieved by assuming Adler's estimate of the CSL mechanism's strength. Quite evidently, the CSL mechanism enters into the expression of the density noise spectrum as an extra thermal-like line broadening contribution. While being formally rather appealing, this elegant result also suggests the strategy to implement in order to observe the collapse model itself, and identifies the challenges that have to be faced, namely a cold enough mechanical system that lets the $\mathcal{Y}$-dependent term dominate over the temperature-determined one. Our numerical estimate shows that, indeed, it is possible to pinpoint the effects of the CSL contribution in a parameter regime currently available in optomechanical labs. Figure~\ref{fig::NSbroadening} shows a typical result achieved by using the parameters stated in Ref.~\cite{Bahrami2014a}.

At the present state, this non-interferometric approach has not been investigated in sufficient detail in the context of MAQRO. While this does not impede the main science goals of MAQRO, we plan nevertheless to investigate this non-interferometric method more closely during the study phase of MAQRO. It may offer the attractive possibility to supplement the results of the other two experiments (subsections \ref{subsec::sci:WAX} and \ref{subsec::sci:DECIDE}).

\subsection{Deviations from quantum physics in wave-packet expansion}
\label{subsec::sci:WAX}
Most forms of decoherence can be described as resulting from the interaction of a quantum system with its environment\cite{Zurek1991a}. Examples are elastic and inelastic scattering as well as emission of massive particles or radiation\cite{Schlosshauer2007a}. All of these interactions result in a change of momentum, eventually leading to dephasing and decoherence of quantum states. In a paper by Collett and Pearle\cite{Collett2003a}, it was shown that decoherence mechanisms assumed in collapse models also lead to momentum transfer. That means, even in the absence of standard decoherence mechanisms, collapse models may result in a random walk due to stochastic momentum transfer. This random walk can, in principle, be observed when comparing the expansion rate of a quantum wave packet with the predictions of quantum theory as well as with the predictions of alternative models. Apart from the original suggestion for such an experiment\cite{Collett2003a}, there have also been more recent suggestions to observe this effect using free-falling or optically trapped, dielectric particles\cite{Kaltenbaek2013a,Bera2015a}.

Even if there is no decoherence, the width of a quantum wave packet will expand over time according to the Schr\"odinger equation. The square of the width of the wave packet $w_s(t)^2$ evolves according to the following relation:
\begin{equation}
w_s(t)^2 = \left\langle\hat{x}^2(t)\right\rangle_s = \left\langle\hat{x}^2(0)\right\rangle + \frac{t^2}{m^2} \left\langle\hat{p}^2(0)\right\rangle.\label{equ::ws}
\end{equation}
Here, the subscript ``s'' denotes evolution according to Schr\"odinger's equation, $m$ is the mass of the particle, the angular brackets denote the expectation value for a given quantum state, $\hat{x}$ denotes the position operator, and $\hat{p}$ denotes the momentum operator. Equation \glei{equ::ws} relates the width of the wave packet at time $t$ with the initial width of the wave packet and the initial width of the momentum distribution.

In the presence of decoherence, the width of the wave packet increases more quickly:
\begin{equation}
w(t)^2 = \left\langle\hat{x}^2(t)\right\rangle = w_s(t)^2 + \frac{2 \Lambda \hbar^2}{3 m^2} t^3.\label{equ::wdec}
\end{equation}
Here, $\Lambda$ is a parameter governing the strength of decoherence mechanisms. The width of the wave packet is not an observable -- it has to be inferred from the statistical distribution of many measurements\cite{DAriano1996a}. If we assume that we perform $N$ measurements of the particle position and if the result of the j-th measurement is $x_j$, for large $N$, the width of the wave packet can be approximated as: 
\begin{equation}
w = \frac{1}{\sqrt{N-1}} \sqrt{\sum^N_{j=1} x^2_j}.\label{equ::wstat}
\end{equation}
Given that the error of each position measurement is $\Delta x_j = \sigma$, the error of our estimate of the width of the wave packet will be:
\begin{equation}
\Delta w = \frac{\sigma}{\sqrt{N-1}} \approx \frac{\sigma}{\sqrt{N}},\label{equ::werr}
\end{equation}
where the approximation holds for large $N$.

The mode of operation of this experiment is to determine the wave-packet size as a function of time $t$, and to compare these measurements with the predictions of quantum physics using equation \glei{equ::wdec}. In this way, we can experimentally determine the decoherence parameter $\Lambda$ and compare it with the predictions of quantum physics. The more $\Lambda$ deviates from the value predicted by quantum physics, the easier it will be to discern by measuring the wave-packet expansion.

For simplicity, let us assume that we have a well isolated quantum system, i.e., quantum physics predicts $\Lambda=0$ or at least much smaller than the deviation we want to measure. The minimum $\Lambda$ we can distinguish experimentally from the case of no decoherence is:
\begin{equation}
\Lambda > \Lambda_\mathrm{min} = 3 m^2 \frac{\sigma w_s(t)}{\sqrt{N-1} \hbar^2 t^3}.\label{equ::lambdaMin}
\end{equation}
We can relate this minimum decoherence parameter to a decoherence rate $\Gamma = r^2_c \Lambda$ by introducing a representative length scale  $r_c=100\,$nm. This is a typical length scale for the experiments in MAQRO and also the same as the length scale chosen in the collapse model of Ghirardi, Rimini and Weber\cite{Ghirardi1986a}:
\begin{equation}
\Gamma_\mathrm{min} = \Lambda_\mathrm{min} r^2_c = 3 m^2 \frac{\sigma w_s(t) r^2_c}{\sqrt{N-1} \hbar^2 t^3}.\label{equ::gammaMin}
\end{equation}

\begin{figure}[htb]
 \begin{center}
  \includegraphics[width=0.55\linewidth]{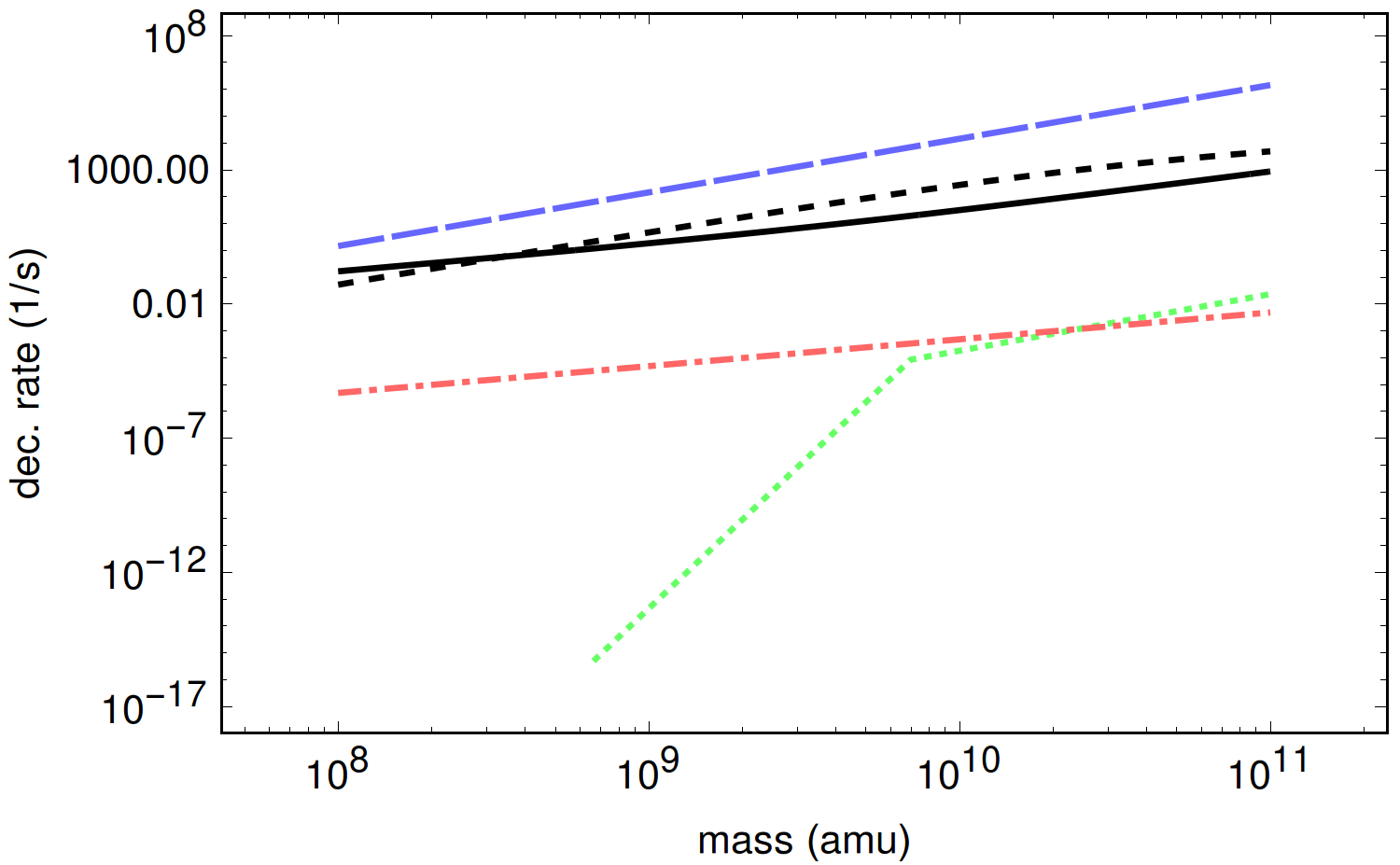}
  \caption{\textbf{Comparison of $\Lambda_\mathrm{min}$} (solid, black) with the decoherence rates predicted for the CSL model with $\lambda = 2.2\times 10^{-17}\,$Hz  (black, dashed), the quantum-gravity model of Ellis et al (blue, long dashed), the model of Di\'osi \& Penrose (red, dot-dashed), and the model of K\'arolyh\'azy (green, dotted). Where models predict a higher decoherence rate than $\Gamma_\mathrm{min}$, one can, in principle, distinguish them from the predictions of quantum physics.\label{fig::WAXminRates}.}
  \end{center}
\end{figure}

In Figure~\ref{fig::WAXminRates}, we compare the predictions of several collapse models with that minimum, discernible decoherence rate $\Gamma_\mathrm{min}$. The figure shows that, by investigating wave-packet expansion, MAQRO can, in principle, perform decisive tests of the CSL model even with the originally suggested parameters\cite{Ghirardi1990a,Collett2003a}, and MAQRO could test the quantum gravity model of Ellis and others\cite{Ellis1984a,Ellis1989a}. However, the plot also illustrates that wave-packet expansion will neither allow testing the model of K\'arolyh\'azy nor that of Di\'osi-Penrose.

In order to estimate the values plotted in Figure~\ref{fig::WAXminRates}, we assumed that we let the wave-packet expand for a maximum of $100\,$s, and that we collect at most $N=24\times 10^{3}$ data points to experimentally estimate the decoherence parameter. The number of data points was chosen in order to limit the integration time to at most four weeks. Moreover, we assumed our test particle to initially be in a thermal state of a harmonic oscillator -- with a mechanical frequency $\omega=10^5\,\mathrm{rad/s}$, an average occupation number of $0.3$, and that we can determine the particle position with an accuracy of $100\,$nm. Because the mechanical frequency for an optically trapped particle only depends on the mass density and the material's dielectric constant, the mechanical frequency is roughly constant for the particles chosen for MAQRO. The occupation number, however, is assumed to be inversely proportional to the mass of the test particle because it depends on the optomechanical coupling achievable.

\begin{figure}[htb]
 \begin{center}
  \includegraphics[width=0.48\linewidth]{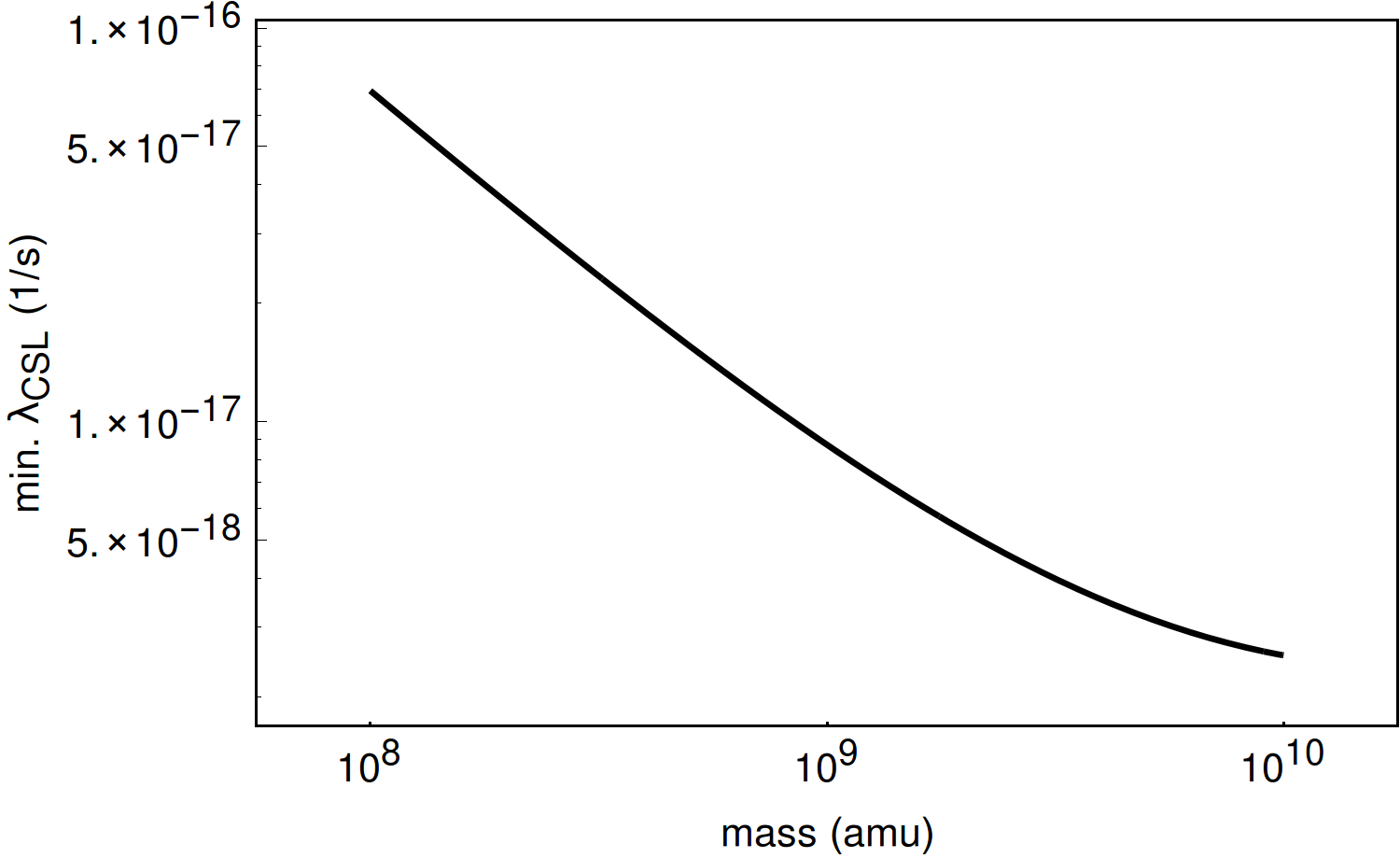}
    \includegraphics[width=0.48\linewidth]{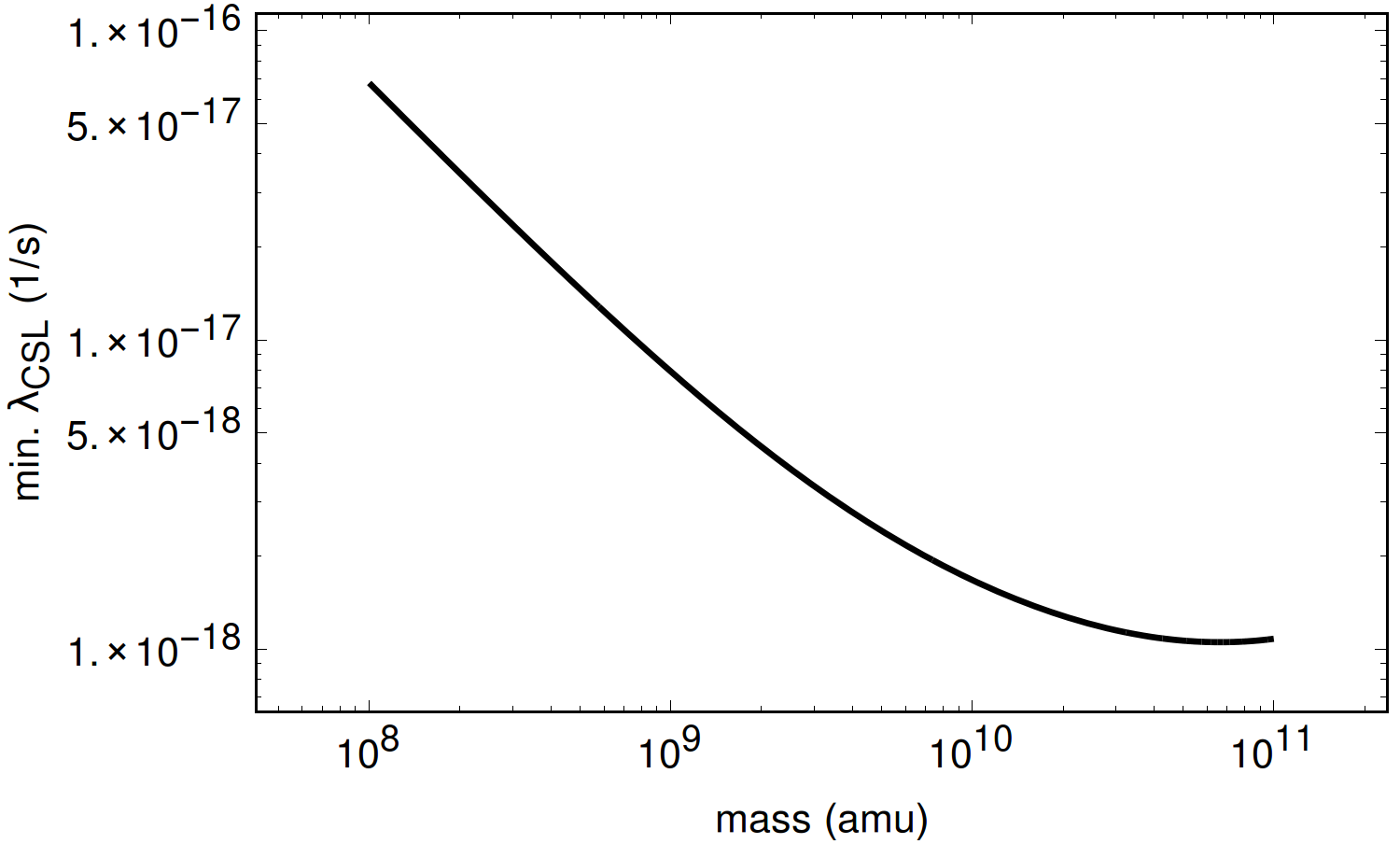}
  \caption{\textbf{Minimum CSL parameter $\lambda_\mathrm{min}$}. The two graphs show the prediction of the minimum CSL parameter $\lambda_\mathrm{min}$ discernible from the case of no decoherence -- for the cases of a test particle of fused silica (left) and of Hafnia ($\mathrm{HfO_2}$) (right).\label{fig::WAXminCSL}}
  \end{center}
\end{figure}

Because testing quantum physics using wave-function expansion was first introduced for the CSL model\cite{Collett2003a}, and because the CSL model represents a rather general, heuristic approach to collapse models, we will now discuss the prerequisites for testing the CSL model in the context of MAQRO. The CSL model depends on two parameters, $a$ and $\lambda$, where $a=100\,$nm defines the typical length scale at which the CSL model predicts a transition from quantum to classical behaviour. For $\lambda$, which predicts the rate of decohering events on the microscopic level, a wide variety of values have been suggested, ranging from $2.2\times10^{-17}\,$Hz\cite{Ghirardi1990a,Collett2003a} to $10^{-8}$Hz\cite{Adler2007a}. The smaller one assumes the value of $\lambda$, the smaller the deviation from quantum physics. Using equation \glei{equ::gammaMin}, we can now estimate the smallest value of $\lambda$ that MAQRO would allow detecting. In particular, we get:
\begin{equation}
\lambda_\mathrm{min} = 4 a^2 \rkl{\frac{m_p}{m}}^2 f\rkl{\frac{r}{a}}^{-1} \Lambda_\mathrm{min} > m^2_p f\rkl{\frac{r}{a}}^{-1} \frac{12 a^2 \sigma w_s(t)}{\sqrt{N-1} \hbar^2 t^3},\label{equ::minCSL}
\end{equation}
where $m_p$ is the proton mass, and \cite{Collett2003a}:
\begin{equation}
f(x) = \frac{6}{x^4} \ekl{1-\frac{2}{x^2}+\rkl{1+\frac{2}{x^2}} e^{-x^2}}.\label{equ::CSLf}
\end{equation}
In Figure~\ref{fig::WAXminCSL}, we plot $\lambda_\mathrm{min}$ as a function of the particle mass for the case of two different nanosphere materials. The plots show that MAQRO should allow testing the CSL model for localization rates $\lambda$ even lower than the originally assumed parameters in Refs.~\cite{Ghirardi1990a,Collett2003a}. Comparing this result with the plot in Figure~\ref{fig::WAXminRates} shows that MAQRO will also allow testing the quantum gravity model of Ellis et al.

\subsection{Decoherence in high-mass matter-wave interferometry}
\label{subsec::sci:DECIDE}
Using matter-wave interferometry with high-mass test particles is the most sensitive tool of MAQRO for testing quantum physics. While the other techniques described earlier allow testing deviations from quantum physics for values of the decoherence parameter larger than $\Lambda_\mathrm{min} \approx 10^{14}\,\mathrm{m^{-2} s^{-1}}$, high-mass matter-wave interferometry will allow MAQRO testing for even smaller deviations.

In the original MAQRO proposal for the M3 call\cite{Kaltenbaek2012b}, the approach suggested was using far-field interferometry based on preparing a double-slit-like quantum superposition where a massive particle is in superposition of being in two clearly separate positions. Since this original proposal, we have adapted MAQRO to use near-field interferometry instead. In particular, the novel approach is based on well-established techniques having been used in a series of high-mass matter-wave experiments\cite{Hornberger2012a}, originally using Talbot-Lau interferometry\cite{Brezger2003a}. Typically, near-field matter-wave interferometry is performed using three gratings. The first grating is used for providing a coherent source of particles. This second grating is the center-piece of the interferometer where the high-mass quantum superposition is prepared. Finally, a third, absorptive grating is used for determining the presence of a periodic interference pattern. 

Over the last two decades, this approach has been adapted for numerous experiments, steadily improving the approach's applicability to ever higher test-particle masses and sizes. For example, one can replace one or more of the gratings with standing-wave, optical gratings instead of nano-fabricated, material gratings. For example, if the first and third grating are absorptive gratings, the second grating can be a pure phase grating (see, e.g., Ref.~\cite{Eibenberger2013a}). In the most recent and, so far, most powerful adaptation of this technique, all three gratings are replaced by optical gratings, implementing an optical time-domain ionizing matter-wave interferometer (OTIMA)\cite{Haslinger2013a}.

\begin{figure}[htb]
 \begin{center}
  \includegraphics[width=0.25\linewidth]{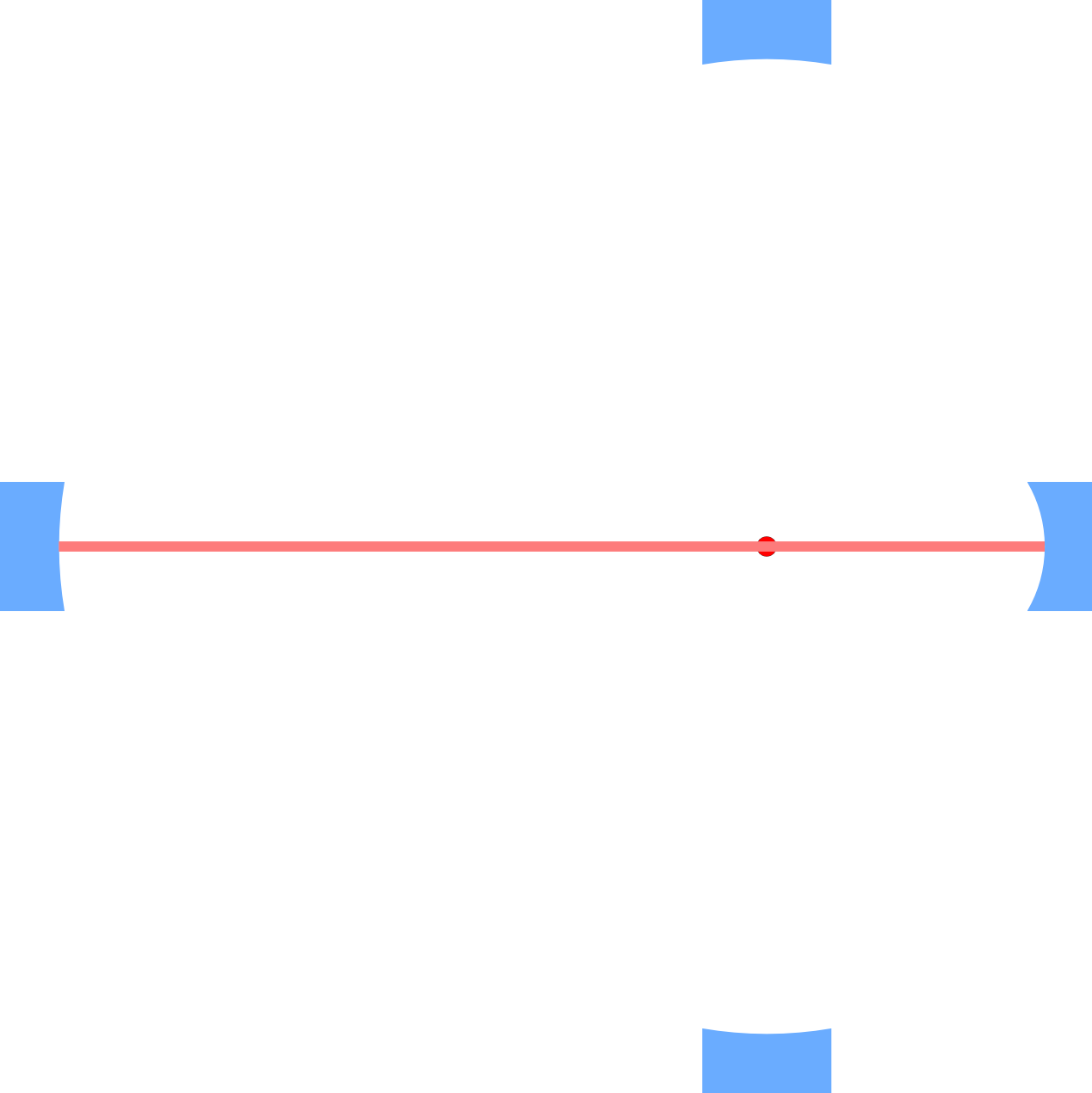}
    \includegraphics[width=0.25\linewidth]{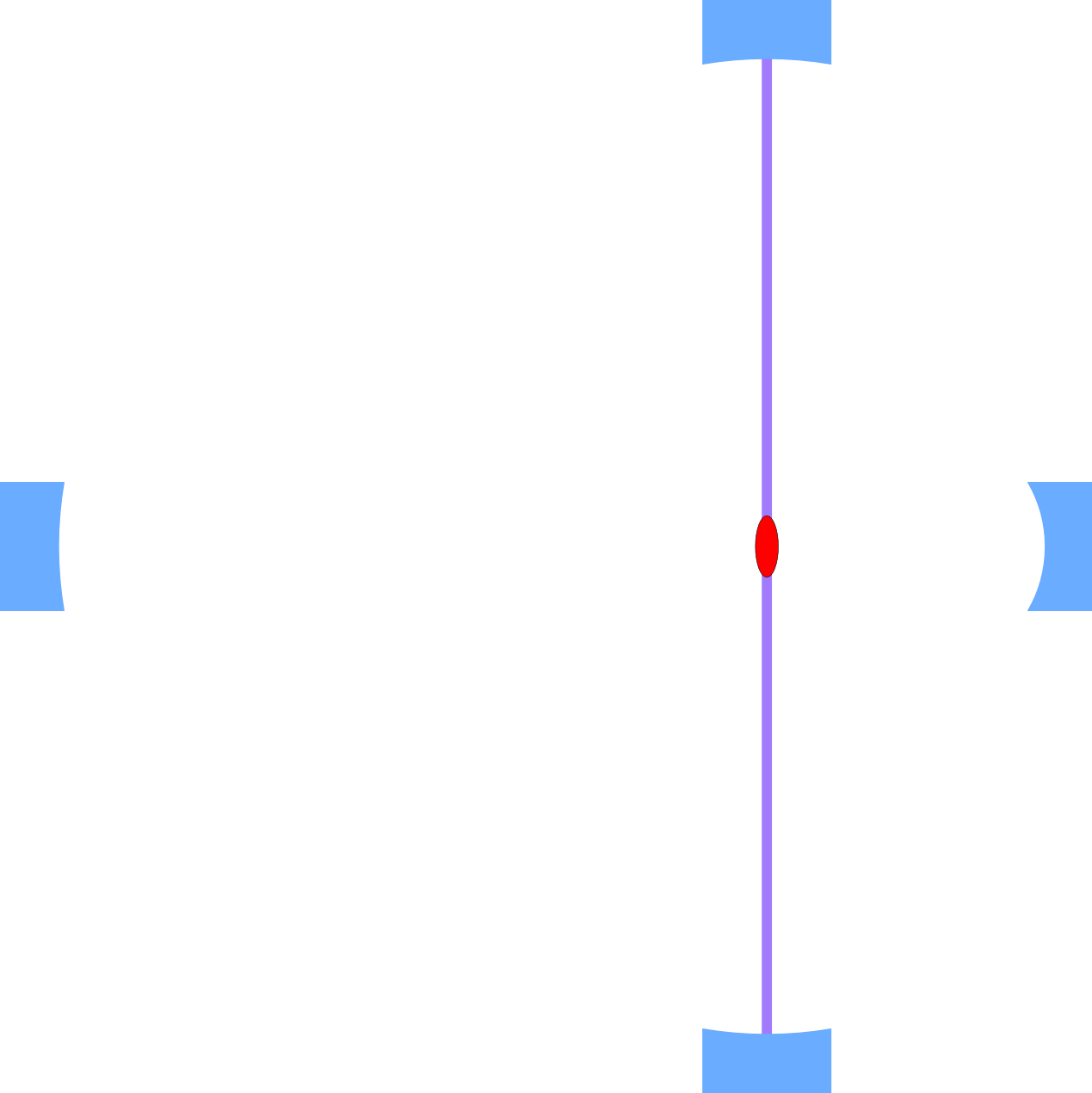}
  \caption{\textbf{Schematic of the novel near-field interferometry approach for MAQRO.}. The approach uses two cavities. First a particle is trapped and its center-of-mass motion 3D-cooled using modes in the first cavity (left). The red dot indicates the particle position. After this preparation, the particle is released, letting the wave-function expand freely for some time $t_1$. After that time, the optical phase grating is applied for a short time in a second cavity (right). The expanded red region illustrates the expanded wave-function.\label{fig::MAQROtalbotScheme}}
  \end{center}
\end{figure}

An alternative approach using only one, pure-phase grating has been proposed recently\cite{Bateman2014a}. Here, we adapt it for use with MAQRO. In particular, instead of using a grating as a coherent source, the source consists of an intra-cavity optical trap used to initially position and to 3D-cool the center-of-mass motion of an individual, trapped particle -- that means, the motion of the particle is cooled in all spatial directions (see Figure~\ref{fig::MAQROtalbotScheme} (left)). After this step of preparation, the particle is released from the trap, and the corresponding wave-function will expand for a time $t_1$. Then a second optical beam, perpendicular to the first one, is switched on. This beam with wavelength $\lambda_g$ forms a standing-wave upon reflection from a mirror. Either one uses another cavity for this or a simple reflection at a mirror (see Figure~\ref{fig::MAQROtalbotScheme}(right)). The optimal option for the wavelength will be discussed in section \ref{subsec::scireq:phase}. This second beam acts as a pure phase grating with grating period $d=\lambda_g/2$. After applying this grating, the state will evolve freely for a time $t_2$, and then all optical fields are switched on in order to measure the position of the particle. The complete process is repeated $N$ times, and the histogram of the particle positions measured can be used to reconstruct the interference pattern.

We will assume a maximum overall time $T=t_1+t_2\approx 100\,$s. This is necessary in order to keep the total integration time for observing an interference pattern within a reasonable time-frame given the limited life time of a space mission. Moreover, longer integration times would be incompatible with the quality of the microgravity environment achievable in MAQRO.

\begin{figure}[htb]
 \begin{center}
  \includegraphics[width=0.5\linewidth]{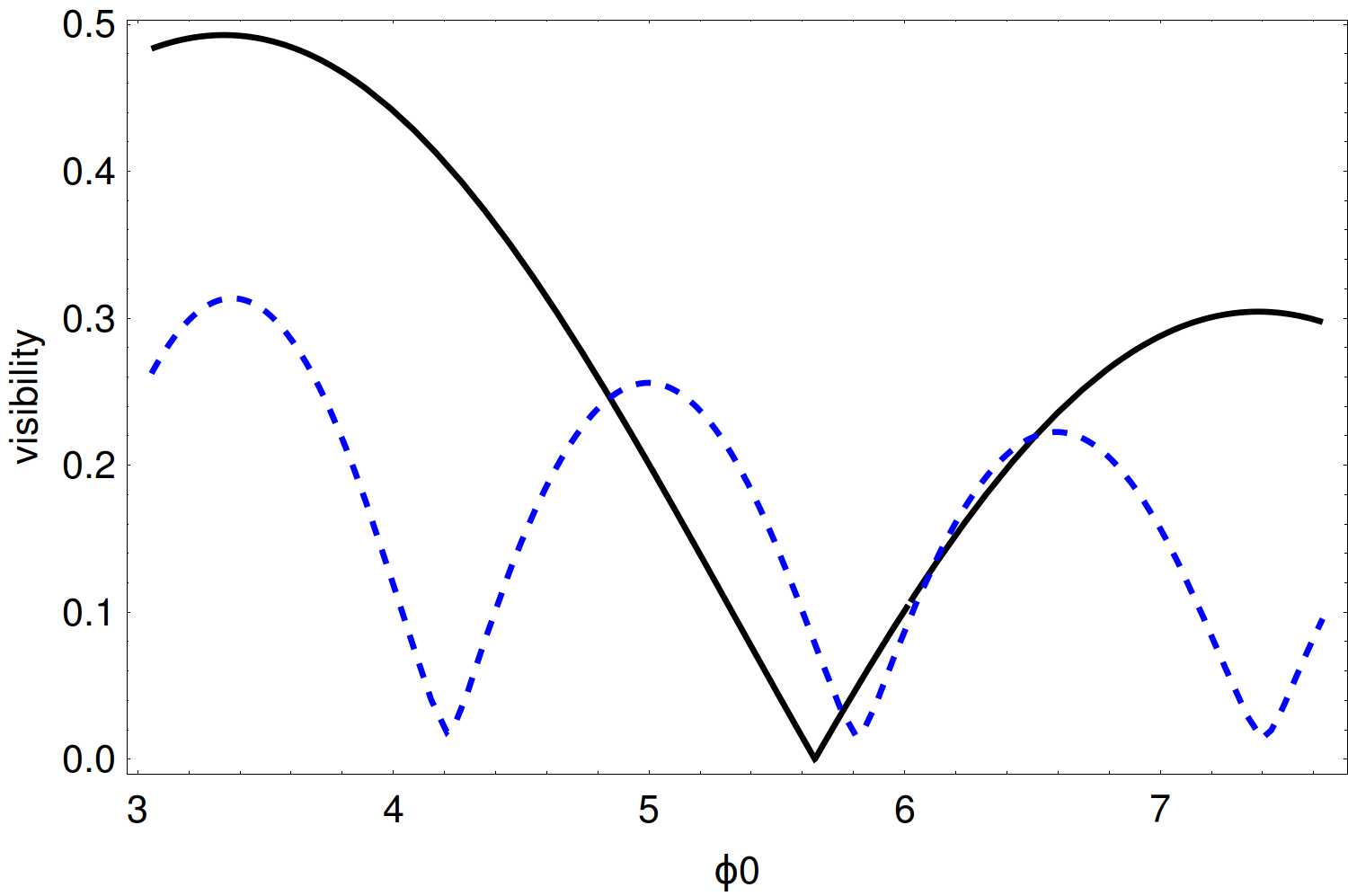}
  \caption{\textbf{Classical vs. quantum interference visibility.} Here, we plot the expected quantum (solid black) vs. the corresponding classical interference visibility (blue, dashed) as a function of $\phi_0$ for a test particle of mass $m=10^9\,$amu, $T=100\,$s, and $\lambda_g=200\,$nm.\label{fig::QMvsClassVis}}
  \end{center}
\end{figure}

We will assume that the initially prepared state is Gaussian, and if we concentrate only on one dimension in the direction we apply the phase grating in, then the corresponding characteristic function is \cite{Bateman2014a}:
\begin{equation}
\chi_0(s,q) = \mathrm{exp}\rkl{-\frac{\sigma^2_x q^2 + \sigma^2_p s^2}{2 \hbar^2}}.\label{equ::characteristic}
\end{equation}
Here, $\sigma_x$ and $\sigma_p$ are the position and momentum uncertainties of the initial state, respectively. Then the interference pattern close to the original position of the particle can be written as (also see Ref.~\cite{Bateman2014a}):
\begin{equation}
P(x) = \frac{m}{\sqrt{2 \pi} \sigma_p T} \sum^\infty_{n=-\infty} \exp\rkl{\ci n k_g x} J_{2 n}\ekl{\phi_0 \sin\rkl{\pi n \kappa}} \exp\ekl{-\frac{1}{2} \rkl{n k_g \sigma_x \frac{\beta}{\alpha}^2}}\exp\ekl{-\frac{\Lambda T \rkl{n \kappa d}^2}{3}}.\label{equ::pattern}
\end{equation}
To enable this compact notation, we have introduced several definitions. Central to this approach is the Talbot time $t_T=(md^2)/h$, where $h$ is Planck's constant, $m$ is the particle mass, and $d$ is the grating period. The Talbot time defines the time scale of the interference. In particular, close to multiples of the Talbot time, the wave-function after applying the phase grating will again have a similar periodic distribution as the grating itself but with the grating period enhanced by a factor $\mu=T⁄t_1$. This is the Talbot effect. In addition, we introduced $k_g=2π⁄\mu d$, $\alpha=t_1⁄t_T$, $\beta=t_2⁄t_T$  and $\kappa=\alpha \beta⁄(α+β)$. $\phi_0$ denotes the phase applied to the quantum state at the antinodes of the phase grating\cite{Bateman2014a}, and $J_n(x)$ is a Bessel function of the first kind.

\begin{figure}[htb]
 \begin{center}
  \includegraphics[width=0.5\linewidth]{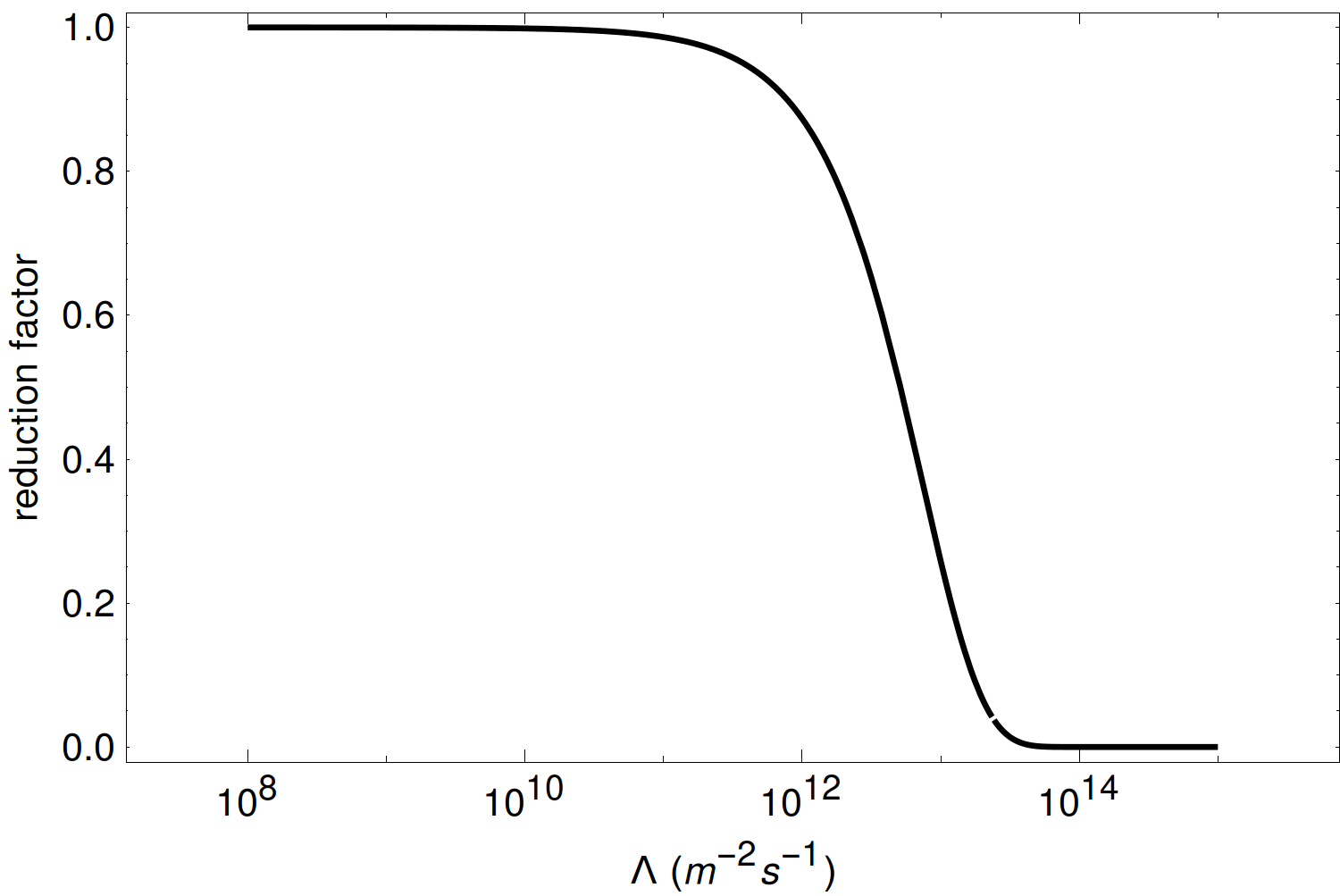}
  \caption{\textbf{Visibility reduction due to decoherence.} Quantum interference visibility reduces as a function of the strength of decoherence, parametrized by the parameter $\Lambda$.\label{fig::VisReduction}}
  \end{center}
\end{figure}

It is important to note that an interference-like pattern can also be observed for purely classical particles. This is due to a moir\'e shadowing effect\cite{Brezger2003a}, and the resulting classical ``interference pattern'' can also be described using equation \glei{equ::pattern} but replacing $\sin⁡(\pi n \kappa)$ with $\pi n \kappa$\cite{Hornberger2009a}. In Figure~\ref{fig::QMvsClassVis}, we plot the corresponding visibilities for the quantum and the classical case in the absence of decoherence. The plot shows a marked difference between the quantum and the classical predictions -- in visibility and in the dependence on $\phi_0$\cite{Bateman2014a}.

\begin{figure}[htb]
 \begin{center}
  \includegraphics[width=0.5\linewidth]{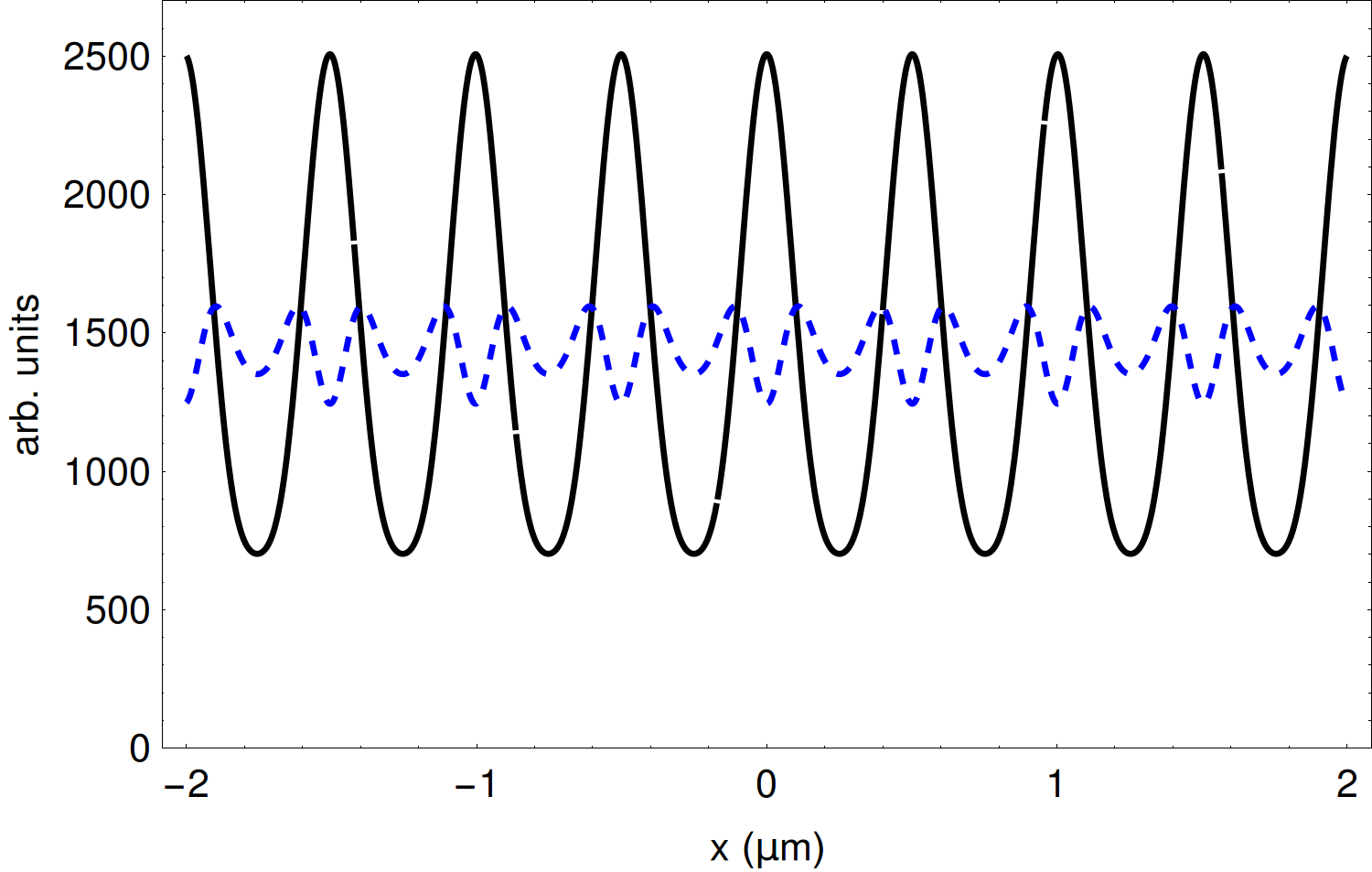}
  \caption{\textbf{Interference patterns.} Expected quantum (black, solid) and classical (blue, dashed) interference patterns.\label{fig::QMvsClassPattern}}
  \end{center}
\end{figure}

In the presence of decoherence, the interference visibility drops as plotted in Figure~\ref{fig::VisReduction}. The plot was calculated for a mass $m=10^9\,$amu, $T=100\,$s and $d=100\,$nm. For smaller masses, we may, in principle, even choose shorter times $T<100\,$s. However, the phase $\phi_0$ experienced by our particles for a given energy $E_G$ of the optical grating (optical power integrated over the time the grating is turned on) decreases with decreasing particle size:
\begin{equation}
\phi_0 = \frac{2 \mathrm{Re}(\alpha) E_G}{\hbar c \epsilon_0 a_G},\label{equ::phi0}
\end{equation}
where $\epsilon_0$ is the vacuum permittivity, $c$ is the speed of light, $a_G$ is the waist of the UV mode, and $\alpha$ is the polarisability of the particle. $\alpha$ is proportional to the particle's mass. Every decrease in mass therefore has to be compensated by higher intensity of UV light in order to achieve the same phase shift. For the smallest particles used in MAQRO, it is even preferable to use IR light instead (see section \ref{subsec::scireq:phase}).

According to Figure~\ref{fig::QMvsClassVis}, the difference between quantum and classical visibility is very pronounced for $\phi_0 \approx 4.2$. For this choice of phase, we plot the expected quantum and classical interference patterns in Figure~\ref{fig::QMvsClassPattern}. As expected, the quantum interference shows significantly higher visibility. The plots also demonstrate the marked difference in the shapes of the quantum and classical predictions (see also Ref.~\cite{Bateman2014a}).

\section{Scientific requirements}
\label{sec::scireq}
Here, we will outline the requirements for realizing the scientific objectives of MAQRO. The requirements for observing high-mass matter-wave interferometry are significantly more stringent than for the other scientific objectives (non-interferometric tests of quantum physics, testing quantum physics by observing wave-packet expansion). For this reason, we focus on the requirements for demonstrating high-mass matter-wave interferometry -- then the requirements for the other scientific objectives will automatically be fulfilled as well.

\begin{table}[hbt]
\begin{tabular}{llll}
  \hline
  \multicolumn{2}{l}{\textbf{Parameter}} & \hspace{0.5cm} & \textbf{Requirement} \\\hline
  \multicolumn{2}{l}{Nominal mission lifetime (without possible extension)} & & 2 years \\
  \multicolumn{2}{l}{Environment temperature} & & $<20\,$K \\
  \multicolumn{2}{l}{Acceleration sensitivity} & & \\
   \hspace{0.5cm} & along UV cavity & & $\lesssim 1\,\mathrm{(pm/s^2)/\sqrt{Hz}}$ \\
    & along IR cavity & & $\lesssim 100\,\mathrm{(pm/s^2)/\sqrt{Hz}}$ \\
    & perpendicular to optical bench & & $\lesssim 5\,\mathrm{(nm/s^2)/\sqrt{Hz}}$ \\
  \multicolumn{2}{l}{Optical-trapping occupation number} & &  \\
    & along cavity & & $\lesssim 10$ \\
    & orthogonal to cavity & & $\lesssim 10^4$ \\
  \multicolumn{2}{l}{Test particles} & & \\
    & Mass & & $10^8\,$amu to $~10^{10}\,$amu \\
    & Charge & & $0\,\mathrm{e^-}$ \\
    & Type & & dielectric, transparent at $1064\,$nm \\
    & Size & & $30\,$nm to $120\,$nm \\
    & Temperature & & $\lesssim 25\,$K \\
  \multicolumn{2}{l}{Period of phase grating} & & $100\,$nm \\
  \multicolumn{2}{l}{Accuracy of position detection} & &  \\
    & along UV cavity & & $20\,$nm \\
    & along IR cavity & & $100\,$nm \\
    & perpendicular to optical bench & & $\ll 60\,\mathrm{\mu m}$ \\
  \multicolumn{2}{l}{Time for on-demand particle loading} & & $\ll 100\,$s \\
  \multicolumn{2}{l}{Measurement time per data point} & & $\lesssim 100\,$s \\
  \multicolumn{2}{l}{Vacuum -- particle density} & & $< 500\,\mathrm{cm^{-3}}$ \\\hline
\end{tabular}
\caption{\textbf{Overview of the scientific requirements of MAQRO.}\label{tab::scireq}}
\end{table}

\subsection{Phase grating}
\label{subsec::scireq:phase}
This requirement only applies for high-mass matter-wave interferometry. As discussed in Section~\ref{subsec::sci:DECIDE}, a pure phase grating with a grating period $d=\lambda_G⁄2$ can be realized by an optical standing wave with wavelength $\lambda_G$. Here, we describe the scientific requirements for implementing this pure-phase grating.

\begin{figure}[htb]
 \begin{center}
  \includegraphics[width=0.5\linewidth]{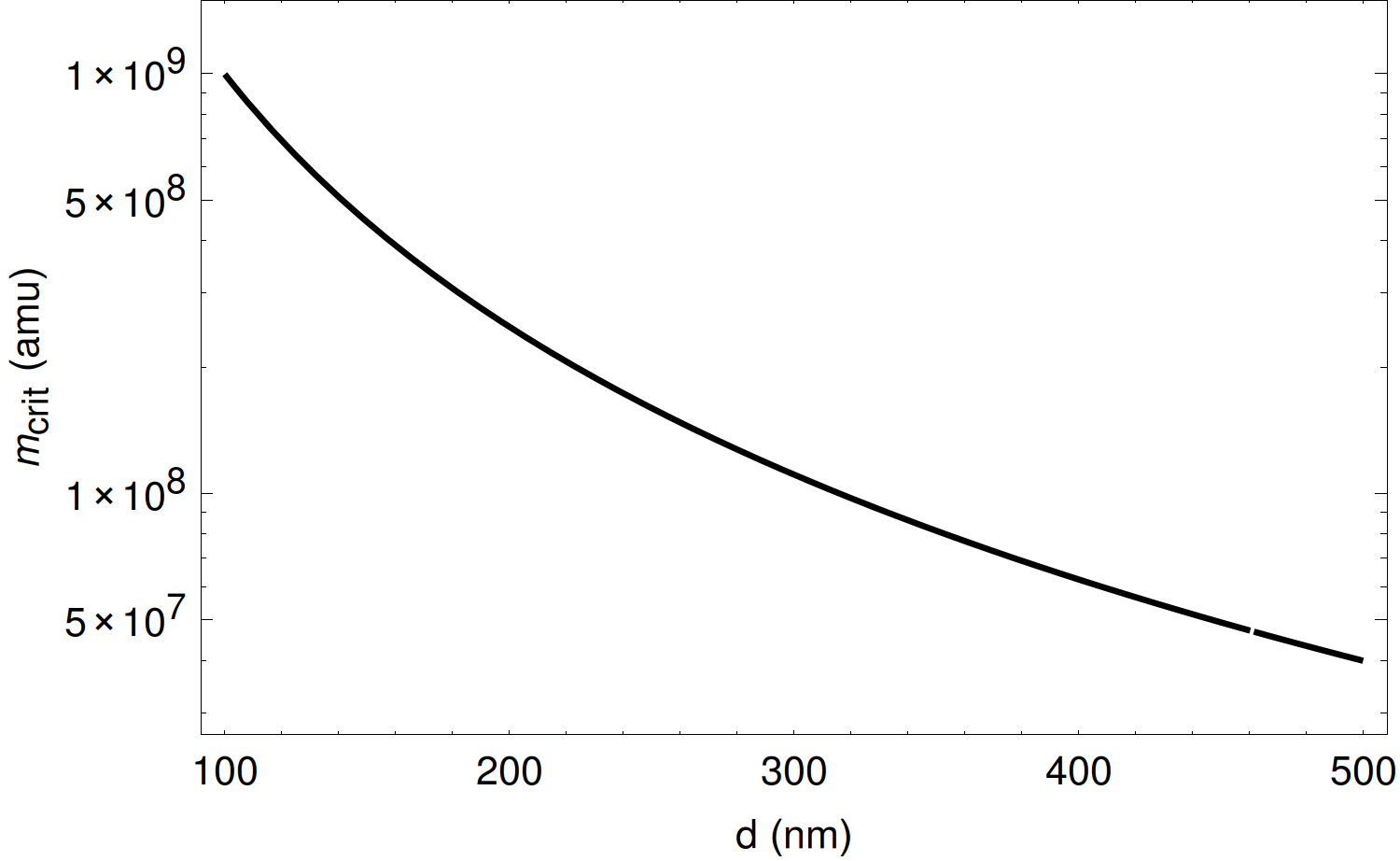}
  \caption{\textbf{Critical mass over grating period.} We plot the approximate upper mass limit for seeing ``useful'' interference as a function of the grating period $d$.\label{fig::critMass}}
  \end{center}
\end{figure}

In matter-wave interferometry based on the Talbot effect, the time scale for the free evolution before applying the phase grating ($t_1$) and the time between this event and the final position measurement ($t_2$) are determined by the Talbot time $t_T=(m d^2)/h$ ($m$: particle mass; $d$: grating period; $h$: Planck's constant). To see reasonable interference visibility, we must have:
\begin{equation}
\kappa = \frac{t_1 t_2}{T t_T} \le \frac{T}{4 t_T},\label{equ::kappaMax}
\end{equation}
where $T=t_1 + t_2$ is the overall \textbf{measurement time per data point}. As mentioned in section~\ref{subsec::sci:DECIDE}, to get reasonable particle statistics and in order to get realistic requirements on the microgravity quality (see section~\ref{subsec::scireq:microgravity}), we have to require $T\le 100\,$s. Because the Talbot time is proportional to the particle mass, this requirement results in an increasingly more stringent upper bound on $\kappa$ for high test masses. On the other hand, $\kappa$ should be on the order of $1$ in order to see a noticeable difference between the quantum prediction of an interference pattern and classically expected moir\'e ``shadow patterns''. In combination with equation~\glei{equ::kappaMax}, this yields a limit on the particle mass:
\begin{equation}
m\lesssim m_\mathrm{crit} \equiv \frac{h T}{4 d^2}.\label{equ::masslimit}
\end{equation}

Figure~\ref{fig::critMass} shows this (rough) mass limit as a function of the grating period chosen. We see that for performing experiments in the mass regime around $10^9\,$amu, the \textbf{grating period} should be $d\le 100\,$nm. We choose $100\,$nm for the grating period in MAQRO because it is the shortest wavelength that will be achievable in space in the foreseeable future.

\begin{figure}[bht]
 \begin{center}
  \includegraphics[width=0.5\linewidth]{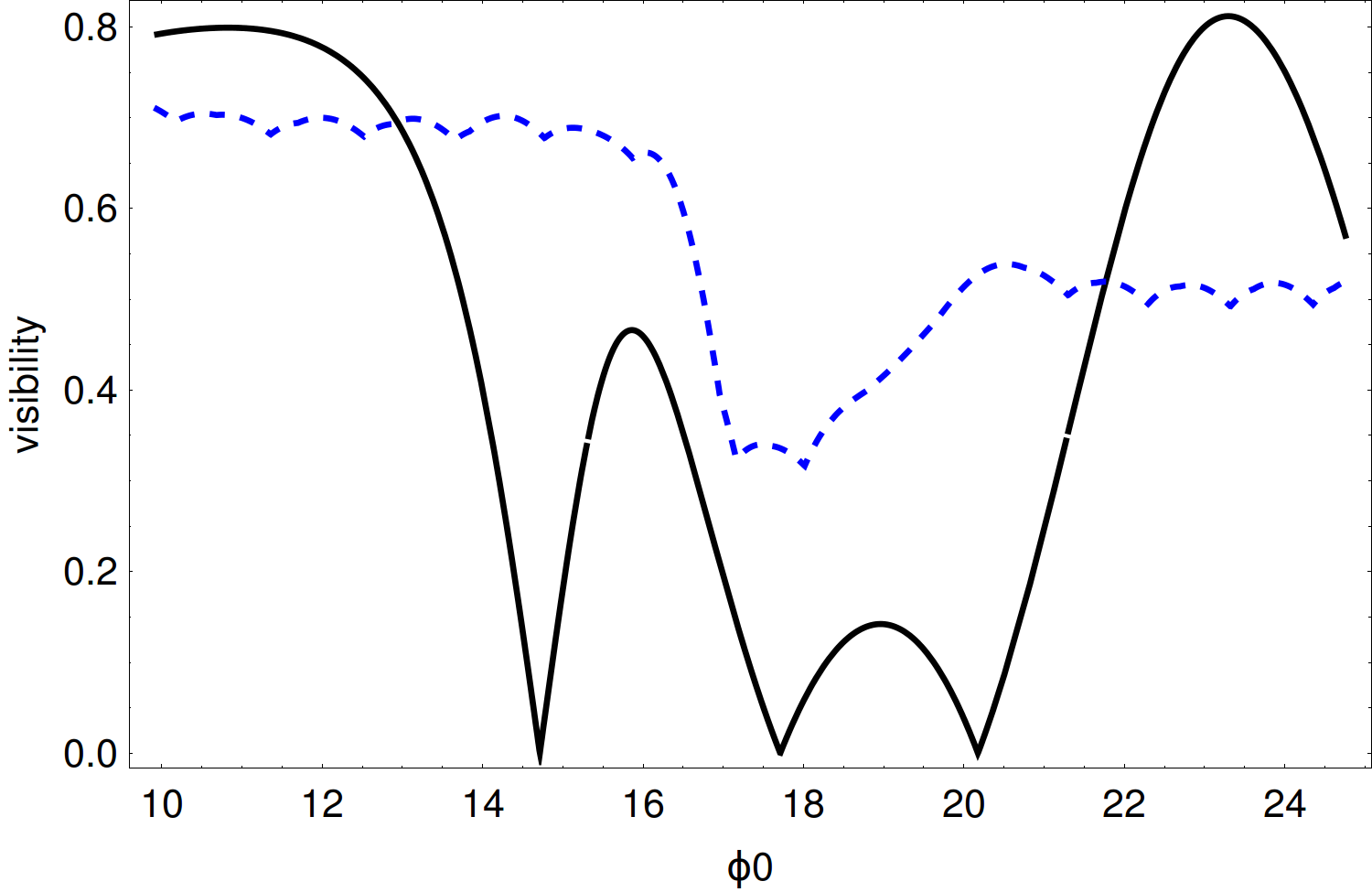}
  \caption{\textbf{Interference visibility for high masses.} Comparison of quantum visibility (solid, black) and classical visibility (dashed, blue) for $m=10^{10}\,$amu.\label{fig::highmassVis}}
  \end{center}
\end{figure}

\begin{figure}[htb]
 \begin{center}
  \includegraphics[width=0.5\linewidth]{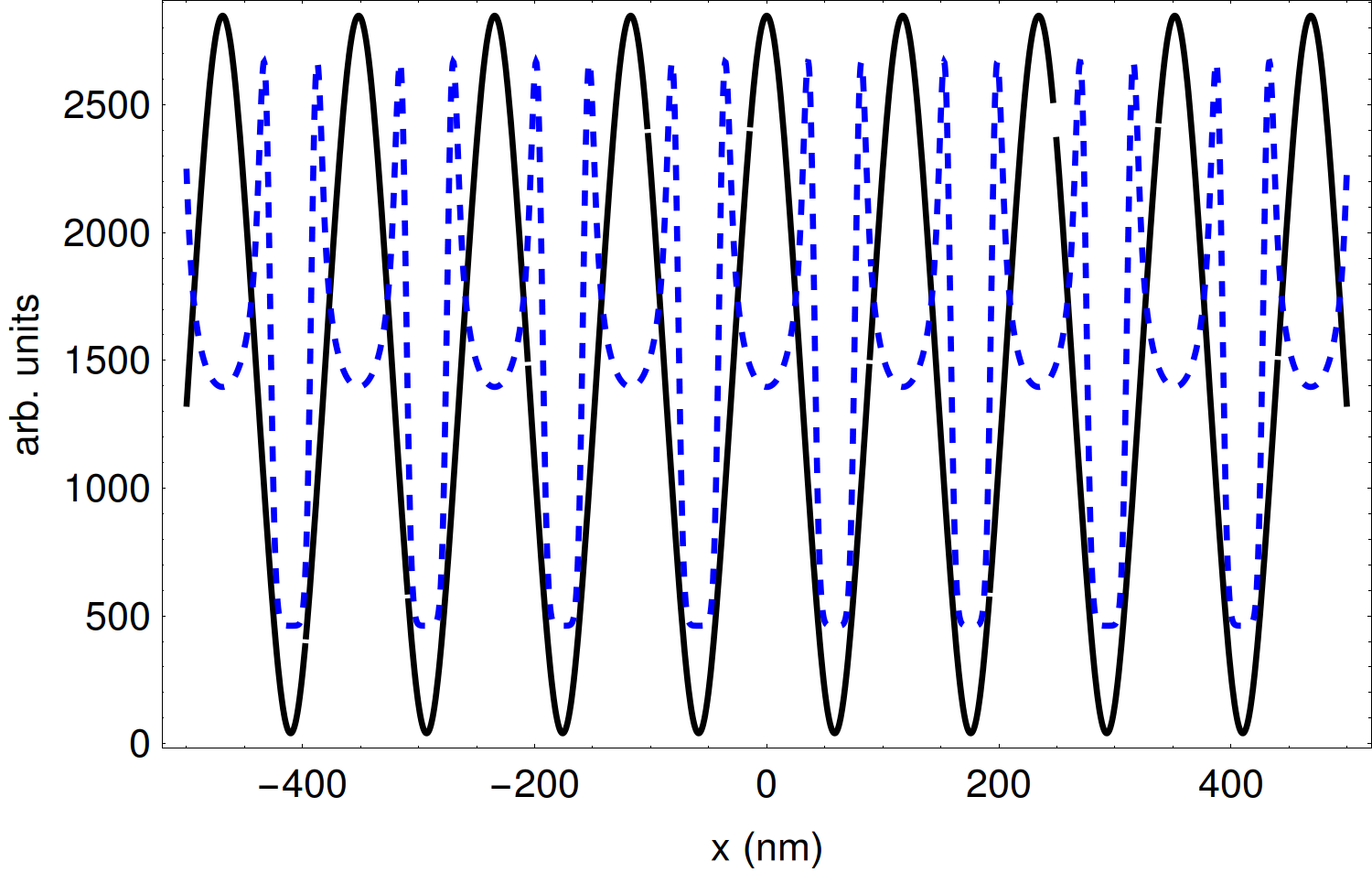}
  \caption{\textbf{Expected high-mass interference patterns.} The black, solid line is the quantum prediction for test particles with $m=10^{10}\,$amu. The blue, dashed line is the classical prediction. The two patterns are qualitatively different.\label{fig::highmassPattern}}
  \end{center}
\end{figure}

While this is not a strict limit, the interference pattern observed will become ever closer to the classically expected one for increasing mass. Figure~\ref{fig::highmassVis} shows that high-visibility interference is still possible for $m=10^{10}\,$amu, and the dependence on $\phi_0$ allows a clear distinction between classical and quantum interference patterns. Figure~\ref{fig::highmassPattern} compares the classical and quantum predictions.

We also mentioned earlier that the power we need to apply for the phase grating becomes higher for smaller particles. For a duration of $1\,\mathrm{\mu s}$ of the grating and a fused-silica particle of mass $m=10^8\,$amu, the optical power would need to be $5\,$mW. For a mass of $m=10^9\,$amu, the required power would still be $0.5\,$mW. For $m=10^8\,$amu, we can instead use a phase grating with $\lambda_G=1064\,$nm. For that wavelength, the necessary power of $6\,$mW is easy to supply -- in particular, if we use a low-finesse cavity for enhancing the power applied.

\subsection{Test particles}
\label{subsec::scireq:particles}
To fulfil MAQRO's scientific goal of testing the predictions of quantum physics and to compare them with the predictions of competing models over a wide parameter space, MAQRO needs to operate with test particles of various \textbf{sizes and materials}. In particular, MAQRO requires particles with different mass densities to test the dependence of the measurements results on particle mass. Collapse models typically depend more strongly on particle mass than quantum physics, which facilitates their experimental distinction.

Known decoherence mechanisms like the scattering, emission and absorption of blackbody radiation depend strongly on the particle size. Performing experiments with particles of different radii will enable tests of such decoherence mechanisms in a new size range while, at the same time, allowing to test alternative theoretical models.

Because MAQRO relies on optically trapping particles, the particles must be \textbf{dielectric and highly transparent}. The particles should also be \textbf{uncharged}. Otherwise, there could be additional, strong decohering mechanisms, and the particles may get lost due to electrostatic interaction with the potentially charged optical bench. The particles do not necessarily need to be spherically symmetric. If they are not, the rotational degree of freedoms need to be cooled in addition to the translation degrees of freedom\cite{Chang2009a,RomeroIsart2010a}.

MAQRO uses scientific heritage from LPF with respect to $\mathbf{1064}\,$\textbf{nm} optics and a $1064\,$nm laser system. For that reason, the test particles need to be transparent at this wavelength. Possible choices for highly transparent materials at this wavelength are various types of fused silica, hafnia ($\mathrm{HfO_2}$) and diamond. The mass density of these materials ranges from $\rho=2200\,\mathrm{kg\,m^{-3}}$ (fused silica) to $\rho=9700\,\mathrm{kg\,m^{-3}}$ (hafnia).

The scientific goal of MAQRO is to perform tests in the \textbf{mass range} from $10^8\,$amu  to $\sim 10^{10}\,$amu. Using fused silica with $\rho=2200\,\mathrm{kg\,m^{-3}}$, we can cover this mass range with a nanosphere \textbf{size range} of $30\,$nm to $120\,$nm. Using other materials, MAQRO can perform tests for even higher particle masses. The size of the test particles will be comparable to the grating period. In order to get large enough phase shifts, the particle sizes will, therefore, have to be chosen to fulfil Mie-resonance conditions. If this is taken into account, then the relatively large size of the particles will not be a concern. This is discussed in detail in the thesis of S. Nimmrichter\cite{Nimmrichter2014a}.

\subsection{Particle loading}
\label{subsec::scireq:loading}
The loading mechanism for loading single, dielectric particles into the optical cavity used for state preparation is a central element of MAQRO. For each measurement, it is required to deliver, on demand a single particle to the optical trap. In order to not significantly prolong the time for a measurement run, the \textbf{time for particle loading} needs to be short compared to the measurement time $T=100\,$s. The particles delivered have to be neutral and should have an internal temperature $T_i\le 25\,$K as described in section \ref{subsec::scireq:decoherence}.

\subsection{State preparation}
\label{subsec::scireq:prep}
A prerequisite of MAQRO is that the motion of the trapped particle can be cooled close to the quantum ground state. This is not necessary for the high-mass interferometry scheme as proposed in Ref.~\cite{Bateman2014a}. For MAQRO, however, it is imperative that the particle remains limited to a defined region around the original trapping position while the wave function expands. On the one hand, this is necessary in order for the particle to stay within the UV beam used for the phase grating. On the other hand, particles lost from the experimental region might get stuck to optical elements on the optical bench. Such a contamination of the optical elements would eventually lead to a reduction in performance of MAQRO.

For these reasons, it is paramount that the motion of the trapped particle is cooled close to the ground state of motion along the cavity axis. Along the axes perpendicular to the cavity axis, the mechanical frequency is much lower but the occupation in this direction should, in energy, also correspond to the occupation along the cavity axis. In order for the particle to stay within a radius of $1\,$mm (the waist of the UV beam), we require an \textbf{occupation number} of $\lesssim 10$ along the cavity and $\lesssim 10^4$ perpendicular to that.

\subsection{Minimizing decoherence effects}
\label{subsec::scireq:decoherence}
As we have stated earlier, in order to be able to see high-mass matter-wave interference in MAQRO, we have to ensure that decohering effects are small enough. In particular, the decoherence parameter $\Lambda$ has to fulfil $\Lambda\le 10^{13}\,\mathrm{m^{-2} s^{-1}}$ (see Figure~\ref{fig::VisReduction}). From this, and assuming fused-silica test particles, one can conclude that the internal temperature of our particles has to fulfil $T_i\le 45\,$K, and the environment temperature also has to fulfil the same requirement $T_e\le 
45\,$K. Given these requirements, decoherence due to scattering, emission and absorption of blackbody radiation will be small enough to observe high-mass matter-wave interference.

\begin{figure}[htb]
 \begin{center}
  \includegraphics[width=0.5\linewidth]{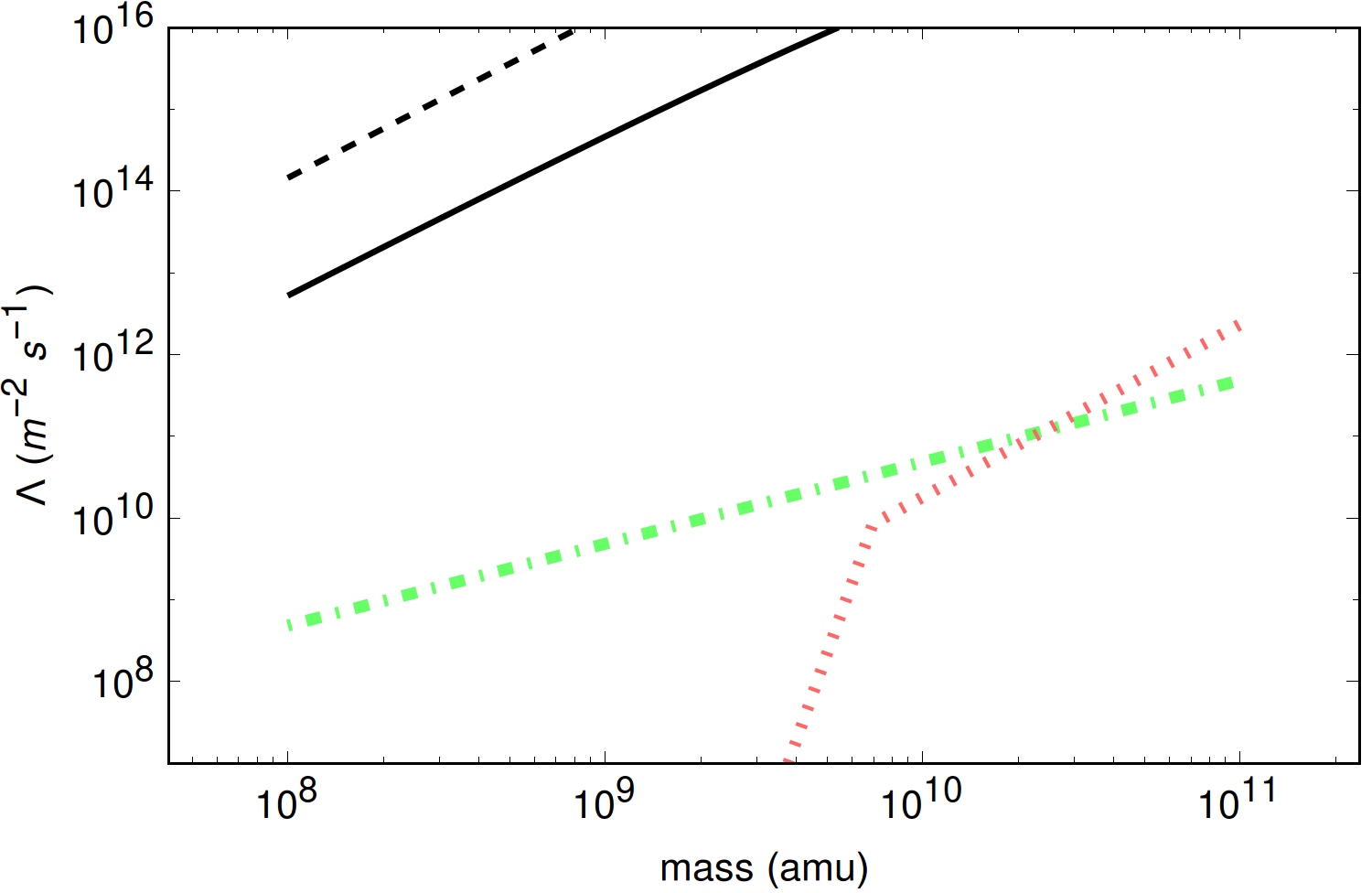}
  \caption{\textbf{Decoherence parameter for various collapse models.} We plot the decoherence parameter $\Lambda$ as a function of mass. CSL model: solid, black; QG model: dashed, black; DP model: dot-dashed, green; K model: dotted, red.\label{fig::collapseLambdas}}
  \end{center}
\end{figure}

However, in order to test for deviations from quantum physics like those predicted by collapse models, the usual decoherence mechanisms should be at most of the same size as the decoherence mechanisms we want to test for. Figure~\ref{fig::VisReduction} shows that MAQRO could, in principle, detect any decoherence mechanisms with a parameter $\Lambda\ge \Lambda_\mathrm{min} =10^{10}\,\mathrm{m^{-2} s^{-1}}$ because they would lead to a noticeable reduction in interference visibility. In order to achieve such low level of decoherence, the requirements on the internal temperature of the test particles and on the environment temperature are accordingly more stringent.

The particle temperature will always be larger than the environment temperature. By limiting the environment temperature to $\le 20\,$K, and the particle temperature to $\lesssim 25\,$K, we can limit the respective decoherence to $\Lambda \lesssim 10^{11}\,\mathrm{m^{-2} s^{-1}}$. If these requirements are not fulfilled but allow for seeing interference in principle, one will have to carefully account for known decoherence mechanisms and check for any additional reduction of interference visibility.

In Figure~\ref{fig::collapseLambdas} we plot $\Lambda$ for various decoherence models. MAQRO allows testing alternative theoretical models if they predict a $\Lambda>\Lambda_\mathrm{min}$. The plots show that MAQRO can test the CSL model and the QG model already for masses starting from $m=10^{8}\,$amu. In order to also test the DP model and the K model, the particle mass has to be on the order of $m=10^{10}\,$amu.

\begin{figure}[htb]
 \begin{center}
  \includegraphics[width=0.5\linewidth]{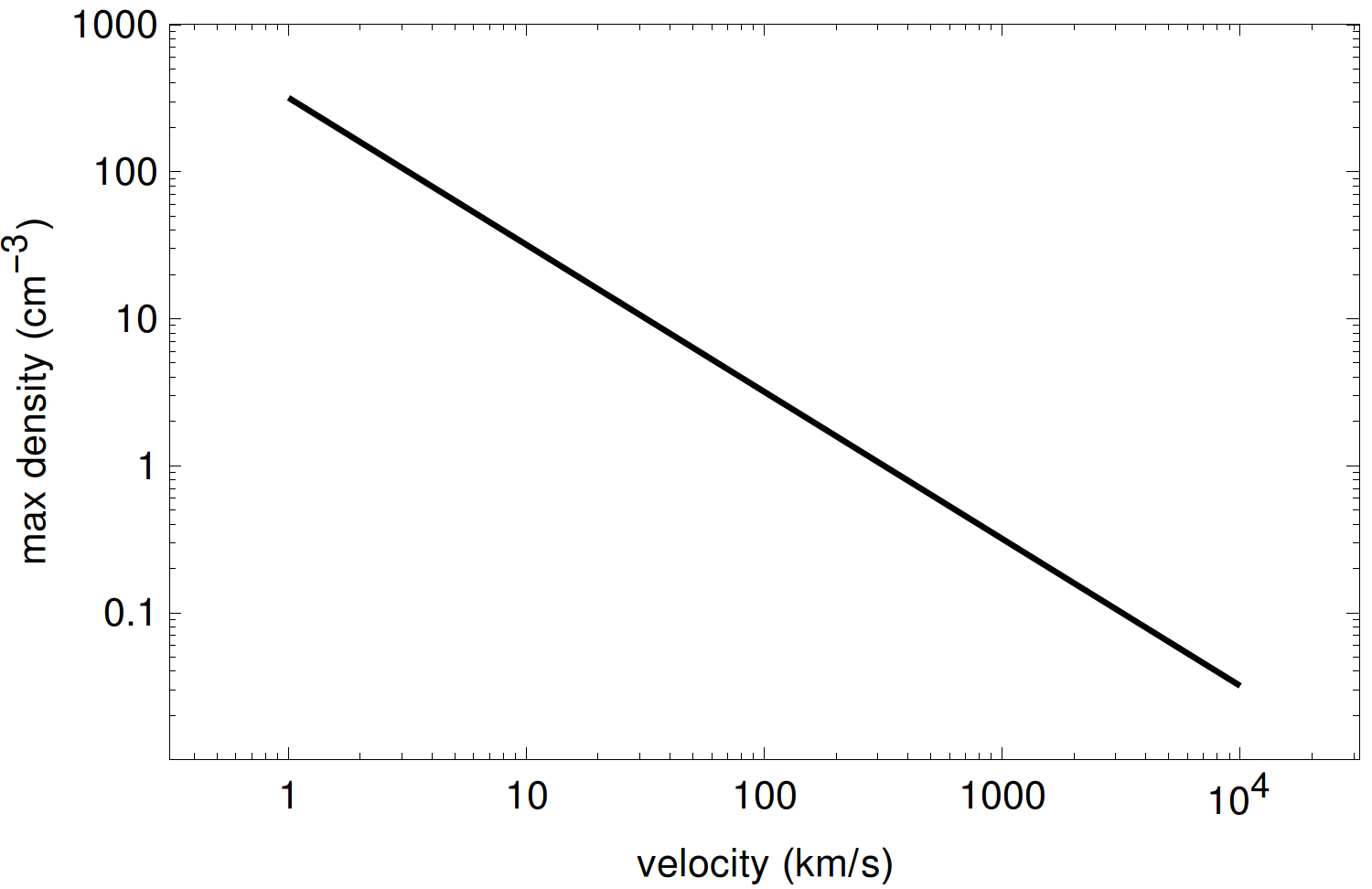}
  \caption{\textbf{Maximum particle density for high-velocity particles.} Conservative estimate of the maximum particle density allowed as a function of particle velocity.\label{fig::solarwind}}
  \end{center}
\end{figure}

An additional decoherence effect may be collisions between the test particles and various atoms or molecules -- i.e., in imperfect vacuum conditions. The de-Broglie wavelength\cite{Broglie1923a} of such particles will always be significantly shorter than the size of our test particles and the size of the quantum states investigated. For that reason, already one or a very few collisions of our test particles with such other particles will decohere our quantum state. The frequency of such collisions can roughly be estimated as:
\begin{equation}
\nu_c = \pi r^2 v_g \rho.\label{equ::gascollisionfreq}
\end{equation}
This scattering cross section assumes that every particle geometrically hitting the test particle will effectively decohere the quantum state. Let us further assume that $T=100\,$s, and that the gas-particle velocity is $700\,$m/s for $T_e=20\,$K (that's a worst-case scenario: hydrogen atoms in equilibrium -- the lower the particle mass, the higher the equilibrium velocity). If we want to have less than one collision during a measurement run, this limits the gas density to $\rho<=500\,\mathrm{cm^{-3}}$. For thermal equilibrium at $T_e=20\,$K, this corresponds to a pressure of $p\le 10^{-13}\,$Pa. For faster particles (e.g., direct exposure to solar wind), this limit is accordingly more stringent as illustrated in Figure~\ref{fig::solarwind} but, at the same time, the particle density is expected to drop for higher particle energies. These requirements may be relaxed upon more detailed investigation of the scattering cross sections of the particles present at the MAQRO orbit. Moreover, the particle density is expected to be reduced due to the wake-shield effect of the spacecraft and the thermal shield.

\subsection{Microgravity environment}
\label{subsec::scireq:microgravity}
During the time the test particle is in free fall, it is subject only to gravitational forces. Due to field gradients, the spacecraft and the test particle will experience slightly different gravitational fields. In addition, the spacecraft itself is the source of a gravitational field. If we assume a spacecraft mass of $250\,$kg, a particle mass of $m=10^{10}\,$amu, and an effective distance of $1\,$m between the two masses, gravitational attraction towards the spacecraft will displace the test particle by  $\sim 80\,\mathrm{\mu m}$ over a time of $100\,$s. While this is significantly less than the wave-packet expansion during that time, it has to be taken into account very accurately. Gravitational fields parallel to the measurement axes defined by the two cavities illustrated in Figure~\ref{fig::MAQROtalbotScheme} have to be known even better. Especially in the direction in which we want to observe high-mass matter-wave interference, the position of the particle has to be known much better than the grating period of $100\,$nm. 

If we are to compensate for the gravitational field of the spacecraft itself or if we want to compensate solar radiation pressure acting on the spacecraft, we have to use micro thrusters. However, such thrusters inevitably have force-noise, which effectively leads to a random walk of the spacecraft. If this random walk is known, then changes of the position of the spacecraft relative to the test particle can be taken into account in the measurement results. If the random walk is not known, then it may blur the interference pattern similar to decoherence. In particular, if we assume white thruster force noise $\mathrm{FN}_0\,\mathrm{N/\sqrt{Hz}}$, then the effect of thruster noise on the interference pattern can be described via an effective ``decoherence'' parameter:
\begin{equation}
\Lambda_\mathrm{th} = \frac{2 \mathrm{FN}^2_0 m^2}{\hbar^2 M^2},\label{equ::thrusternoise}
\end{equation}
where $M$ is the mass of the spacecraft, and $m$ is the mass of the test particle. As an example, for $M=250\,$kg, $m=10^{10}\,$amu, and for a thruster force noise of $\mathrm{FN}_0=100\,\mathrm{nN/\sqrt{Hz}}$ as in LPF, we get $\Lambda_\mathrm{th}=8\times 10^{15}\,\mathrm{m^{-2} s^{-1}}$. This shows that thruster noise is a critical issue. As mentioned earlier, this is not a problem if the random walk of the spacecraft is known to high enough precision. The precision necessary is not the same in all three spatial directions.

Parallel to the UV cavity (and the interference pattern), the effective ``decoherence'' due to the random walk has to fulfil $\Lambda \le  \Lambda_\mathrm{min}$. In terms of accuracy for acceleration measurements along this axis this corresponds to $\le 1\,\mathrm{(pm s^{-2})/\sqrt{Hz}}$. Parallel to the IR cavity, the requirement is defined by the position accuracy of $100\,$nm we need for accurately measuring wave-function expansion (see section~\ref{subsec::sci:WAX}). This results in $\le 100\,\mathrm{(pm s^{-2})/\sqrt{Hz}}$  accuracy for acceleration measurements. Perpendicular to the IR and the UV cavity, the requirement is more relaxed because the position only has to be known much better than the waist of the IR cavity mode ($\sim 60\,\mathrm{\mu m}$). This results in $\le 5\,\mathrm{(nm s^{-2})/\sqrt{Hz}}$ for acceleration measurements

\subsection{Position detection}
\label{subsec::scireq:position}
The period of the interference patterns to be observed will be only slightly larger than the grating period of $100\,$nm. For that reason, in order to resolve these patterns, we need to detect the position of the test particles with accuracy much better than $100\,$nm along the direction of the UV cavity. Along the IR cavity, the position accuracy only needs to be $100\,$nm in order to achieve high enough accuracy for measuring wave-packet expansion (see section~\ref{subsec::sci:WAX}). In the direction perpendicular to the UV and the IR cavities, the accuracy has to be much better than the IR cavity waist ($\sim 60\,\mathrm{\mu m}$) to enable taking into account the IR wave-front curvature.

\section{Proposed scientific instrument}
\label{sec::instr}
To fulfil the stringent requirements on the environment temperature and the particle density of the residual gas, MAQRO is divided into two subsystems. The ``outer subsystem'' (see section~\ref{subsec::instr:Outer}) is placed outside the spacecraft and isolated from the spacecraft via thermal shields. The inner subsystem (see section~\ref{subsec::instr:Inner}) contains most optical and electronic equipment. Optical fibres and an electric harness provide the interface between the two. In Table~\ref{tab::techreq}, we provide an overview of the technical requirements of MAQRO.

\begin{table}[hbt]
\begin{tabular}{llll}
  \hline
  \multicolumn{2}{l}{\textbf{Parameter}} & \hspace{0.5cm} & \textbf{Requirement} \\\hline
  \multicolumn{2}{l}{Nominal mission lifetime (without possible extension)} & & 2 years \\
  \multicolumn{2}{l}{Environment temperature} & & $<20\,$K \\
  \multicolumn{2}{l}{Acceleration sensitivity} & & \\
   \hspace{0.5cm} & along UV cavity & & $\lesssim 1\,\mathrm{(pm/s^2)/\sqrt{Hz}}$ \\
    & along IR cavity & & $\lesssim 100\,\mathrm{(pm/s^2)/\sqrt{Hz}}$ \\
    & perpendicular to optical bench & & $\lesssim 5\,\mathrm{(nm/s^2)/\sqrt{Hz}}$ \\
  \multicolumn{2}{l}{Optical-trapping occupation number} & &  \\
    & along cavity & & $\lesssim 10$ \\
    & orthogonal to cavity & & $\lesssim 10^4$ \\
  \multicolumn{2}{l}{Period of phase grating} & & $100\,$nm \\
  \multicolumn{2}{l}{Accuracy of position detection} & &  \\
    & along UV cavity & & $20\,$nm \\
    & along IR cavity & & $100\,$nm \\
    & perpendicular to optical bench & & $\ll 60\,\mathrm{\mu m}$ \\
  \multicolumn{2}{l}{Time for on-demand particle loading} & & $\ll 100\,$s \\
  \multicolumn{2}{l}{Measurement time per data point} & & $\lesssim 100\,$s \\
  \multicolumn{2}{l}{Vacuum -- particle density} & & $< 500\,\mathrm{cm^{-3}}$ \\
  \multicolumn{2}{l}{IR-cavity finesse} & & $\gtrsim 3\times 10^4$ \\  
  \multicolumn{2}{l}{IR+UV-cavity finesse for IR} & & $\lesssim 30$ \\
  \multicolumn{2}{l}{IR+UV-cavity finesse for UV} & & no UV cavity \\\hline
\end{tabular}
\caption{\textbf{Overview of the technical requirements of MAQRO.} See also Table~\ref{tab::scireq}.\label{tab::techreq}}
\end{table}

\subsection{Outer subsystem}
\label{subsec::instr:Outer}
The outer subsystem can be divided into several assemblies -- they are listed in Table~\ref{tab::outer}, along with links to detailed descriptions and an assessment of the technology readiness level (TRL) at the time of the M4 mission proposal.

\begin{table}[hbt]
\begin{tabular}{llll}
  \hline
  \textbf{Assembly name} & \textbf{Link to description} & \textbf{Current TRL} & \textbf{Heritage} \\\hline
  Thermal-shield structure & \ref{subsubsec::Outer:Thermal} & 5 & JWST, Gaia \\
  CMOS camera & \ref{subsubsec::Outer:CMOS} & 6 & JWST \\
  Optical-bench assembly & \ref{subsubsec::Outer:Bench} & 6 & LPF \\
  High-finesse IR cavity assembly & \ref{subsubsec::Outer:IRcavity} & 4-5 &  \\
  Low-finesse IR+UV cavity assembly & \ref{subsubsec::Outer:IRUVcavity} & 3 &  \\
  Loading mechanism & \ref{subsubsec::Outer:LM} & 3 &  \\
  Accelerometer & \ref{subsubsec::Outer:Acc} & 5 & GOCE, Microscope, \ldots  \\\hline
\end{tabular}
\caption{\textbf{Overview of the assemblies comprising the outer subsystem.} \label{tab::outer}}
\end{table}

\subsubsection{Thermal-shield structure}
\label{subsubsec::Outer:Thermal}
\begin{figure}[htb]
 \begin{center}
  \includegraphics[width=0.9\linewidth]{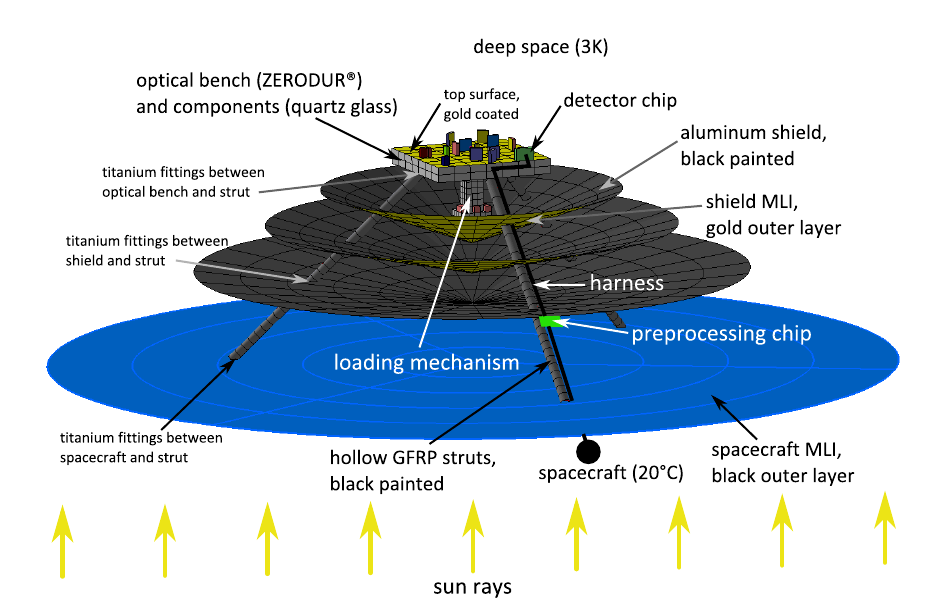}
  \caption{\textbf{CAD drawing of heat-shield geometry.} The structure is attached to the spacecraft facing away from the sun. Three glass-fiber reinforced plastic (GFRP) struts hold three consecutive shields isolating the optical bench from the spacecraft surface (Image source: Ref.~\cite{Pilan-Zanoni2015a}).\label{fig::ShieldGeometry}}
  \end{center}
\end{figure}

This outer subsystem contains as few sources of dissipation as possible to achieve optimal passive cooling by radiating directly to deep space. The design also allows direct venting into space, to achieve an extremely high vacuum level. This concept was originally developed for the M3 mission proposal of MAQRO\cite{Kaltenbaek2013b}, based on related approaches in JWST, GAIA and the Darwin mission proposal\cite{Leger2007a}. The design was refined in an ESA-funded study\cite{Kaltenbaek2012a} and in increasingly detailed thermal simulations\cite{Hechenblaikner2014a,Pilan-Zanoni2015a}. Figure~\ref{fig::ShieldGeometry} shows the shield geometry.

As we stated in section~\ref{subsec::scireq:decoherence}, to see matter-wave interference in MAQRO, the environment temperature has to fulfil $T_e\le 45\,$K. In order to use the interferometer to test for small deviations from the predictions of quantum physics, the environment temperature has to be even lower: $T_e\le 20\,$K. In a thermal study, finite-element simulation was used to demonstrate that these conditions can be fulfilled using the thermal-shield concept of MAQRO\cite{Hechenblaikner2014a}. In particular, it was shown that all elements on the optical bench could passively be cooled to $T_e\sim 27\,$K, and that the immediate volume around the trapped test particle (the ``test volume'') could reach an even lower temperature $T_e\sim 16\,$K. This thermal study confirmed that the shield geometry was near optimal. In particular, more than three consecutive shields would not bring a significant advantage, while reducing the number of shields to two would lead to a significant increase in the temperature achievable. These results could be further improved in a more detailed thermal analysis. In particular, this was achieved by using reflective instead of refractive optics\cite{Pilan-Zanoni2015a} -- yielding a temperature of $T_e\sim 25\,$K for the optical bench and $T_e\sim 12\,$K for the test volume.

The design of the heat shield is based broad technological heritage and the use of space-proof materials. That means, all structural components of the heat shield are space-proof. For this reason, we assess at least \textbf{TRL~5} for this assembly.

\subsubsection{CMOS camera}
\label{subsubsec::Outer:CMOS}
Optical detection of the position of the test particles plays a central role in MAQRO. To this end, several techniques are combined. One of these techniques is to detect scattered light. For this purpose, we can use technological heritage for a CMOS camera from the James Webb Space Telescope (JWST)\cite{Loose2006a,Bai2008a}. In particular, this technology has been designed in order to allow a separation of the CMOS detector chip from the preprocessing chip\cite{Loose2006a}. This way, the detector chip with low dissipation can be placed on the optical bench while the preprocessing chip (higher dissipation) can be placed further away from the sensitive experimental region. This is illustrated in Figure~\ref{fig::ShieldGeometry}. For this technology, we estimate \textbf{TRL~6} or higher.

\subsubsection{Optical-bench assembly}
\label{subsubsec::Outer:Bench}
In Figure~\ref{fig::MAQROtalbotScheme}, we assumed that we would potentially use two orthogonal cavities which we denote here as the IR (high-finesse) cavity and a low-finesse IR+UV cavity. The latter was assumed to potentially be a dual-wavelength cavity for $1064\,$nm and for $\sim 200\,$nm. However, a more detailed analysis shows that we will not be able to use a $\sim 200\,$nm cavity due to reasons of thermal stability. This is discussed in more detail in subsection \ref{subsubsec::Outer:IRUVcavity}. Nevertheless, we will denote the cavity as IR+UV cavity to distinguish it from the high-finesse IR cavity.

\begin{figure}[htb]
 \begin{center}
  \includegraphics[width=0.6\linewidth]{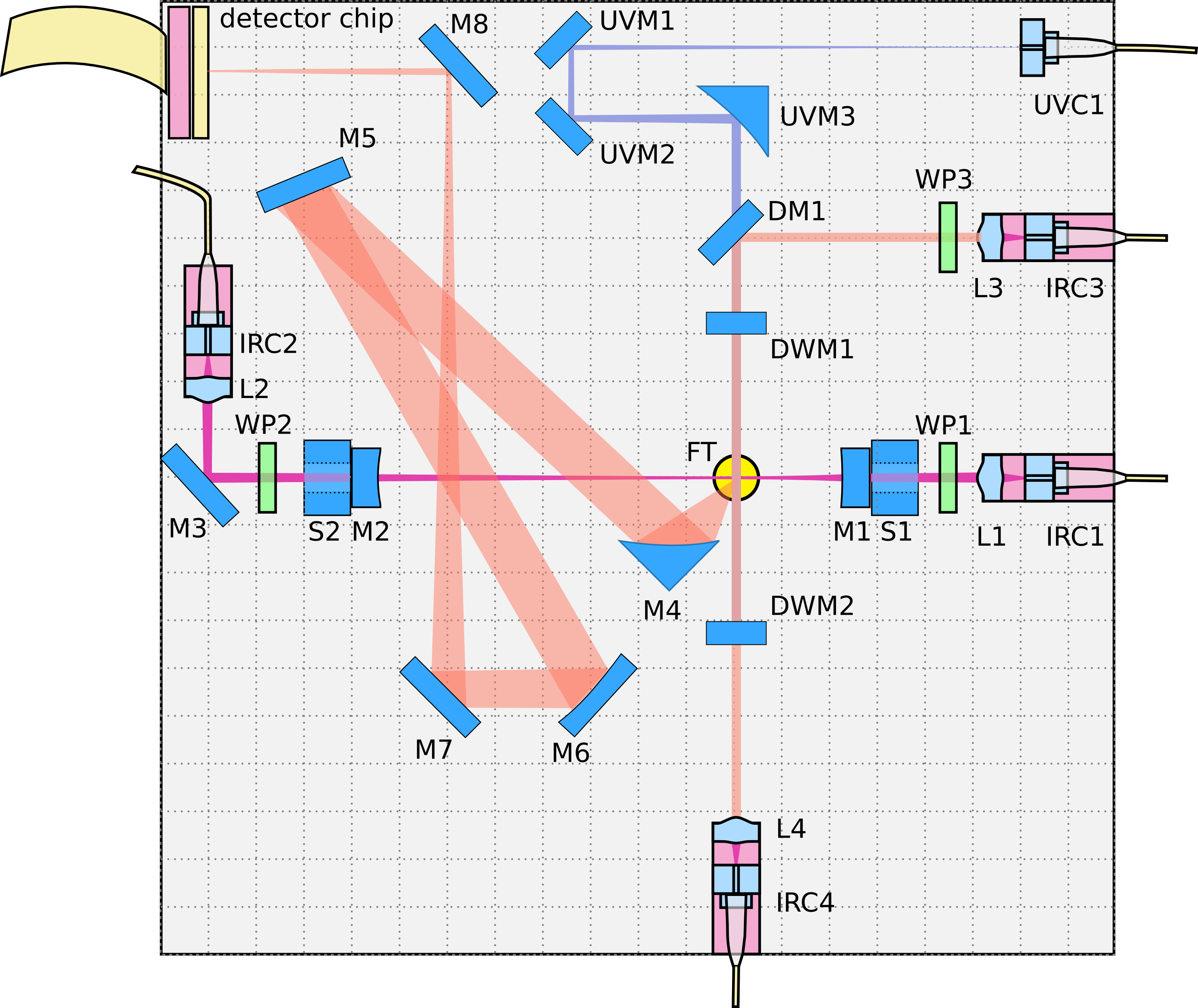}
  \caption{\textbf{Top view of the optical bench.} The optical bench is $20\times 20\,\mathrm{cm^2}$ large. UVM: UV mirrors; M: IR mirrors; DM: dichroic mirrors; DWM: dual-wavelength mirrors; UVC: UV couplers; IRC: IR couplers; WP: quarter-wave plates; L: lenses; FT: base-plate feed-through S: spacers holding cavity mirrors. The mirrors M1 and M2 form a high-finesse IR cavity containing several modes (violet beam path originating at IRC1). DWM1 and DWM2 form a low-finesse cavity for IR light. The IR beam is indicated in light red, originating from IRC3 and coupled in again at IRC4. The UV beam originates at UVC1. The red-shaded, broad path indicates scattered-light imaging.\label{fig::opticalBench}}
  \end{center}
\end{figure}

Based on these considerations, Figure~\ref{fig::opticalBench} shows the optical assembly on top of the optical bench. As we sketched earlier in Figure~\ref{fig::MAQROtalbotScheme}, the main elements are two orthogonally oriented cavities: a high-finesse cavity for $1064\,$nm light formed by the mirrors M1 and M2. For increased stability and for easier alignment, these mirrors are mounted on blocks of ULE material with a centre hole (``spacers'' S1 and S2). A second cavity (low-finesse, dual-wavelength for $\sim 200\,$nm and $1064\,$nm) is formed by the dual-wavelength mirrors DWM1 and DWM2.

Four IR fibre couplers IRC1 to IRC4 supply the optical bench with IR light and/or couple it back in again for further use. UVM3 and M4 are parabolic mirrors. Mirrors M4-M8 optically image light scattered by nanoparticles onto the CMOS detector chip. The light is focused on the detector by the concave mirror M6. Using reflective optics is preferred over refractive optics for thermal considerations (see subsection~\ref{subsubsec::Outer:Thermal}).

IR and UV light are combined using the dichroic mirror DM1. DWM1 is highly transparent and DWM2 is highly reflective for $200\,$nm light. At the same time, both DWM1 and DWM2 should be reflective enough to form a low-finesse IR cavity. At the exit of the cavity, the IR light is coupled back into IRC4. The UV light expands freely from the UV coupler UVC1 and is collimated by UVM3 to a beam with $1\,$mm waist. UV light reflected at DWM2 is coupled back into UVC1 again.

The region denoted as FT (feed-through) is a hole through the optical-bench base plate. It allows test particles to be passed from the loading mechanism below the optical bench (see subsection~\ref{subsubsec::Outer:LM}) to the trapping region within the IR cavity.

There exists direct technological heritage for all parts of the optical-bench assembly except for the high-finesse IR cavity and for the IR+UV cavity. For this reason, we assess the TRL of the optical-bench assembly (without the cavities) to be \textbf{TRL~6} or higher. However, since the optical bench will be operated at cryogenic temperatures instead of room temperature as in LPF, this will need to be carefully considered and tested. Our assessment of the technological readiness of the cavity assemblies is described in subsections \ref{subsubsec::Outer:IRcavity} and \ref{subsubsec::Outer:IRUVcavity}.

\subsubsection{High-finesse IR cavity assembly}
\label{subsubsec::Outer:IRcavity}
As described in section~\ref{subsec::scireq:prep}, the preparation of quantum states in MAQRO requires cooling the centre-of-mass motion of optically trapped test particles close to the quantum ground state. To this end, MAQRO will apply a combination of intra-cavity side-band cooling and feed-back cooling\cite{Gieseler2012a,Kiesel2013a,Kubanek2009a,Yin2011a}. This requires good optomechanical coupling as well as a high-finesse cavity. The cavity of MAQRO has a cavity length of $97\,$mm. We chose this value for the cavity to be as long as possible given the size of the optical bench. This way, we minimize the solid angle covered by the ``hot'' cavity mirrors from the point of view of the test particle. The reasoning behind this is to optimize the thermal environment for passive cooling. Because of the large length of the cavity, it has to be asymmetric in order to achieve high enough optomechanical coupling. The precise value of $97\,$mm results from choosing standard radii of curvature of $30\,$mm and $75\,$mm for the cavity mirrors. Given this cavity geometry, we require a minimum finesse of $3\times 10^4$ to achieve cooling close to the quantum ground state and to achieve high enough intra-cavity power and a longitudinal mechanical frequency on the order of $\omega_{m,L} = 10^5\,\mathrm{rad/s}$.

In a recent project (MAQROsteps, Project Nr. 840089) funded by the Austrian Research Promotion Agency (FFG), R. Kaltenbaek and his team implemented an adhesively bonded high-finesse IR cavity for optomechanical experiments in ultra-high vacuum. They used space-proof gluing technology and ultra-low-expansion (ULE) material to implement a stably bonded cavity with a finesse of $\mathcal{F} = 10^5$. Their efforts effectively increased the TRL of this technology to TRL 4-5 (relevant environment with respect to vacuum level but not with respect to environment temperature, no radiation and vibration tests). The cavity implemented only had a cavity length of $13\,$mm. Until mid-2015, they will use a similar approach to demonstrate an adhesively bonded cavity with the same geometry as needed for MAQRO.

\subsubsection{Low-finesse IR+UV cavity}
\label{subsubsec::Outer:IRUVcavity}
Originally, we intended using a dual-wavelength cavity for $1064\,$nm and $\sim 200\,$nm to benefit from intra-cavity power enhancement for the $200\,$nm light and to achieve good position read out using IR light. However, practical limitations prevent the use of a UV cavity, and the finesse of the cavity for the IR wavelength has an upper limit. 

The reason is that the phase grating has to be applied during a very short time ($\sim 1\,\mathrm{\mu s}$) after a long time of free expansion $t_1$. During this time of free expansion, the IR and UV lasers cannot be locked to the cavity. Therefore, the IR and UV beams could not be turned on again for a short time without first locking the laser to the cavity again.

\begin{figure}[htb]
 \begin{center}
  \includegraphics[width=0.4\linewidth]{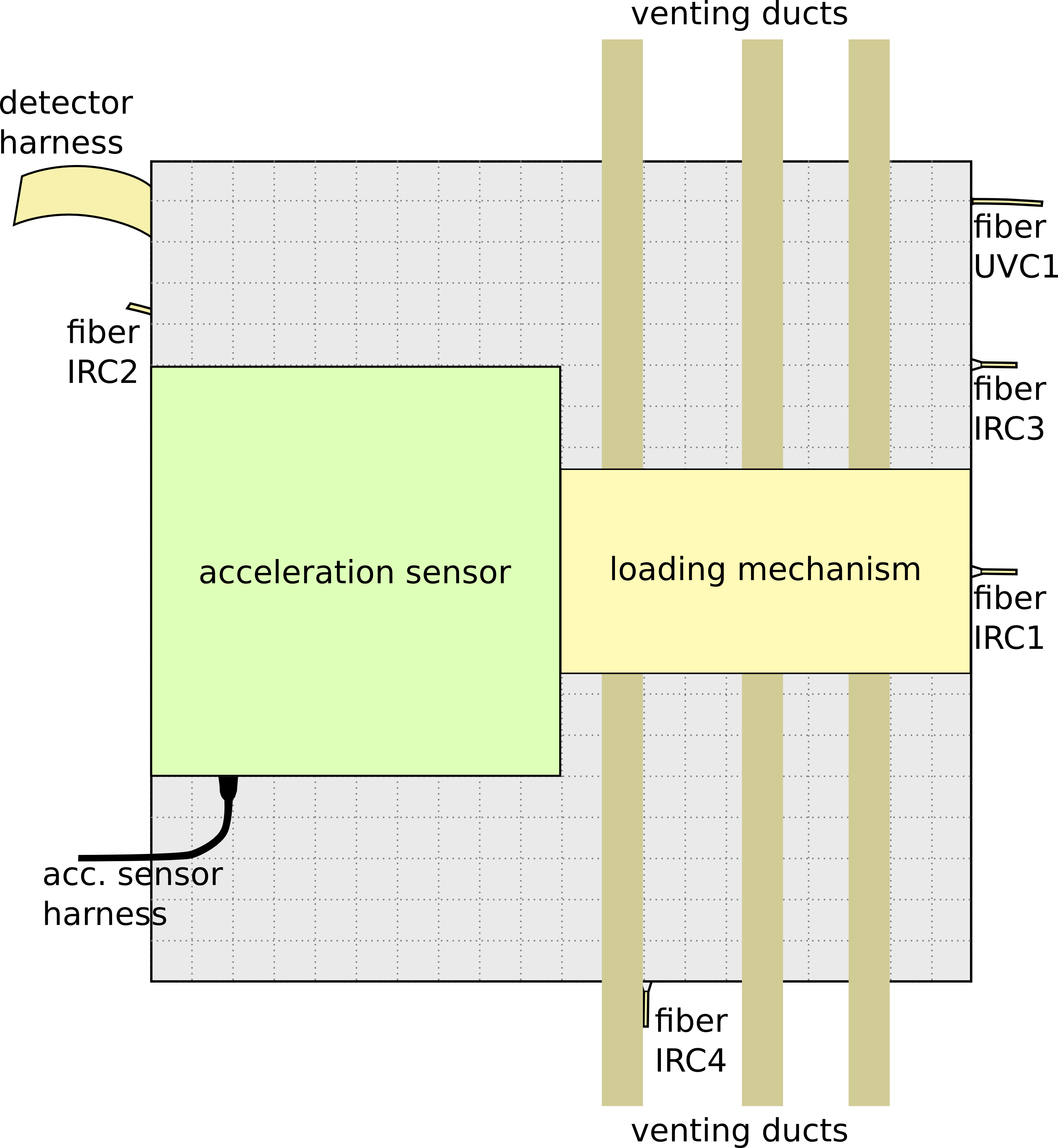}
  \caption{\textbf{Bottom view of the optical bench.} The image illustrates where venting ducts could be placed to minimise the amount of buffer gas potentially leaking to the experimental region. The figure also shows the position of the external acceleration sensor and fibers from the top of the optical bench.\label{fig::LMouterBottom}}
  \end{center}
\end{figure}

If we assume that the cavity length $L$ changes by $\delta L$, and if we assume that we were on resonance before that change and are still on resonance afterwards, then we get a lower limit on the cavity linewidth $\kappa$:
\begin{equation}
\kappa = \frac{\pi c}{2 L \mathcal{F}} > \frac{\delta L}{L} \nu = \frac{\delta L}{L} \frac{c}{\lambda}.\label{equ::kappa}
\end{equation}
Since the optical bench will consist of ULE material (SiC or Zerodur), the relative length change can be assumed to be about $\delta L / L\sim 10^{-6}$ if the temperature is kept stable to $1\,$K. In that case, we get an upper limit of $\sim 30$ for the finesse of the IR+UV cavity for $1064\,$nm and $\sim 6$ for $\sim 200\,$nm light. For this reason, we assume that we will not use a cavity for the $\sim 200\,$nm light and a low-finesse cavity for $1064\,$nm light. Currently \textbf{TRL~3}.

\subsubsection{Loading mechanism}
\label{subsubsec::Outer:LM}
\begin{figure}[htb]
 \begin{center}
  \includegraphics[width=0.4\linewidth]{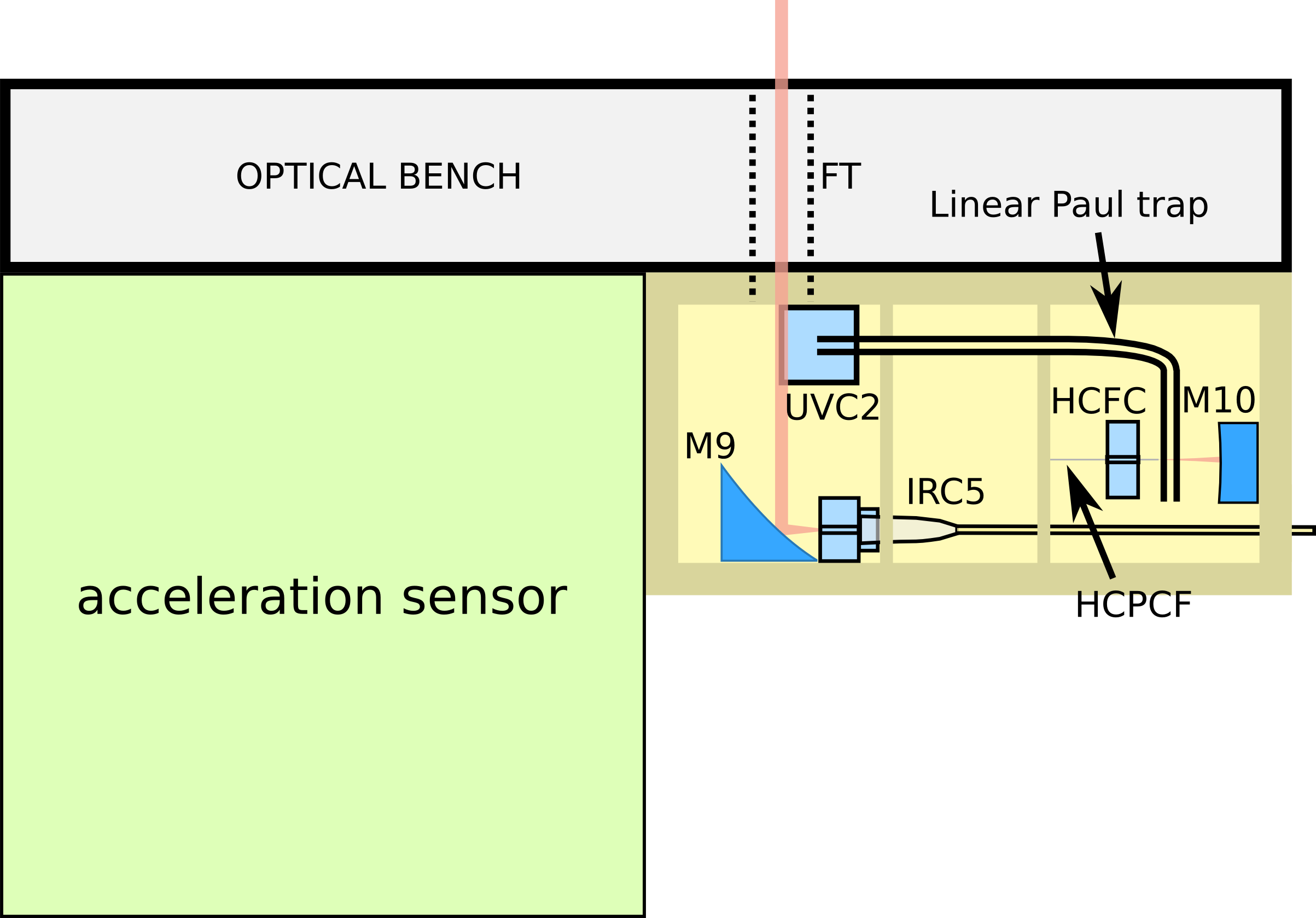}
  \caption{\textbf{Side view of the loading mechanism.} The image illustrates the three sub-divisions of the loading-mechanism chamber. The HCPCF is mounted on a fiber coupler (HCFC) close to the four rod-like electrodes of a linear Paul trap. At this position, the test particles are handed over from the guiding fiber to the Paul trap. This is also where the buffer gas will leave the chamber via the HCPCF. The particles are then guided close to a UV coupler where UV light is used to discharge them. Finally, they will enter an IR beam propelling the particles to the top of the optical bench.\label{fig::LMouterSide}}
  \end{center}
\end{figure}

The main part of the loading mechanism is located in the inner subsystem (see subsection~\ref{subsubsec::Inner:LM}). While that inner part is responsible for dispensing particles from a particle source, and characterising them, the central tasks of the outer part of the loading mechanism are to guide the particles from the inside of the spacecraft to the optical bench, to discharge the particles and to propel them into the optical trap.

In order to transport the test particles from the spacecraft to the optical bench, we will use a hybrid combination of optical trapping and guiding as well as linear Paul trapping. To this end, we use several hollow-core photonic-crystal fibres (HCPCF) with a core diameter of $\sim 10\,\mathrm{\mu m}$. As far as possible, each of these fibres should run independently along one of the struts of the thermal-shield structure. This way, if one fibre were to be damaged for some reason, the chance would be higher for the other fibres not to be affected.

The HCPCFs guiding the test particles will also contain buffer gas to sympathetically cool the particles. The external loading mechanism is contained in a closed chamber that is internally divided into sub chambers (see Figure~\ref{fig::LMouterSide}). Each of these sub chambers will be vented to space in order to prevent buffer gas from reaching the experimental platform (see Figure~\ref{fig::LMouterBottom}).

The amount of gas leaking along the HCPCF outside the spacecraft is small: for example, the pressure inside a $10^3\,\mathrm{cm^3}$ chamber with buffer gas at room pressure and a single HCPCF leading from the chamber would only loose a negligible amount of pressure over the lifetime of the mission. Nevertheless, we have to ensure that the buffer gas does not contaminate the vacuum in the experimental region.

During the early part of the development phase of MAQRO, we will perform finite-element simulations of the behaviour of the buffer gas inside a HCPCF along the length of the fibre and as it exits the fibre at the end. Important questions will be (1) whether sympathetic cooling via the buffer gas allows achieving low enough test-particle temperatures, (2) how much pressure the buffer gas will exhibit on the transported test particles as it exits the fibre end, (3) how badly the buffer gas will contaminate the UHV environment of the optical bench, and (4) the ideal configuration of venting ducts. During a later time of the development period, we plan to investigate these questions experimentally in a representative test environment.

Figures \ref{fig::LMouterBottom} and \ref{fig::LMouterSide} illustrate the general idea of the loading mechanism based on two candidate technologies to be investigated during the development phase. Moreover, the figure shows the position of a UV coupler close to the end of the guiding linear Paul trap. The $200\,$nm light used for the phase grating will also be used in the loading mechanism to discharge the test particles. Finally, an important part of the loading mechanism is a collimated IR beam ($1\,$mm waist) used to propel the particles to the trapping region on top of the optical bench. The same beam will be used at the end of each measurement to dispose of the test particle. We estimate the current TRL as \textbf{TRL~3}.

\subsubsection{Accelerometer}
\label{subsubsec::Outer:Acc}
\begin{figure}[htb]
 \begin{center}
  \includegraphics[width=0.3\linewidth]{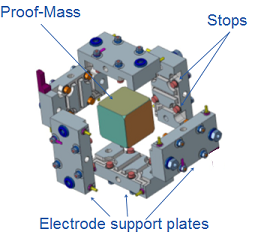}
  \caption{\textbf{Sensor core of the external accelerometer.} The figure shows test mass and electrode housing. Size of sensor core: $\le 10\times 10 \times 10\,\mathrm{cm^3}$, mass: $\le 2\,$kg. Image credit: ONERA.\label{fig::accOuter}}
  \end{center}
\end{figure}

A central prerequisite of MAQRO is to prevent random relative motion between the test particle and the spacecraft (see section~\ref{subsec::scireq:microgravity}). This results in stringent requirements on the accuracy for measuring accelerations of the spacecraft. While there will be an accelerometer at the centre-of-mass of the spacecraft (subsection~\ref{subsubsec::Inner:Acc}), this will not provide direct information about the relative local acceleration between test particle and optical bench. Using a model of the spacecraft to infer that information inevitably reduces the accuracy of the information gained. To achieve the required accuracy of $\le 1\,\mathrm{(pm/s^2)/\sqrt{Hz}}$, MAQRO features a second acceleration sensor close to the test particle (see Figure~\ref{fig::accOuter} as well as Figures \ref{fig::LMouterBottom} and \ref{fig::LMouterSide}).

The ONERA sensor to be used will harness a cubic test mass. Based on past experience of ONERA, in the cryogenic environment close to the optical bench, the sensor sensitivity should fulfil the requirements of MAQRO. The control unit and a power-conversion unit will be placed inside the spacecraft with a distance $\le 2\,$m from the sensor core. Tests on separating the core from the control unit and placing the core in a cryogenic environment were already performed. We estimate \textbf{TRL~5}.

\subsection{Inner subsystem}
\label{subsec::instr:Inner}
The inner subsystem can be divided into several assemblies -- they are listed in Table~\ref{tab::inner}, along with links to detailed descriptions and an assessment of the technology readiness level (TRL) at the time of the M4 mission proposal.

\begin{table}[hbt]
\begin{tabular}{llll}
  \hline
  \textbf{Assembly name} & \textbf{Link to description} & \textbf{Current TRL} & \textbf{Heritage} \\\hline
  IR laser system & \ref{subsubsec::Inner:Laser} & 6 & LPF \\
  UV source & \ref{subsubsec::Inner:UV} & 3 & \\
  IR-mode generation & \ref{subsubsec::Inner:modegeneration} & 3 & LPF \\
  IR-mode locking & \ref{subsubsec::Inner:modelocking} & 3 & LPF  \\
  Data-acquisition subsystem & \ref{subsubsec::Inner:DAQ} & 6 & JWST, LPF \\
  Loading mechanism & \ref{subsubsec::Inner:LM} & 3 &  \\
  Accelerometer & \ref{subsubsec::Inner:Acc} & 5 & GOCE, Microscope, \ldots  \\\hline
\end{tabular}
\caption{\textbf{Overview of the assemblies comprising the inner subsystem.} \label{tab::inner}}
\end{table}

\subsubsection{IR laser system}
\label{subsubsec::Inner:Laser}
For the IR laser system, MAQRO relies on technological heritage from LPF and LISA\cite{Trobs2006a}. We should essentially be able to use the very same laser technology. In particular, this is a highly stable continuous-wave (CW) $1064\,$nm NPRO (non-planar ring oscillator) laser. For MAQRO, we will also need such a laser and keep it locked to the high-finesse cavity on the optical bench. Using an EOM, we will lock a side-band of this laser to the UV+IR cavity. The laser needs to be tunable over at least one full free spectral range (FSR) of the high-finesse IR cavity ($1.5\,$GHz). Due to the LPF heritage, we estimate at least \textbf{TRL~6}.

\subsubsection{UV source}
\label{subsubsec::Inner:UV}
For the phase grating, we need a CW source of $\sim 200\,$nm light with a pulse duration $\le 1\,\mathrm{\mu s}$  and a peak power $\le 0.5\,$mW. While this is not available off the shelf, the necessary amount of delta-development to adapt existing technology for that purpose should be feasible within the development phase of MAQRO.

In particular, there are two possible approaches: (1) frequency double a $\sim 400\,$nm fundamental beam, e.g., using novel developments in cavity-assisted second-harmonic generation using whispering-gallery-mode $\beta$-Barium-Borate resonators\cite{Lin2013a}. Over the last years, single-frequency $\sim 400\,$nm laser diodes in Littrow configuration\cite{Hildebrandt2003a} have become readily available commercially. A fall-back option for space applications to produce the $\sim 400\,$nm pump light is sum-frequency generation using $1064\,$nm light in combination with a $670\,$nm InGaAsP laser diode. Another option (2) to generate the $200\,$nm light is to frequency quintuple $1064\,$nm light adapting the scheme presented in Ref.~\cite{Vasilyev2011a}.

All elements of this have been demonstrated in the lab. We estimate the current readiness to be \textbf{TRL~3}.

\subsubsection{IR-mode generation}
\label{subsubsec::Inner:modegeneration}
In order to optically trap our test particle in the high-finesse IR cavity and to cool its center-of-mass motion in all spatial directions, we intend to use several IR modes. In particular, two TEM00 modes will be used to trap the particle and to cool its motion along the direction of the high-finesse IR cavity\cite{Kiesel2013a}. To also cool the motion of the particle along the two dimensions perpendicular to the cavity mode, we will use higher-order TEM01 and TEM10 modes\cite{Yin2011a}.

The two TEM00 modes are separated in frequency by one free spectral range (FSR) of the high finesse cavity ($\mathrm{FSR} = c/(2 L)\approx 1.5\,$GHz), where $L=97\,$mm is the cavity length. The TEM01 and TEM10 modes are close to each other in frequency and about $1.2\,$GHz from the fundamental TEM00 mode. One can generate the required optical frequency shift of $\sim 1.2\,$GHz from the fundamental modes by using GHz electro-optic modulators (EOMs) for the GHz phase modulation. The modulation frequencies can be separated from the fundamental mode by using a temperature stabilized Fibre-Bragg grating. To generate the required resonance frequencies for the TEM10 and TEM01 modes, one can then use an acousto-optic modulator (AOM) for a frequency shift in the MHz range. AOM and EOM technology is readily available in space as technological heritage from LPF. Spatially, the TEM01 and TEM10 modes can be filtered from the generated light fields by the optical cavity directly. We will also investigate the more efficient conversion of the light fields to these mode shapes by holograms.

In order to combine and later separate again the various laser modes, the two TEM00 modes will be prepared in orthogonal polarization. The higher-order spatial modes will be combined (and separated) based on spatial-mode filtering techniques.

All these techniques are currently being used in the lab. We estimate the current readiness to be \textbf{TRL~3}.

\subsubsection{IR-mode locking}
\label{subsubsec::Inner:modelocking}
The IR laser can be locked to the high-finesse IR cavity by using standard Pound-Drever-Hall (PDH) locking techniques\cite{Pound1946a,Drever1983a}. Since the other optical modes for intra-cavity cooling in the high-finesse IR cavity are derived via EOMs and AOMs from the fundamental laser mode (see subsection~\ref{subsubsec::Inner:modegeneration}), they also follow any changes of the cavity resonance frequency. In addition, these higher-order modes can in turn be locked to the cavity via PDH locking. We will also use an EOM to generate a mode to be locked to the UV+IR cavity. To this end, each of the modes to be locked to the cavities will separately be frequency modulated in the MHz range to allow for the generation of distinct PDH error signals from the light reflected from the cavity.

PDH locking is a standard technique. Its TRL is at least \textbf{TRL~3}.

\subsubsection{Data-acquisition subsystem}
\label{subsubsec::Inner:DAQ}
With this general term we encompass a host of sensors and devices providing information about the performance of the instrument and delivering the measurements results. All these devices are readily available in a laboratory environment and, in particular, given the technological heritage from LTP, we estimate the current TRL to be \textbf{TRL~6}. 

\subsubsection{Loading mechanism}
\label{subsubsec::Inner:LM}
Significant progress has been achieved over the last years towards loading nanoparticles into optical traps at ultra-high vacuum (UHV). Common methods include sprays creating microdroplets of liquid solution containing particles (e.g., Refs.~\cite{Kiesel2013a,Gieseler2012a}) at comparatively high ambient pressure and then using vacuum pumps to reduce the ambient pressure once particles are optically trapped. In this approach, it is paramount to actively cool the motion of the trapped particles in order to achieve low ambient pressure\cite{Gieseler2012a,Gieseler2013a}. Other possible approaches include using hybrid optical and Paul traps in combination with particles launched from a ceramic piezoelectric speaker\cite{Millen2014b}, or the transfer of trapped particles at high pressure (as discussed above) to hollow-core photonic-crystal fibres (HCPCFs). The idea is that it may be possible to guide the particles inside a HCPCF from a high to a low pressure environment\cite{Benabid2002a,Schmidt2013a,Grass2013a}.

\begin{figure}[htb]
 \begin{center}
  \includegraphics[width=0.3\linewidth]{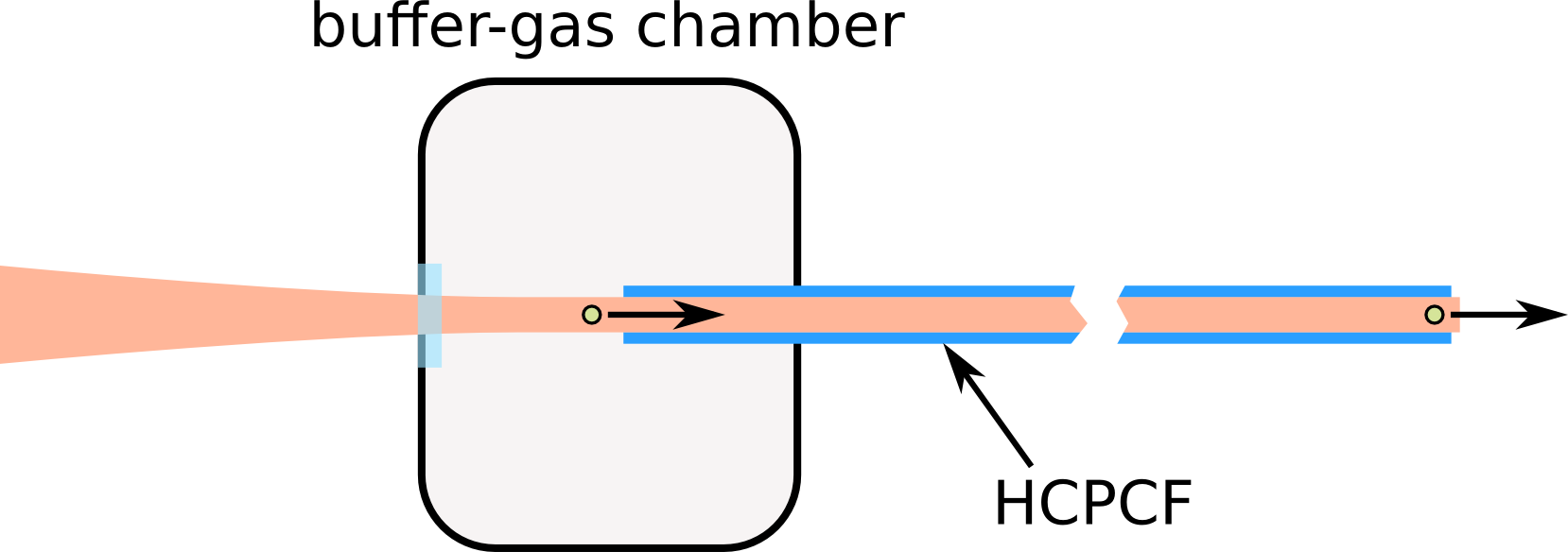}
  \caption{\textbf{Transfer of particles through hollow-core photonic-crystal fiber (HCPCF).} After loading and characterization of a nanoparticle (indicated by a green dot) inside a buffer-gas chamber, the particle is optically guided into and along a HCPCF outside the spacecraft.\label{fig::buffergas}}
  \end{center}
\end{figure}

In the initial phase of payload development, we will investigate those methods and alternative methods for directly loading optical traps in UHV. The latter methods rely on using ultra-sonic vibrations of a carrier substrate to desorb nanoparticles from the surface. In particular, we will investigate the use of GHz surface-acoustic waves on piezoelectric materials and the use of MHz bulk vibrations in thin-rod piezoelectric materials. After these initial studies, the most promising of the technologies or a combination thereof will be chosen and adapted for use in MAQRO.

Common to most approaches will be the initial optical trapping of particles inside the spacecraft in a buffer-gas environment (see Figure~\ref{fig::buffergas}). While the most natural choice for the buffer gas is Helium because it remains in gas phase even at the low temperatures at the optical bench, this choice will need to be investigated more closely in the initial development phase of the loading mechanism. Gas will only be supplied to the chambers after commissioning.

The initial optical trap will be used to characterize the particles trapped in order to quantify the size and mass of the particle as well as the charge it carries. Then the particle will be guided outside the spacecraft via a combination of hollow-core fibres and linear Paul trapping. The Paul trapping is necessary to guide and additionally constrain the particle trajectories without having to use too high optical powers. Strong beam intensities would prevent the particles from sympathetic cooling in the presence of the buffer gas. Using amplitude and frequency modulation of the guiding beam, we can shuttle and radially cool the particle (paper in preparation). The linear Paul trap can be realized via four rod-like electrodes encompassing the HCPCF.

As we described in section~\ref{subsec::scireq:decoherence}, the test particles are required to have a low internal temperature. Inside the MAQRO spacecraft, we do not have the means to cool the particle temperature to that degree. Instead, our approach is to sympathetically cool the particles using the buffer gas. While the buffer gas itself will be approximately at room temperature inside the spacecraft, the gas will quickly cool as it passes along the HCPCF outside the spacecraft. As the HCPCF approaches the optical bench, we expect the temperature of the buffer gas to eventually assume the environment temperature in that region ($\le 25\,$K).

In order to avoid confusing different particle materials, there should be at least one buffer-gas loading chamber per particle material to be used. In a given chamber, there may, however, be various sizes of nanoparticles. Before the particles are guided to the experiment, their size will be determined by observing the light scattered from the particles \& by measuring the mechanical frequency of the trapped particles in the hybrid optical + Paul trap. We estimate of the loading mechanism to be \textbf{TRL~3}.

\subsubsection{Accelerometer}
\label{subsubsec::Inner:Acc}
\begin{figure}[htb]
 \begin{center}
  \includegraphics[width=0.4\linewidth]{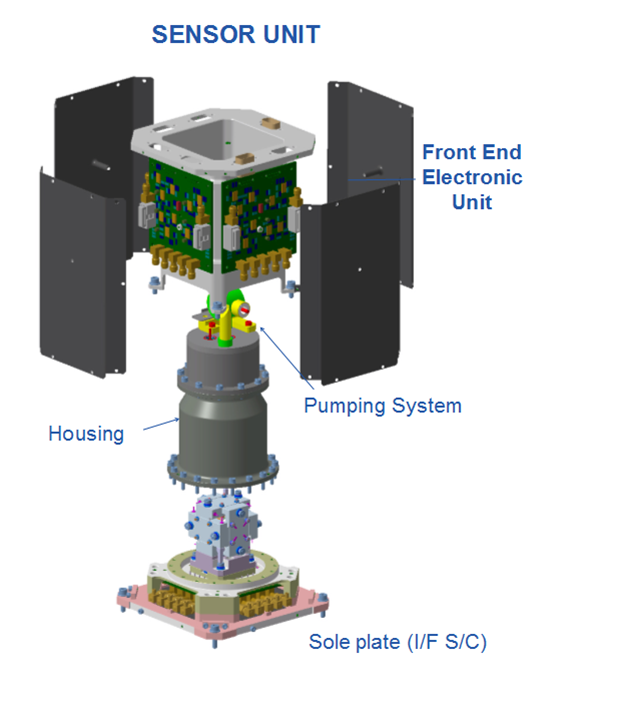}
  \caption{\textbf{Internal accelerometer sensor unit.} The internal sensor combines the sensor core and the control unit. Size: $20\times 20\times 20\,\mathrm{cm^3}$, mass: $8\,$kg. Image credit: ONERA.\label{fig::accInner}}
  \end{center}
\end{figure}

The payload of MAQRO contains a highly sensitive accelerometer positioned at the centre of mass of the spacecraft. While the main task of the outer accelerometer (see subsection~\ref{subsubsec::Outer:Acc}) is to monitor accelerations of the optical bench, the task of the inner accelerometer is to provide the necessary data to precisely control the micro-propulsion system of MAQRO for the drag-free and attitude-control system (DFACS).

Figure~\ref{fig::accInner} shows the accelerometer inside the MAQRO spacecraft, based on Ref.~\cite{Lenoir2011a}. The sensitivity of this sensor is $2\,\mathrm{(pm/s^2)/\sqrt{Hz}}$ at $0.01\,$Hz on a $4\times 10^{-6}\,\mathrm{m/s^2}$ range. Both accelerometers have an acquisition and a science mode. During the acquisition mode, higher accelerations are allowed during the time before the spacecraft enters science operation. Given the rich heritage of ONERA accelerometers, we estimate at least \textbf{TRL~5}.

\subsection{Critical Issues}
\label{subsec::instr:Critical}
Since the original proposal for the M3 opportunity\cite{Kaltenbaek2012b}, MAQRO has made significant progress towards maturing the payload technologies and concepts as well as to address critical issues. In the following, we will provide a list of critical issues, and how they will be addressed in MAQRO.

\subsubsection{Nanoparticle temperature}
\label{subsubsec::Crit:Ti}
As we described in section~\ref{subsec::scireq:decoherence}, the internal temperature of the test particle must not be much higher than the environment temperature. Since the test particles in MAQRO are optically trapped for state preparation and on other occasions, any realistic particle with non-zero absorption will heat. In the original proposal of MAQRO, we suggested to solve this issue by finding better materials. While this can definitely help reducing this problem, it is unlikely to fully solve this issue any time soon, and any solution would be very material specific while MAQRO should perform tests with a variety of nanosphere materials. For this reason, we chose a different approach for M4.

While we still propose using low-absorption materials, we plan to overcome this critical issue by a combination of several techniques: (1) using each particle only once and keep it optically trapped only for a short time, (2) use charged particles and a combination of optical trapping and Paul trapping or only Paul trapping whenever possible, and (3) use buffer gas in HCPCFs to sympathetically cool the test particles during transport.

Using this combination, we are confident that it will be possible to address and solve this issue.

\subsubsection{Preparation of macroscopic superpositions}
\label{subsubsec::Crit:superpos}
In the original proposal, we suggested using extreme UV with very low power but with a wavelength of only $30\,$nm to prepare the macroscopic superpositions needed to observe double-slit-type interference\cite{Kaltenbaek2012a,Kaltenbaek2012b}. Even for the low powers needed, this technology will not exist within a foreseeable time in space. In addition, this approach led to free-fall times well beyond $100\,$s, which poses another host of problems -- e.g., requirements on the thrusters \& limitations on the particle statistics achievable during the mission life time.

For M4, we fully revised this approach to use well established technology for observing matter-wave interference with massive particles. This approach does not only NOT require extreme UV light, it also brings the benefit of much shorter free-fall times and higher interference visibilities. See subsection~\ref{subsec::sci:DECIDE}.

\subsubsection{Loading mechanism}
\label{subsubsec::Crit:LM}
MAQRO puts exceedingly strict requirements on the mechanism for loading nanoparticles into the optical trap. While there are several candidate technologies in existence, none of them is directly adaptable to MAQRO. For this reason, the costs for payload development include the costs for an intense development phase.  During that phase, four candidate technologies will be closely investigated and experimentally tested. At the end, the best combination of these technologies will be implemented for MAQRO. Given recent developments related to the candidate technologies, we are confident that a solution similar to the one described in this proposal can be implemented within time for MAQRO.

\subsection{Operations and measurement technique}
\label{subsec::instr:Operations}
\begin{figure}[htb]
 \begin{center}
  \includegraphics[width=0.8\linewidth]{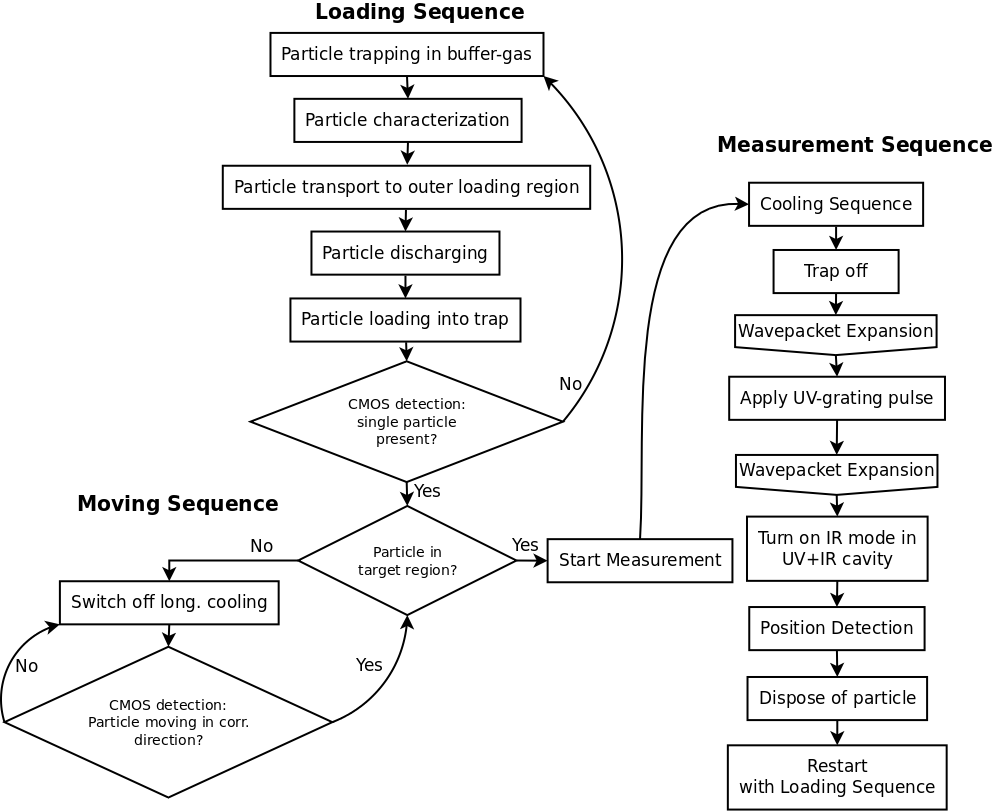}
  \caption{\textbf{Flow chart of the measurement procedure of MAQRO.} Details are described in the main text.\label{fig::measProcedure}}
  \end{center}
\end{figure}

Here, we will provide an overview of the science operations of MAQRO. The measurement sequence can roughly be divided in three distinct sequences: (1) the loading sequence, (2) a moving sequence for transporting a particle to the target region (the crossing point of the high-finesse IR cavity and the low-finesse UV+IR cavity, and (3) the actual measurement sequence. Figure~\ref{fig::measProcedure} provides a flow chart of the overall measurement procedure.

\subsubsection{Loading sequence}
\label{subsubsec::OP:DAQ}
As described in subsection~\ref{subsubsec::Inner:LM}, particles will initially be prepared in chambers flooded with buffer gas. In those buffer-gas chambers, a particle will be trapped and characterized. This can already be performed up front before a particle is needed for the experiment. Once this is accomplished, the particle is transferred to a loading region below the optical bench outside the spacecraft using optical transport in a HCPCF assisted by linear Paul trapping. This is described in subsection~\ref{subsubsec::Outer:LM}. Before the particle is loaded into the intra-cavity optical trap, it needs to be discharged. The transport to the optical trap operates in free space via radiation pressure. Monitoring via the CMOS camera on the optical bench allows verifying the successful completion of the loading sequence. If it was not successful, the whole procedure has to be repeated until successful. If it was successful, we turn on the side-band cooling along the cavity axis as well as the feed-back cooling for the transverse directions.

\subsubsection{Moving sequence}
\label{subsubsec::OP:LM}
Loading the particle into the optical trap this way does not guarantee that the particle will be at the correct position along the cavity mode of the high-finesse IR cavity. The correct position is defined via the crossing point of that cavity with the UV+IR cavity. The necessary accuracy of the positioning is determined by the requirement that it should be much better than the mode waist of the UV+IR cavity, that means, $\ll 1\,$mm.

By monitoring the scattered light with the CMOS imaging system, we can keep track of the particle position with $\mathrm{\mu m}$ resolution. To move the particle along the cavity axis, we simply turn off the cooling in this direction. The particle motion will heat due to laser noise and light scattering and, if needed, due to purposeful heating by frequency modulation at twice the longitudinal trap frequency. This heating will lead to the particle moving out of the trap along the cavity axis. By observing the CMOS signal, we can determine whether the particle moves in the correct direction. If not, we turn on longitudinal cooling again and restart the moving sequence. If the particle is moving in the correct direction, we only have to wait until it is at the correct position, and then switch on the longitudinal cooling once more.

\subsubsection{Measurement sequence}
\label{subsubsec::OP:Acc}
As soon as the particle is at the correct position, we can use three-dimensional intra-cavity cooling to cool the centre-of-mass motion of the particle close to the quantum ground state along the cavity axis and to low occupation numbers in the transverse directions\cite{Yin2011a}. Also see the scientific requirements in Table~\ref{tab::scireq}. When the cooling sequence is completed, all optical fields are switched off, and the wavepacket will expand for a time $t_1$, which is chosen depending on the nanoparticle and the phase $\phi_0$ that will be applied. The next step is to turn on the UV beam for a time $\sim 1\,\mathrm{\mu s}$ to apply the pure phase grating. After applying the phase grating, the particle will again propagate freely for a time $t_2 = T-t_1$. Finally, the IR field in the UV+IR cavity is switched on in order to measure the position of the test particle via cavity readout.

After completing the measurement, the particle is no longer needed, and an IR beam orthogonal to the optical bench is applied to propel the particle away from the spacecraft and into space to prevent contamination of the scientific instrument with stray nanoparticles.

\section{Proposed mission configuration and profile}
\label{sec::conf}
From the scientific requirements (section~\ref{sec::scireq}, Table~\ref{tab::scireq}) it is apparent that MAQRO requires extremely high vacuum conditions, cryogenic temperatures (realized via passive cooling) and very stringent microgravity requirements. A mission to L1/L2 is ideally suited and allows fulfilling these requirements.

\subsection{Orbit requirements}
\label{subsec::conf:orbitreq}
\begin{figure}[htb]
 \begin{center}
  \includegraphics[width=0.6\linewidth]{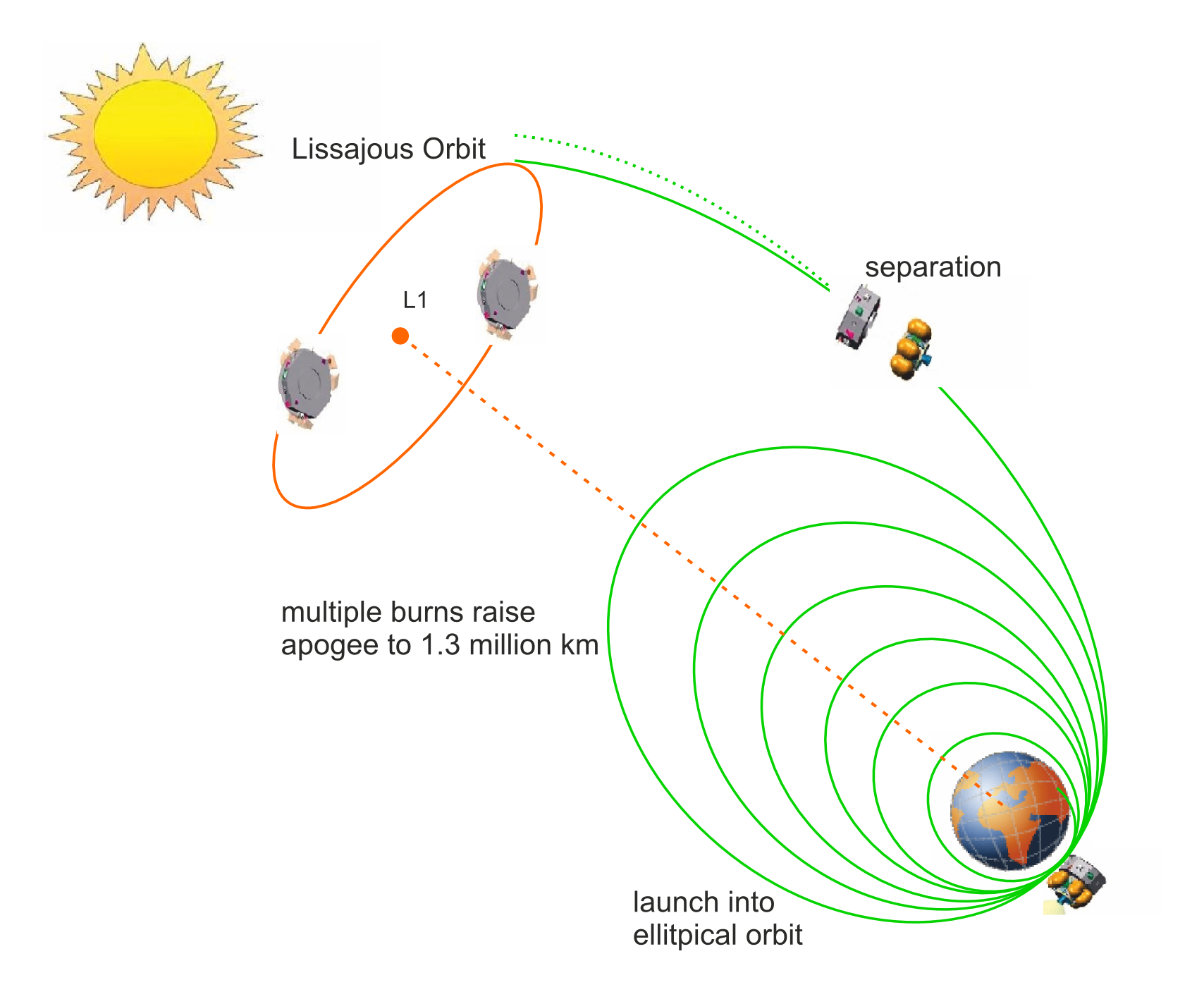}
  \includegraphics[width=0.3\linewidth]{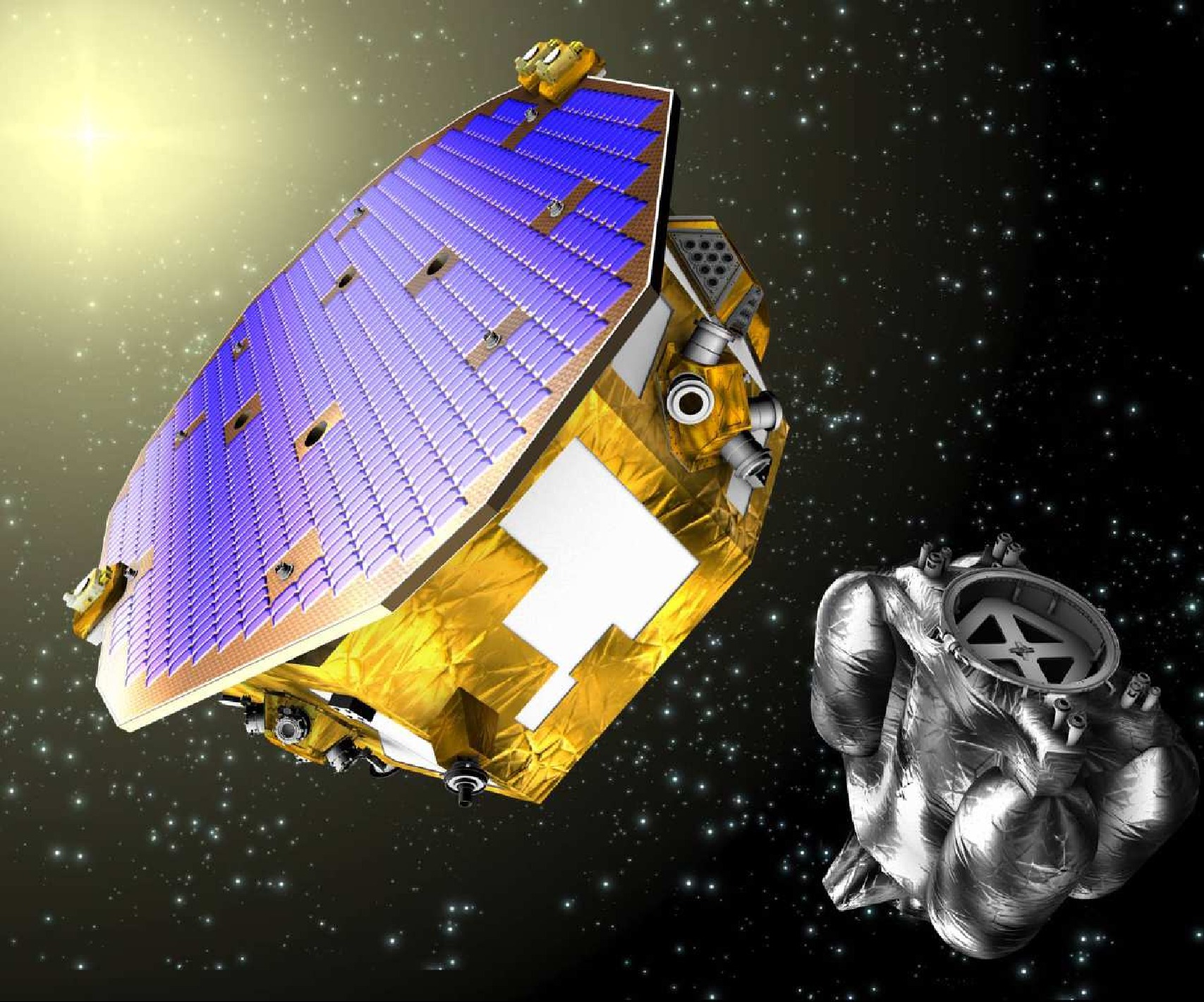}
  \caption{\textbf{(left)} Sketch of the transfer to a halo orbit around L1. \textbf{(right)} Artist's impression of the LPF spacecraft separating from the propulsion module (image credit: ESA).\label{fig::orbit}}
  \end{center}
\end{figure}

Following LISA Pathfinder's example, the MAQRO space-craft is injected into a halo orbit around the sun/earth Lagrange point L1 (L2 would be a feasible alternative), following the initial injection into elliptical earth orbit and 8 apogee raising orbits. For an orbit around L2, similar considerations are applicable. This configuration corresponds to the Vega mission scenario for an L1/L2 orbit given in the call annex.

\subsection{Alternative orbits}
\label{subsec::conf:altorbit}
\begin{figure}[htb]
 \begin{center}
  \includegraphics[width=0.5\linewidth]{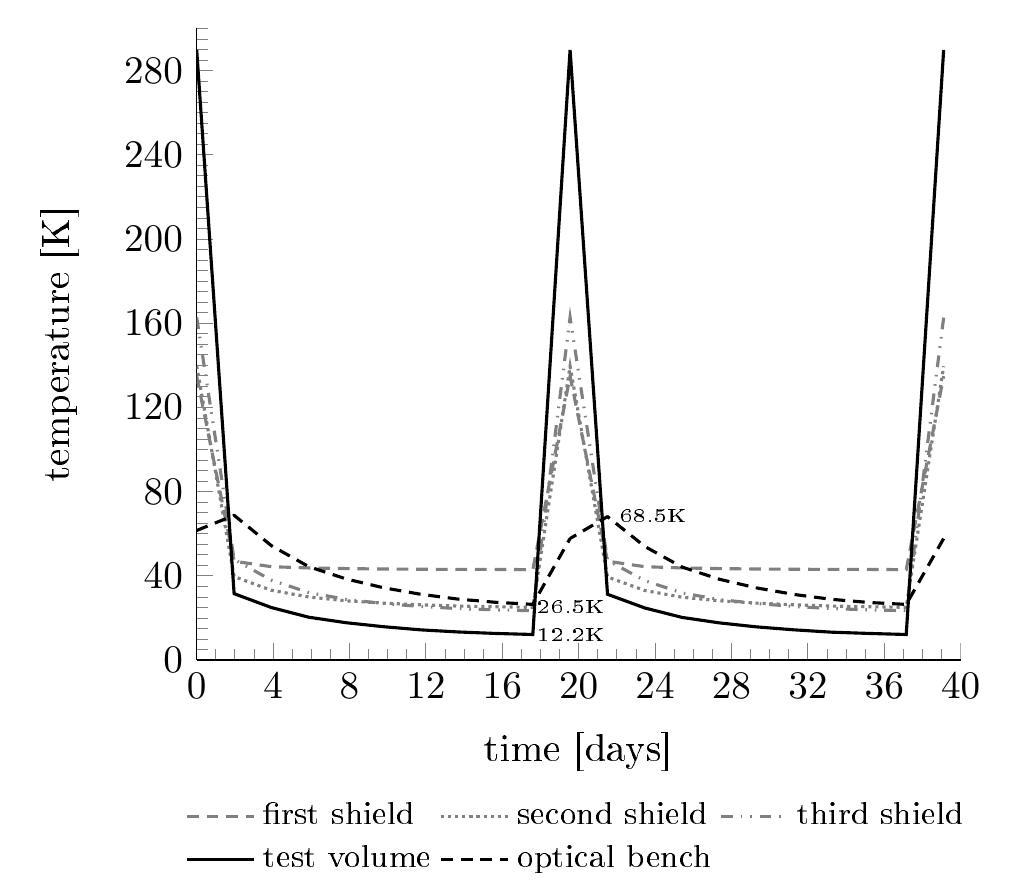}
  \caption{\textbf{Temperature of the heat-shield structure in HEO configuration.} Close to perigee, the thermal-shield structure heats up and then needs more than two weeks to cool down again to fulfil the requirements of MAQRO. Figure from Ref.~\cite{Pilan-Zanoni2015a}.\label{fig::HEOtemp}}
  \end{center}
\end{figure}

For the original M3 proposal of MAQRO\cite{Kaltenbaek2012b}, we investigated the possibility of using a HEO. More recent investigations showed\cite{Pilan-Zanoni2015a} that a HEO is no feasible alternative. Apart from possible issues with repeatedly crossing the van-Allen belt, a main issue for MAQRO would be thermal considerations. Figure~\ref{fig::HEOtemp} shows results reported in Ref.~\cite{Pilan-Zanoni2015a} for the heat-shield temperature over time in the course of orbital evolution. These results show that there would only be a short time window during which the optical bench reaches a temperature compatible with the requirements of MAQRO. The acquisition of a full interferogram would therefore take about 10 to 30 orbits increasing the necessary mission life time to more than 10 years.

A feasible alternative to an L1/L2 orbit would be the orbit suggested for the ASTROD I mission proposal although it would have to be investigated in more detail how the necessary pointing of the optical telescope of ASTROD I would influence the performance of MAQRO.

\subsection{Mission lifetime}
\label{subsec::conf:lifetime}
\begin{figure}[htb]
 \begin{center}
  \includegraphics[width=0.5\linewidth]{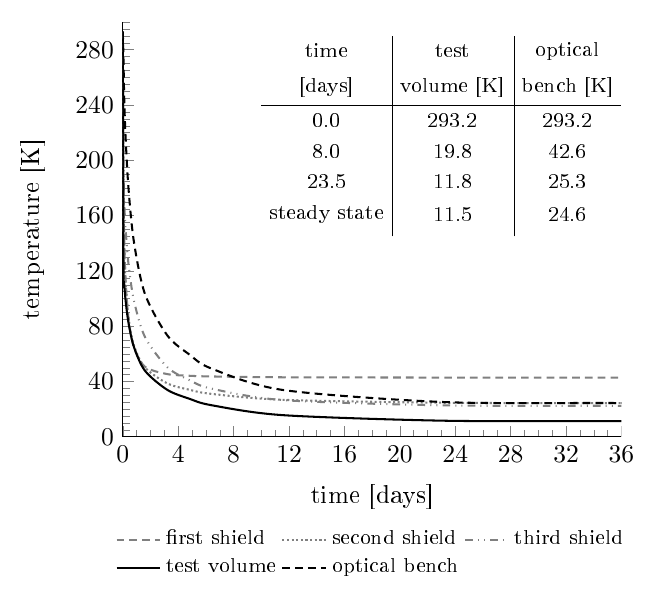}
  \caption{\textbf{Cool-down of parts of the thermal-shield structure over time.} Starting from room temperature, the optical bench reaches steady state after about 25 days. Due to its small volume and low heat capacity, the test volume temperature drops more rapidly. Figure from Ref.~\cite{Pilan-Zanoni2015a}.\label{fig::cooldown}}
  \end{center}
\end{figure}

The total mission time will be 24 months. Multiple burns (9 for Vega and 15 for Rockot) raise the apogee to $1.3\times 10^6\,$km during 15 days. The following transfer to L1 ($1.5\times 10^6\,$km from earth) takes 30 days. After science payload commissioning (including an optional bake-out), the heat-shield structure and the optical bench will need to passively cool for about 25 days to reach operating temperature (see Figure~\ref{fig::cooldown}). After that, the science operation is scheduled to last for $\sim 21$ months yielding an overall lifetime of 24 months. The ultimate upper limit on the mission life time will be determined by the amount of fuel available for the cold-gas micro-thrusters as well as by the amount of buffer gas and test particles available for performing experiments.

\subsection{System requirements and spacecraft key issues}
\label{subsec::conf:sys}

\subsubsection{Payload mass budget}
\label{subsubsec::sys:mass}
\begin{table}[hbt]
\begin{tabular}{lllll}
  \hline
  \textbf{Launch composite} & \textbf{LPF mass (kg)} & \textbf{CBE MAQRO mass (kg)} & \textbf{Maturity margin} & \textbf{CBE + margin} \\\hline
  Payload & 178 & 100 (+5) & 30\% & 131 (+8)\\
  Science spacecraft (wet, w.o payload) & 257 & 278 & 5\% & 292\\
  Propulsion module (dry) & 214 & 214 & 5\% & 225\\
  Launch composite dry total & 649 & 592 & & 648 (+8)\\
  Consumables & 1250 & 1250 & & 1250\\
  Launch composite wet total & 1899 & 1842 (+5) & & 1898 (+8) \\\hline
\end{tabular}
\caption{\textbf{Total mass budget of MAQRO compared to LPF.} \label{tab::massBudget}}
\end{table}

The MAQRO mass budget is closely based on LPF. As the space-craft platform of MAQRO is identical to the one of LISA Pathfinder, we shall focus on the MAQRO payload and compare it to LTP, the LISA Pathfinder payload. It is apparent that by omitting the heavy inertial sensor from LPF, the payload mass of MAQRO is dramatically reduced, however, several modification have to be taken into account. Table~\ref{tab::massBudgetDetails} shows a detailed list of the mass budgets for the MAQRO payload, and Table~\ref{tab::massBudget} shows the total mass.

\begin{table}[hbt]
\begin{tabular}{llllr}
  \hline
  \textbf{Items \& units} & \textbf{CBE mass (kg)} & \textbf{Maturity margin} & \textbf{CBE + margin} \\\hline
  Accelerometer sensor units & 4 & 5\% & 4.20\\
  Accelerometer control units & 12 & 5\% & 12.60\\
  Power control unit & 3 & 5\% & 3.15\\
  Heat shield (incl. struts, inserts and & \multirow{2}{*}{10} & \multirow{2}{*}{50\%} & \multirow{2}{*}{15.00} \\
  launch locks, protective cover and margin & & & \\
  Shield baking: heater + power unit (optional) & (+5) & 50\% & (+7.50)\\
  Optical bench & 6 & 30\% & 7.80\\
  CMOS readout electronics & 2 & 100\% & 6.00\\
  IR spatial mode generation and filtering & 3 & 100\% & 6.00\\
  Buffer-gas chambers + gas + loading & 20 & 100\% & 40.00\\
  Laser assembly (IR) & 12 & 10\% & 13.20\\
  Laser assembly (UV) & 6 & 100\% & 12.00\\
  Phase-meter & 4 & 25\% & 5.00\\
  Payload processor (DMU) & 8 & 10\% & 8.80\\
  Diagnostic elements (sensors, \ldots) & 10 & 50\% & 15.00\\
  Assembly \& interface equipment & 7 & 20\% & 8.40\\
  Harness & 13 & 20\% & 15.60\\\hline
  \textbf{Total:} & 100 (+5) & \textbf{30\% avg.} & 131 (+8) \\\hline
\end{tabular}
\caption{\textbf{Total mass budget of MAQRO compared to LPF.} \label{tab::massBudgetDetails}}
\end{table}

For the spacecraft, we will assume the same mass as for LPF, including $51\,$kg for the cold-gas micro-propulsion system. We add an additional $21\,$kg of additional fuel to the spacecraft mass budget to account for the longer life time of MAQRO compared to LPF. Given these estimates the overall mass of MAQRO including generous margins is nearly identical to the mass of LPF. However, it may be possible to reduce the mass budget of MAQRO by removing or simplifying the disturbance reduction system (DRS) of LPF in the case of MAQRO.

\subsubsection{Power budget}
\label{subsubsec::sys:power}
\begin{table}[hbt]
\begin{tabular}{lrllrlrlr}
  \hline
  &&\hspace{0.2cm}&& \textbf{Power} &\hspace{0.2cm}& \textbf{Maturity} &\hspace{0.2cm}& \textbf{CBE +} \\
  \textbf{Items \& units LTP} & \textbf{Power (W)} & & \textbf{Items \& units LTP} & \textbf{CBE (W)} & & \textbf{margin} & & \textbf{margin} \\
  \hline
  Inertial sensor FEE & 40 & & ONERA accelerometers (total) & 20 & & 5\% & & 21.00\\
  Charge management (UV-lamps) & 8 & & & & & & & \\
  Data and diagnostic & 30 & & Data and diagnostic & 30 & & 10\% & & 33.00\\
  Phase-meter & 18 & & Phase-meter + CMOS camera & 18 & & 25\% & & 22.50\\
  Laser assembly IR & 45 & & Laser assembly IR & 45 & & 10\% & & 49.50\\
   &  & & Laser assembly UV & 20 & & 25\% & & 25.00\\
   &  & & Voltage supply for Paul traps & 10 & & 25\% & & 12.50\\\hline
  \textbf{Total (science mode)} & \textbf{141} & & \textbf{Total (science mode)} & \textbf{143} & & \textbf{avg. 15\%} & & \textbf{166.00} \\\hline
  Total (maximal) & 163 & & Total (maximal) & 143 & & 15\% & & 166.00\\
  Total (minimal, only DMU) & 30 & & Total (minimal, only DMU) & 30 & & 10\% & & 33.00\\
  & & & Optional heater for shield & 242 & & 20\% & & 290.40\\\hline
  & & & \textbf{Total (DMU + heater)} & \textbf{272} & & \textbf{avg. 15\%} & & 323.40\\\hline
\end{tabular}
\caption{\textbf{Power budget for MAQRO.} We distinguish between two operation modes: normal operation (science mode) and operation during the optional baking-out of the heat-shield structure. \label{tab::powerBudgetDetails}}
\end{table}

In Table~\ref{tab::powerBudgetDetails}, the power budget of the MAQRO payload is compared to LTP to demonstrate that the power requirements are essentially the same. The power requirements for other parts of the science space-craft are not listed as they are assumed to be identical. It is therefore possible to conclude that the Pathfinder solar array of $\sim 680\,$W of Pathfinder is sufficient for the needs of MAQRO.

Note that a bake-out mechanism for the outermost heat shield (+optical bench) can optionally be included for MAQRO. The heater requires $\sim 242\,$W for bake-out at $300\,$K. Before commissioning, LTP and MAQRO only require $30\,$W. Nevertheless, this high power may render it unfeasible to perform a bake out unless the necessary power can temporarily be allocated from the science spacecraft. This option will be investigated more closely in the future.

\begin{table}[hbt]
\begin{tabular}{lll}
  \hline
  \textbf{Payload power required} & \textbf{LTP} & \textbf{MAQRO} \\\hline
  Payload, maximal & 163 & 166 \\\hline
  \textbf{Total power required} & \textbf{LPF} & \textbf{MAQRO} \\\hline
  Spacecraft in transfer orbit & 638 & 638 \\
  Spacecraft in science mode & 613 & 638 \\\hline
\end{tabular}
\caption{\textbf{Overview of the total power budget for LTP and MAQRO.} \label{tab::powerBudget}}
\end{table}

The comparison of the total power budget in Table~\ref{tab::powerBudget} shows that the maximum power consumption of MAQRO in science mode is identical to the power consumption of the spacecraft in transfer orbit. The slight increase in power consumption of MAQRO in science mode with respect to LISA Pathfinder in science mode should be possible using the $680\,$W power supplied by the solar array. If necessary, some of the equipment can be turned off during the long free-fall times.

\subsubsection{Link budget}
\label{subsubsec::sys:link}
Communication for MAQRO will be on X-band using low gain hemispherical and medium gain horn antennas, just as in Pathfinder. A communication bandwidth of $60\,$kbps fulfils the downlink bandwidth requirements for MAQRO. Therefore $\sim 6\,$W of transmitted RF-power are sufficient to establish the required downlink rate for on-station nominal operation. As in Pathfinder, it is suggested to use the $35\,$m antenna of the ground station Cebreros in Spain. Table~\ref{tab::linkBudget} provides an overview of the link budget.

\begin{table}[hbt]
\begin{tabular}{lllllllll}
  \hline
  & & \textbf{Downlink} & \hspace{0.2cm} & \textbf{Nominal TM} & \hspace{0.2cm} & \textbf{Uplink} & \hspace{0.2cm} & \textbf{Nominal TC} \\
  \textbf{Operation mode} & \textbf{Antenna} & \textbf{rate (ks/s)} & & \textbf{margin (dB)} & & \textbf{rate (ks/s)} & & \textbf{margin (dB)} \\\hline
  On station nominal & Medium gain & 120 & & 10 & & 2 & & 38 \\
  On station emergency & Low gain & 1 & & 10 & & 2 & & 20 \\  
  LEO phase nominal & Low gain & 120 & & 36 & & 2 & & 71 \\
  LEO phase w/c range & Low gain & 120 & & 16 & & 2 & & 44 \\\hline
\end{tabular}
\caption{\textbf{Overview of the total power budget for LTP and MAQRO.} \label{tab::linkBudget}}
\end{table}

\subsubsection{Spacecraft thermal design}
\label{subsubsec::sys:thermalDesign}
Part of this will be standard thermal-control tasks, to keep the overall S/C and its external and internal units \& equipment within the allowable temperature ranges by a proper thermal balance between isolating and radiating outer surfaces, supported by active control elements such as heaters. This will be based on LPF heritage. In addition, for the MAQRO mission, the thermal design has to focus on a proper thermal I/F design from the warm S/C to the extremely cold external subsystem, the heat-shield structure. Because the heat-shield structure is supported by an already very stable S/C and because it is coupled well to the extremely stable $3\,$K environment of deep space, the heat-shield structure can be kept at an extremely stable temperature.
To achieve a good thermal stability for the equipment inside the S/C, similarly to the LISA Pathfinder S/C, the MAQRO S/C internal dissipation fluctuations have to be minimized and the S/C interior has to be isolated from the solar array because it inherently introduces solar fluctuations into the S/C. In order to achieve the required $\le 25\,$K for the optical bench, the (warm) mechanical I/F should be designed as cold as possible, e.g., $270\,$K, and the S/C surfaces facing towards the external payload should be covered by a high-efficient multi-layer insulation (20 layers) and the outermost layer should have a high emissivity $> 0.8$. This measure will serve for radiative pre-cooling of the outer thermal shield of the payload.

\subsubsection{Attitude and orbit control}
\label{subsubsec::sys:attitude}
Star trackers and Solar sensor are used to determine the attitude. The cold-gas thrusters are exclusively used for attitude control after the propulsion module has been ejected, i.e., there are no reaction wheels. 
The attitude and control system (AOCS) for the science module is used whenever no science activity is carried out. It is referred to as MPACS on LPF, and can be used in a similar way for MAQRO. Likewise, a similar (or possibly even simplified) version of the drag-free attitude and control system (DFACS) can be adapted from the LISA Pathfinder concept.

\subsubsection{Redundancy considerations}
\label{subsubsec::sys:redund}
Because the spacecraft is identical to LPF, we profit from the redundancy scheme of LPF. For example, the thrusters are operated in hot redundancy, and also the IR laser diodes feature redundancy. For MAQRO, we will include multiple buffer-gas tanks, each with at least one HCPCF leading to the outer loading mechanism (see sections \ref{subsubsec::Outer:LM} and \ref{subsubsec::Inner:LM}), and we will use redundant UV diodes based on the idea of redundant pump diodes in LTP. For the two cavities on the optical bench, the role of input and output can be exchanged as a form of cold redundancy. In addition, we will always guide two UV hollow-core fibres in parallel to the optical bench. For the purpose of the UV grating and for the discharging mechanism, the small displacement between the two fibres is negligible. 

\subsubsection{Vacuum requirements}
\label{subsubsec::sys:vacuum}
These have been discussed in detail in the M3 mission proposal of MAQRO and in the corresponding published version\cite{Kaltenbaek2012b}. There, we showed that the vacuum requirements on the optical bench can be fulfilled on the external platform. The conclusion was that the low temperature of the optical bench will essentially lead to freezing out of outgassing processes. Given the optimized design of the thermal shield and optical bench\cite{Hechenblaikner2014a,Pilan-Zanoni2015a}, the temperature is even a bit lower than assumed in Ref.~\cite{Kaltenbaek2012b}. 

Outgassing from other, hotter parts of the spacecraft will not affect the experimental region because no part of the spacecraft is in the direct field of view from the optical bench. Particles outgassing from hotter regions of the spacecraft will have high enough velocities to overcome the gravitational attraction of the spacecraft.

This leaves us with three effects that may affect the collision rate of the test particle with residual gas or other particles:
\begin{itemize}
\setlength{\itemsep}{0pt}
\setlength{\parskip}{0pt}
\item \textbf{Interplanetary particle density}\\
Around L1/L2, this should readily be compatible with the requirements of MAQRO, i.e., the particle density should be $\le 500\,\mathrm{cm^{-3}}$\cite{Biermann1957a}.
\item \textbf{Solar wind}\\
At 1 AU, we expect a particle density of $\sim 10\,\mathrm{cm^{-3}}$ with velocities $\le 500\,\mathrm{km/s}$. Because the spacecraft will partially shield the solar wind, the particle density will be even less. If we assume $\sim 1\,\mathrm{cm^{-3}}$, the conservative limit given in Figure~\ref{fig::solarwind} shows that this is within the MAQRO requirements.
\item \textbf{Leakage of buffer gas to the experimental region}\\
Using venting ducts as shown in Figure~\ref{fig::LMouterBottom}, it should be possible to keep the amount of buffer gas reaching the experimental region within requirements. This will have to be investigated in more detail in the future.
\end{itemize}

\subsubsection{Heat-shield structure}
\label{subsubsec::sys:shield}
A detailed discussion of the thermal considerations for the thermal-shield structure is given in section~\ref{subsubsec::Outer:Thermal} and in Refs.~\cite{Hechenblaikner2014a,Pilan-Zanoni2015a}. Here, we will focus on estimating the mass of the structure. 

In order to conservatively estimate the mass of the shield structure, we will assume that the shields extend even a bit further than detailed in section~\ref{subsubsec::Outer:Thermal}. In particular, we assume that they extend far enough to cover the shield even from radiation from the sun at an angle of 45 degrees. The shields' diameters will still be smaller than the spacecraft diameter. In Ref.~\cite{Pilan-Zanoni2015a}, we even investigated the case where the shields extend beyond the spacecraft and are exposed to direct radiation from the sun. Given appropriate coating, even this extreme case should be possible.

Under this assumption, and assuming that the points of the conical shields are $10\,$cm, $15\,$cm and $20\,$cm distant from the spacecraft and have opening angles of $7.5^\circ$, $15^\circ$ and $22.5^\circ$, the areas of the three shields are $0.9\,\mathrm{m^2}$, $0.6\,\mathrm{m^2}$ and $0.4\,\mathrm{m^2}$. For an estimate of the mass, let us assume the specific density of Aluminium and a thickness of $1\,$mm for the shields. Then the sum mass of the three shields is $m_\mathrm{layers} \approx 5\,$kg. The hollow struts are made from carbon-fibre reinforced plastics (CFRP) of very low thermal conductivity and expansion, as well as good mechanical stability. They are $40\,$cm long, $2\,$cm in diameter and have a wall thickness of $1.6\,$mm, giving a combined weight of less than one $1\,$kg: $m_\mathrm{struts} \approx 0.6\,$kg. The struts are fitted to the bushings inserted into the base-plate of the optical bench. Each of the three inserts has approximately $0.2\,$kg of weight giving a total of $m_\mathrm{inserts} \approx 0.6\,$kg. Assuming a slightly higher weight $m_\mathrm{mount} \approx 1\,$kg for mounting the struts to the spacecraft, we get an overall mass of $m_\mathrm{shield} \approx 7\,$kg for the thermal-shield structure (without the optical bench and harness).

\subsubsection{Protective cover \& shield bake out}
\label{subsubsec::sys:cover}
During transfer to L1 and before ejection of the propulsion module the thermal shield is covered by an additional protective cover. The weight of the cover is estimated to be $5\,$kg, based on an aluminium cylinder with $1\,$m diameter, $0.16\,$mm wall thickness and a height of $0.5\,$m.
 
Vacuum quality and outgassing is a key aspect of MAQRO. From our analysis in Ref.~\cite{Kaltenbaek2012b}, we found that outgassing is practically completely frozen out at temperatures as low as $\sim 30\,$K. Nevertheless, mainly as a means of risk mitigation for as yet unaccounted effects, it would be very useful to consider bake-out of the thermal shield and the exterior optical bench before commissioning. For that purpose, heaters could be attached to the outermost shield and the optical bench. Considering that the outer shield area is approximately $\sim 0.43\,\mathrm{m^2}$ and that the effective area of the optical bench is $\sim 0.1\,\mathrm{m^2}$, we obtain a total radiative surface of $\sim 0.53\,\mathrm{m^2}$ with an emissivity close to $1$. This requires a heating power of $P\approx 242\,$ W if we bake-out at $300\,$K and a heating power of $P\sim 759\,$W if we bake-out at $400\,$K. The latter is not possible given the solar array of MAQRO. Even baking out at $300\,$K takes a vast amount of power. A solution may be to use smaller shields as originally proposed in the M3 proposal or to temporarily allocate power from the science spacecraft during the bake-out procedure.

\subsection{Communication, mass-data storage \& ground segment}
\label{subsec::comm}
\begin{table}[hbt]
\begin{tabular}{lll}
  \hline
  \textbf{Mission requirement for} & \hspace{0.4cm} & \textbf{Suggested solution} \\\hline
  Launcher & & Vega  \\
  Spacecraft platform & & LPF platform with science spacecraft and  \\
  & & propulsion module \\
  Mission lifetime & & 24 months \\
  Communication & & 60 kbps TM, 2 kbps TC,\\
  & & communication for 8 hrs/day \\
  Mass data storage on-board & & Solid state mass memory (SSMM) of 1 GB  \\
  Ground-segment assumptions & & Cerebros, Spain (35 m) \\\hline
\end{tabular}
\caption{\textbf{Main mission requirments.} \label{tab::commBudget}}
\end{table}

A communication window of 8 hours per day, as in the Pathfinder mission, is sufficient to transfer science data to ground. Data are received by the $35\,$m Cerebros antenna and transferred to ESOC for further processing. Considering a maximal rate of $20\,\mathrm{kbit/s}$ of science and attitude control data during experimental runs, the data recorded during 24 hours of science runs can be transferred to ground at $60\,\mathrm{kbit/s}$ in the 8 hour communication window each day. The on-board computer architecture should provide the means to continuously store science data for a period of up to three days in a solid state mass memory (SSMM), which implies a minimal capacity of $\sim 650\,$Mbytes, which is easily achieved with any modern mass memory (capacity up to 2 TeraBit).
A brief overview over the main requirements for the mission profile is given in Table~\ref{tab::commBudget}.

\subsection{Science operations \& archiving}
\label{subsec::OPnArch}
Data for MAQRO are received by the $35\,$m Cerebros antenna in Spain and then routed to the European Space Operations Center (ESOC) in Darmstadt. The mission operations center (MOC) there ensures that the spacecraft meets its mission objectives, and it operates and maintains the necessary ground segment infrastructure. 
Because of the L1/L2 orbit, there will only be 8h of ground station contact per day at a downlink rate of $60\,$kbps. The payload is commanded via Payload Operation Requests (POR) stored in the mission timeline. Real-time commanding only occurs during commissioning and contingency events. 

The Science \& Technology Operations Center (STOC), located in Madrid, is responsible for the planning of the payload operations, data analysis, and mission archive. Scientific advisers and investigators will collaborate with the core STOC team.

Volume requirements for data archiving and distribution are rather low for MAQRO. The total data received over 2 years is estimated to be well below $1\,$TB, including diagnostic and house-keeping data.

\subsection{Mission phases}
\label{subsec::miss}

\subsubsection{Launch}
\label{subsubsec::miss:launch}
The spacecraft is injected into a low orbit by the Launch Vehicle. Separation from the upper stage may occur in sunlight or eclipse. Following separation, the Chemical Propulsion Subsystem initializes, following an initialisation sequence controlled by On Board Software (OBSW). During this period, which may partly be in eclipse, there is no attitude control, and the spacecraft tumbles uncontrolled, with power mainly or solely from the battery (a $600\,$Wh battery fulfils the needs of MAQRO). Once sensors and actuators are available, a transition to Sun Acquisition mode is autonomously performed. After the initial injection into elliptical earth orbit, the propulsion module is used to transfer the spacecraft to L1 via 8 apogee raising orbits.

\subsubsection{Commissioning}
\label{subsubsec::miss:comm}
Shortly before reaching the final on-station orbit around L1, the Propulsion Module (PRM) and the protective cover of the thermal-shield structure are separated from the Science Module (SCM). After separation, the spacecraft is spin-stabilised sun pointing. The nominal attitude profile is maintained using the micro-propulsion subsystems. Based on technological heritage from LISA Pathfinder, Microscope and GAIA, MAQRO will use cold-gas thrusters providing up to $100\,\mathrm{\mu N}$ variable thrust for the full mission life time.

\subsubsection{Passive cooling \& calibration}
\label{subsubsec::miss:cool}
Directly after commissioning, when the protective cover over the thermal-shield structure is removed, the structure will start to passively cool via radiation to deep space. This cooling period takes about 25 days (see Figure~\ref{fig::cooldown}). This time can, at the same time, be used for testing and calibration. In particular, we can perform tests of the following components:
\begin{itemize}
\setlength{\itemsep}{0pt}
\setlength{\parskip}{0pt}
\item IR and UV laser systems
\item Locking the cavities
\item CMOS system
\item Internal and external accelerometers
\item Loading and characterizing nanoparticles in the buffer-gas chambers
\item Use accelerometers to measure possible acceleration due to gas leakage
\item Test runs of switching combinations of thrusters on and off, influence on spacecraft attitude
\item Transferring particles to optical bench
\item Discharging particles
\item Loading particles into optical trap
\item Measuring particle position
\item Releasing and recapturing particle
\item Application of UV phase-grating on particle
\item Disposing of nanoparticles
\item Monitor development of heat-shield temperature over time, determine cooling rates, compare with simulations
\end{itemize}

Once enough time has passed to achieve the operating temperature, and after the initial tests are completed, MAQRO can start science operation.

\subsubsection{Science operation}
\label{subsubsec::miss:sciOP}
The first experiments to run on MAQRO will be to observe wave-packet expansion to determine the level of decoherence present in our system (see section~\ref{subsec::sci:WAX}). We will perform tests for at least 3 different particle materials of different mass density. For each particle type, we will perform the experiment with at least 5 different radii. All these tests, including possible repetitions should be completed within the first 10 months after commissioning.

If these first experiments demonstrate that everything works, and that decoherence present is small enough, we can switch to the second and most important stage of MAQRO: observing high-mass matter-wave interference (see section~\ref{subsec::sci:DECIDE}). If it should be clear already earlier that the prerequisites for performing these experiments are fulfilled, this second sequence of tests can already be started earlier on in the mission. The main goals of the mission should be achieved within the first 20 months after commissioning, leaving some time to perform additional experiments or to repeat experiments to increase statistical significance.

If the MAQRO instrument is still operating after the nominal mission life time and an extension of the life time is granted, additional experiments may be performed to increase the scientific output of the mission; for example, performing experiments on wave-packet expansion or high-mass interferometry repeatedly using the same test particle \& inferring the influence of particle heating and thermal radiation on the measurement results. Moreover, parameters can be varied in finer steps, or effects like micro-thruster noise on the measurement results can be investigated, and it would be possible to precisely determine the quantum state prepared by performing time-of-flight quantum-state tomography~\cite{RomeroIsart2011a}.

\subsubsection{Spacecraft disposal}
\label{subsubsec::miss:dispose}
In general, the halo orbits around L1/L2 are unstable and there is no direct need for spacecraft disposal. In order to enable a safe disposal of the spacecraft after the end of science operations and shortening the drift time, we can either use part of the mission lifetime and the corresponding fuel, or we can add some extra amount of fuel for a disposal after the nominal lifetime scheduled. This will have to be investigated in more detail in the definition phase of the mission.

\section{Conclusions \& Outlook}
We have presented an updated version of the proposal for a medium-sized space mission, MAQRO, originally proposed in 2010. This proposal was submitted in response to the ESA M4 call for a mission opportunity for a medium-size space mission. The main scientific objective of the mission is testing quantum theory using high-mass matter-wave interferometry in combination with novel techniques from quantum optomechanics. The update includes several significant changes with respect to the original proposal in order to address novel developments as well as critical issues in the original mission proposal. In particular, we presented an update of the thermal shield design allowing to perform high-mass matter-wave interference on a separate platform outside the spacecraft in order to fulfil the strict temperature and vacuum requirements of MAQRO. We introduced a novel type of matter-wave interferometer adapted for a microgravity setting as well as novel schemes for loading test particles into the central optical trap to meet the stringent requirements of MAQRO. 

This novel approach promises to overcome principal limitations of ground-based experiments and to resolve technical limitations of the earlier proposal by harnessing state-of-the-art space technology, well-established techniques of matter-wave interferometry and recent developments in quantum optomechanics using optically trapped dielectric particles. MAQRO will offer the unique opportunity to investigate a yet untested parameter regime allowing to probe for a quantum-to-classical transition and for possible novel effects at the interface between quantum and gravitational physics. Moreover, the high sensitivity of the MAQRO instrument might even allow testing a specific type of low-energy dark-matter models\cite{Riedel2013a,Bateman2015a}.

The present proposal highlights the rapid progress in recent years to achieve quantum control over macroscopic optomechanical systems and to harness space as an intriguing new environment for tests on the foundations of physics. MAQRO may prove a pathfinder for quantum technology in space, opening the door for a range of future applications in high-sensitivity measurements using techniques from quantum optomechanics and matter-wave interferometry.

\begin{acknowledgments}
AB acknowledges financial support from NANOQUESTFIT, INFN, and the COST Action MP1006. AR is supported by the German Space Agency (DLR) through Grant No.~DLR 50WM1136. LN acknowledges support through ERC-QMES (no. 338763). RK acknowledges support from the FFG Austrian Research Promotion Agency (no. 3589434).
\end{acknowledgments}

\bibliographystyle{apsrev4-1}
\bibliography{MAQRO_M4}  

\end{document}